\documentclass[preprint,10pt,authoryear]{elsarticle}

\usepackage{amssymb}
\usepackage{graphicx}
\usepackage{todo}
\usepackage{booktabs}
\usepackage{multirow}
\usepackage{tabularx}
\usepackage{subfig}
\usepackage[ruled,lined,linesnumbered]{algorithm2e}
\usepackage{amssymb}
\usepackage{amsmath}
\usepackage{url}
\usepackage{pbox}
\usepackage{adjustbox}
\usepackage{color, colortbl}
\usepackage{algorithmic}
\usepackage{rotating}
\usepackage[ruled,lined,linesnumbered]{algorithm2e}
\usepackage{tikz}
\usepackage{float}
\usepackage[roman]{parnotes}
\makeatletter
\def\parnoteclear{%
	\gdef\PN@text{}%
	\parnotereset
}
\makeatother

\def\checkmark{\tikz\fill[scale=0.4](0,.35) -- (.25,0) -- (1,.7) -- (.25,.15) -- cycle;}

\makeatletter
\newcommand*{\rom}[1]{\expandafter\@slowromancap\romannumeral #1@}
\makeatother

\definecolor{LightGray}{gray}{0.9}
\definecolor{MidGray}{gray}{0.7}
\definecolor{DarkGray}{gray}{0.5}

\makeatletter
\def\parnoteclear{%
	\gdef\PN@text{}%
	\parnotereset
}
\makeatother

\restylefloat{figure}

\newcommand{\specialcell}[2][c]{%
	\begin{tabular}[#1]{@{}c@{}}#2\end{tabular}}

\makeatletter
\renewcommand\paragraph{\@startsection{paragraph}{4}{\z@}%
	{-2.5ex\@plus -1ex \@minus -.25ex}%
	{1.25ex \@plus .25ex}%
	{\normalfont\normalsize\bfseries}}
\makeatother

\begin{document}

\begin{frontmatter}



\title{Graph Based Recommendations: From Data Representation to Feature Extraction and Application}


\author{Amit Tiroshi, Tsvi Kuflik, Shlomo Berkovsky, Mohamed Ali (Dali) Kaafar}

\address{}

\begin{abstract}
Modeling users for the purpose of identifying their preferences and then personalizing services on the basis of these models is a complex task, primarily due to the need to take into consideration various explicit and implicit signals, missing or uncertain information, contextual aspects, and more. In this study, a novel generic approach for uncovering latent preference patterns from user data is proposed and evaluated. The approach relies on representing the data using graphs, and then systematically extracting graph-based features and using them to enrich the original user models. The extracted features encapsulate complex relationships between users, items, and metadata. The enhanced user models can then serve as an input to any recommendation algorithm.
%In our case, the approach has been demonstrated using Random Forests, Gradient Boost and SVM.
The proposed approach is domain-independent (demonstrated on data from movies, music, and business recommender systems), and is evaluated using several state-of-the-art machine learning methods, on different recommendation tasks, and using different evaluation metrics. The results show a unanimous improvement in the recommendation accuracy across tasks and domains. In addition, the evaluation provides a deeper analysis regarding the performance of the approach in special scenarios, including high sparsity and variability of ratings.

\end{abstract}

\begin{keyword}

Recommender systems, graph-based recommendations, feature extraction, graph metrics.

\end{keyword}

\end{frontmatter}


%\textbf{Highlights}
%\begin{itemize}
%\item Generic framework for enhancing recommender systems through extracting features from a graph-based data representation.
%\item Thorough evaluation of the framework using a number of datasets, recommendation tasks, and machine learning mechanisms.
%\item Investigation of the impact of sub-graph selection and data representation of the accuracy of the recommendations.
%\end{itemize}

\section{Introduction}
\label{sec:introduction}
Recommender systems aim at helping users find relevant items among a large variety of possibilities, based on their preferences \citep{adomavicius2005toward}.
In many cases, these personal preferences are inferred from patterns that emerge from data about the users' past interactions with the system and with other users, as well as additional personal characteristics available from different sources. These patterns are typically user-specific and are based on the metadata of both the users and items, as well as on the interpretation of the observed user interactions \citep{kobsa2001generic,zukerman2001predictive}.
Eliciting user preferences is a challenging task because of issues such as changes in user preferences, contextual dependencies, privacy constraints, and practical data collection difficulties \citep{ricci2011introduction}. Moreover, the collected data may be incomplete, outdated, imprecise, or even completely inapplicable to the recommendation task at hand. In order to address these issues, modern recommender systems attempt to capture as much data as possible, and then, apply data mining and other inference techniques to elicit the desired preferences \cite{cantador2015cross}. Several techniques can be applied for the pattern-mining task, among which are techniques originated in machine learning and statistics, e.g., clustering and regression, or those that evolved in information retrieval and user modeling \citep{Mobasher07}.

Regardless of the technique exploited by a recommender system, it is inherently bound by the available user data and the features extracted/elicited from it. One major question that arises in this context is \textit{how to engineer\footnote{Feature engineering is sometimes also referred to in the literature as feature extraction, generation, and discovery, depending on the field of research. In this paper, it broadly refers to the task of adding new features to a dataset, regardless of the manner in which it is done (e.g., manual vs. automatic).} meaningful features from often noisy user data?} Features may be manually engineered by domain experts. This approach is considered expensive and non-scalable because of the deep domain knowledge that is necessary, the creativity required to conceive new features, and the time needed to populate and evaluate the contribution of the features. A notable example of this challenge is provided by the Netflix Prize winning team, in their recap: ``while major breakthroughs in the competition were achieved by uncovering new features underlying the data, those became rare and very hard to get'' \citep{koren2009bellkor}.

An alternative to manual feature engineering is automatic feature engineering, which is a major area of research in machine learning \citep{guyon2006feature}, particularly in the domains of image recognition \citep{nixon2008feature,due1996feature} and text classification \citep{scott1999feature}.
%
%The importance of automatic feature engineering has been recognized in the context of general purpose machine learning, as summarized in a recent paper: \textit{``At the end of the day, some machine learning projects succeed and some fail. What makes the difference? Easily the most important factor is the features used. ...Feature engineering is more difficult because it's domain-specific, while learners can be largely general-purpose. ...one of the holy grails of machine learning is to automate more and more of the feature engineering process. One way this is often done today is by automatically generating large numbers of candidate features and selecting the best by (say) their information gain with respect to the class. ...there is ultimately no replacement for the smarts you put into feature engineering.''}
%\citepp{domingos2012few}.
So far, automatic feature engineering has mainly focused on
either algebraic combinations of existing features, e.g.,
summation or averaging of existing features \citep{markovitch2002feature}, finding domain specific feature generators, e.g., for character recognition in image processing \citep{nixon2008feature,due1996feature}, or eliciting latent
features as in the SVD \citep{klema1980singular} and PCA \citep{wold1987principal} methods. The algebraic approaches for automatic feature engineering manage to produce large quantities of features; however, the relationships between the engineered features and the underlying patterns in the data are often not interpretable \citep{kotsiantis2006data}. For example, if averaging the ratings for items with the sum of some other arbitrary feature improves predictions, the reasons for this improvement will not necessarily be clear. Similarly, the latent feature discovery techniques do not provide sufficient insight regarding the representation or meaning of those features \citep{koren2009matrix}.

In this work, a novel framework is proposed that uses graph-based representation properties to generate additional features from user modeling data of recommender systems, with the objective of improving the accuracy of the generated recommendations. The proposed framework is underpinned by the idea of examining a tabular recommender system's data from the graph theory-based perspective, which represents entities and their relationships as a graph and allows the extraction of a suite of new features computed using established graph-based metrics. The extracted features encapsulate information about the relationships between entities in the graph and lead to new patterns uncovered in the data. In most cases, they are also interpretable; for example, a node's degree (number of edges to other nodes) represents the importance of the node in the graph, while the path length between two nodes communicates their relatedness (the shorter the path - the more related are the nodes). The approach is domain-independent and can be applied automatically.

The proposed framework offers several benefits for automatic
feature extraction. Given a new dataset, it is usually impossible to determine a-priori which graph representations will yield the most informative set of features for the recommendation generation.
Thus, the proposed framework provides a systematic method for
generating and assessing various graph representations, their
contribution to the newly extracted features, and, in turn, to the accuracy of the generated recommendations. Additionally, since the number of nodes and relationship types in each graph representation is different, an exhaustive method of distilling
the possible graph metrics from each representation is proposed.

Two case studies are conducted to gather extensive empirical evidence and demonstrate how graph features supplement existing feature sets, improve the accuracy of the recommendations, and perform adequately as stand-alone out-of-the-box features. The case studies answer the following questions:
\begin{itemize}
	%\item How does the use of graph features affect the performance of recommendation generation and user feedback prediction in different domains and tasks?	
    \item How does the use of graph features affect the performance of rating predictions and recommendation generation in different domains and tasks?
	%\item How do features extracted from different sub-graphs and graph representations affect the recommendation results?
    \item How are the recommendations affected by the sub-graph and its representation used to generate the graph features?
%	\item How do graph features perform at different levels of user/item feedback variability and sparsity?
\end{itemize}

Multiple datasets, multiple machine learning mechanisms, and multiple evaluation metrics are used across the case studies, in order to demonstrate the effectiveness of the approach. Overall, the results show that graph-based representation and automatic feature extraction allow for the generation of more precise recommendations.
%Contributions to scenarios of high variability and sparsity are also analysed and demonstrated.
A comparison across various graph schemes is conducted and the justification for systematic feature extraction is established. Hence, this work concludes the line of research presented earlier in \citep{tiroshi2013cross,tiroshi2014improving,tiroshi2014graph} and provides a complete picture that validates the applicability of the proposed graph-based feature generation approach to recommender systems.

The rest of the paper is structured as follows. Next, the necessary background is provided, and related work is described. Then, the graph representation and graph-based feature extraction process is formalized, and its advantages and disadvantages are discussed. Two case studies demonstrating the contribution of the graph-based features to the recommendation process are then presented. Through these, the overall performance of the
framework, as well as the performance of certain graph representations and feature subsets, is evaluated.
%Also analyzed is the performance of the framework under the cold-start, data variability, and data sparsity conditions.
Finally, the implications of the findings are discussed, together with the suggested future work.

\section{Background and Related Work}
\label{sec:background}
Graphs have been exploited in recommender system for many tasks, mainly due to their ability to represent many entities of different types and their relationships in a simple data structure that offers a broad variety of metrics and reasoning techniques. In this section we provide a general background on the use of graphs in recommender systems, followed by specific aspects of graph representation in recommender systems and feature engineering.

\subsection{Graph-Based Recommender Systems}
\label{sec:rel_graph_rec}

In recent years, especially since social networks were identified as a major source for freely available personal information, graphs and networks data structures have been used as tools for user modeling, especially since they combine different entities and links into one simple structure capturing the links between the entities. This section aims at giving the readers an idea about how graph techniques are used in graph-based user modeling and recommender systems. Given the vast amount of studies (a search for ``graph-based'' and ``recommender systems'' in Google Scholar yielded 225 results for 2016 alone), this is only a brief presentation of recent studies and not an in-depth survey.

What was clearly noticeable was that most of the graph-based representations were defined for a specific problem, in specific domains, and in many cases they applied variants of random walk as the only graph feature used for recommendations. \citep{PhamLCZ15} suggested to use a simple graph representation for recommending groups to users, tags to groups, and events to users, using a general graph-based model called HeteRS, while considering the recommendation problem as a query dependent node proximity problem.
%Lee et al. [2015] also used personalized graph walk for inferring users’ interests in groups of items (where the graph contained users, groups of similar items and features – search keywords and items categories).
\citep{PortillaRAA15} applied random walk for predicting YouTube videos watching, on a graph composed of videos as nodes and the link representing the appearance of videos in the recommendation lists. \citep{Wu:2015} suggested the use of a heterogeneous graph for representing contextual aspects in addition to items and users, and used random walk for context-aware recommendation. \citep{LeeKP15} applied random walk for finding top-$K$ paths from an origin user node to an item node in a heterogeneous graph, as a way for identifying the best items for recommendation. Still, these works used the PageRank algorithm for the purpose of generating the recommendations. \citep{LeePKL13} used an enhanced version of personalised PageRank algorithm to recommend items to target users and proposed to reduce the size of the graph by clustering nodes and edges. \citep{ShamsH16a} also applied personalised PageRank over the user/item graph augmented with pairwise ranking for items recommendation.

In addition to the wide use of random walk based algorithms, there is a variety of task-specific representations and metrics. It is interesting to note that even for a specific task, a variety of approaches was suggested. For instance, for song/playlist recommendations,
\citep{BenziKBV16} combined graph-based similarity representation of playlists and songs with classical matrix factorization to improve the recommendations. \citep{OstuniNSOS15} took a different approach and suggested to use tags and sound description represented as a knowledge graph, from which similarity of nodes was extracted using a specific metric they defined.
\citep{Mao0HZ16} suggested using graph representation for music tracks recommendations, where they represented by graphs the relative preferences of users, e.g., pair-wise preference of tracks. They used the graph as a representation for user preferences for tracks and calculated the probability of a user liking a track based on the probability that s/he likes the in-linked tracks.

Some researchers suggested to use graph representations as an alternative to the classical collaborative and hybrid recommenders. \citep{Moradi2015462} used clustering of graph representation of users and items for generating a model for item- and user-based collaborative filtering.
\citep{BaeHPY15} used graphs for representing co-occurrence of mobile apps, as logged from users mobile devices, and the similarity of user graphs was used for finding a neighborhood and generation recommendations. \citep{CordobesCALMOPS15} also addressed the app recommendation problem and explored the potential of graph representation for several variants of recommendation strategies for recommending apps to users through banners on webpages.
%They used graph representation for classical collaborative filtering as well as additional methods that assigned different weights to the edges. Nodes were Apps and similarity was measured by the similarity of the nodes according to their shared neighbors and links weights. (in a way this approach is somewhoat similar, in a smaller scope).
\citep{ParkPJL15} proposed a graph representation for linking item based on their similarity; hence, having a graph that links items while the weight on the edges represents their similarity. Users were linked to items they rated, such that items most similar to the items rated by the users could be recommended. \citep{LeeL15a} suggested an approach for graph-based representation of the user-item matrix, where links among items represent the positive user ratings, and use entropy to find the items to recommend to users, thus, introducing serendipity into the recommendation process. \citep{HongJ16} used affinity between users for creating a user graph, where users are nodes and edges represent affinity for the purpose of group recommendations of
movies \cite{said2011group}.
%In some cases, graph representation was used for content-based recommendation, like the work of Zoidi et al. (2015) that reviewed labels propagation approaches over graph representation where nodes are connected based on their similarity in a feature space, and discussed their potential use for assigning labels to multimedia items (nodes).

A highly relevant line of work focuses on enriching recommender systems dataset with information extracted from graph representation of the data, which is called MetaPaths. A good recent example study is the work of \citep{VahedianBM16}. The author suggested to enrich a classical recommender systems dataset (in their case DBLP authors/papers dataset) with what the so-called metapath data – links extracted from citations network. They added this information to the existing set of features, then applied classical matrix factorization, and showed an improvement to the results using only the original data. Our framework can be considered as a generalized variant of \citep{VahedianBM16}, where a specific set of metrics was extracted from the graph representation of the data and matrix factorization was applied for recommendation generation purposes. The studies presented in this work used a variety of metrics, datasets, and recommendation methods.

Additional applications of graphs for recommendations include domains of cultural heritage, tourism, social networks, and more. \citep{Chianese2016} used graphs for representing context evolution in cultural heritage: nodes modeled states and transitions between the nodes were based on observation of user behavior \cite{bohnert2008using}. \citep{ShenDG16} used graphs for representing tourist attractions and their similarity, where different graphs could represent content-based, collaborative, and social relationships. \citep{Jiang2016} used graph techniques for trust prediction in social networks. \citep{GodoyC16} reviewed the use of folksonomies, which can be naturally seen as user-item-tag graphs, in recommender systems. As we see, graphs-based approaches in user modeling and recommender systems have become highly popular and there is a growing numbers of tools that enable analysis of large graphs. We refer an interested reader to \citep{BatarfiSFNBBS15} and \citep{ZoidiFNP15} for recent and encompassing reviews of the area.
%, as reviewed by Batarfi et al. (2015) and even tutorials, like the one offered by Beutel et al. (2015) offered a tutorial focused about using graph representation and graph-related metrics for variety of user modeling/recommendation tasks.

\subsubsection{Similarity Measurement Using Graphs and their Application}
Previous research on recommender systems that use graph
representations focused on measuring the similarity of two
entities in the data (user-to-item, user-to-user, or
item-to-item), and tried to associate this with a score or rating
\citep{amatriain2011data}. Graph-based similarity measurement is
based on metrics extracted from a graph-based representation
\citep{desrosiers2011comprehensive}. Two key approaches for
measuring similarity using graphs are path-based and random
walk-based.

In the path-based similarity, the distance between two graph
nodes can be measured using the \textit{shortest path} and/or the total
\textit{number of paths} between the two. The definition of the
shortest path may include a combination of the number of edges
transitively connecting the two nodes in question and the weights
of these edges if exist, e.g., if a user is connected to an
item and the user's rating for the item as the edge label.
Shortest paths can then be computed for a user node and an item
node in question, in order to quantify the extent to which the
user prefers the item. The ``number of paths'' approach works
similarly, by calculating the number of paths between the two
nodes as a proxy for their relatedness (the more paths, the more
related they are). However, this approach is more computationally
intensive.

Random walks can be used to compute similarity by
estimating the probability of one node being reached from another
node, given the available graph paths. The more probable it is
that the target node can be reached from the source node,
the higher is the relatedness of the two nodes.
Random walks can be either unweighted (equal probability of edges)
or weighted (edges having different probabilities based on their
label, e.g., rating) \citep{desrosiers2011comprehensive}.

Examples of recommendation studies in which the approaches detailed above were applied can be found in \citep{li2009recommendation,losch2012graph,KonstasSJ09}, as well as in Section \ref{sec:rel_graph_rec}. \citep{li2009recommendation} reducted the recommendation
problem was to a link prediction problem.
That is, the problem of finding whether a user would like an item
was cast as a problem of finding whether a link exists between
the user and item in the graph. A similarity measure between user
and item nodes was computed using random walks. Items were then
ranked based on their similarity scores, such that top scoring
items were recommended to users. Using classification accuracy metrics,
this approach was shown to be superior to other non-graph
based similarity ranking methods.

A similar walking distance metric was used in
\citep{losch2012graph}, complemented by graph structure metrics
such as the number of sub trees. These metrics were used for the
purpose of link prediction and property value prediction in RDF
semantic graphs, using a learning technique based on an SVM.
%An example of such an RDF-related task is assigning an object to a class.
Experimental results showed that
the graph features varied in their performance based on
the graph structure on which they operated, for example, full
versus partial subtrees. It was also noted that the newly defined
features were not dataset-specific, but could be applied to any
RDF graph
%, while previously known structure-based features were dataset-specific.
The graph structures in the context of
RDF are less applicable to those used in the approach
proposed in this work, because the recommendation dataset graphs
do not follow a hierarchical model of RDFs.
In the presented approach, any feature value is
connected to other features values based on co-occurrence in the
dataset, without the need for matching a predefined
structure or scheme.

Finally,  \citep{KonstasSJ09} developed a graph-based approach
for generating recommendations in social datasets like Last.fm.
The work focused on optimizing a single graph algorithm (random
walk with restarts) and its parameters, such as the walk restart.
The reported results show an improvement in recommendations
using the random walk approach, compared to the baseline
collaborative filtering. In the presented work, random walks on
a graph, although with static parameters, are represented by the
PageRank score feature. The above studies are also extended in
this work by generalizing the adoption of graph metrics beyond
random walks and their use for similarity measurements, and they
are not bound to specific graph structures, such as RDF	trees.

\subsubsection{Representing social data and trust using graphs}
Other studies involving graph approaches in recommender systems primarily
addressed the context of representing social, semantic, and
trust data. In some studies, only the graph
representation was used as the means to query the data, e.g.,
neighboring nodes and the weights of edges connecting to them
\citep{ma2009learning}, while others utilize both the graph
representation and graph-based reasoning methods
\citep{massa2007trust,quercia2014shortest}.

A survey of connection-centric approaches in recommender systems
\citep{perugini2004recommender} exemplifies how the data of an
email network \citep{schwartz1993discovering} and of a
co-occurrence in Web documents \citep{kautz1997referral} can be
represented in graphs. The graph representation of the email
interactions between users defines each user as a node and edges
connect users, who corresponded via email. In the case of
Web documents, people are again represented as
nodes and edges connect people, who are mentioned in the same
document. When these graphs are established, they can be
used to answer recommendation-related queries. In the email graph,
a query regarding the closeness of users can be answered using a
similarity or distance metric, such as those mentioned in the
previous section. In the Web co-occurrence graph, a query
regarding people sharing interests can be answered by counting
their common neighbors (assuming the co-occurrence in the type of
Web documents collected is an indicator of shared interests).

Other graph representation variants are hypergraphs
\citep{berge1973graphs}. They differ from
graphs by allowing an edge, denoted by a hyperedge, to connect
with multiple nodes.
%This definition has implications on many graph and methods.
Hypergraphs have been proposed in the context of recommendation generation,
for the purpose of representing complex associations, such as social
tagging \citep{jaschke2007tag,berkovsky2007providing,bu2010music,tan2011using}, where a
tag is attached to an item by a user. If the tag, user, and item
are represented by nodes, at least two edges are required to represent the association between the three entities\footnote{A single edge between the user and item can be labeled with the chosen tag, but then the reuse of tags by other users or for other items becomes less comprehensible.}. This association can be represented by a hyperedge connecting the three nodes. In these studies, similarity metrics, e.g., a modified hypergraph PageRank, are then composed
based on this structure and used for the recommendation
generation. Results presented in \citep{jaschke2007tag} show that
the similarity metrics from hypergraphs led to better
recommendations than variations that did not utilize the
properties of the hypergraph representation.

Prior works focusing on the means of incorporating trust between
users for the sake of improving the recommendations were surveyed
in \citep{o2005trust}. For example, \citep{ma2009learning}
proposed a graph representation encapsulating trust between users.
The representation modeled users as the graph nodes and the trust
relationships between them were reflected by the weights on the edges.
Data extracted from the graph, e.g., who trusts whom and to what
extent, was used in the recommendation process, and it was shown
to improve the generated recommendations. However, the graph was
used only to represent the data and propagate the trust scores.

Another usage of graphs for recommendation purposes is in the case
of geospatial recommendations. Quercia et al. used
graphs to find the shortest path between geographical locations,
while also maximizing the enjoyment of the path for the user \citep{quercia2014shortest}.
Locations were represented as nodes and connected to each other
based on geographical proximity. Nodes were also ranked based on how
pleasant (beautiful, quiet, happy) the locations were. Finally, a
route that optimizes the shortness and pleasantness was computed based
on a graph method and recommended to the user. In this work, both graph-based
representation and graph theory methods are used for recommendation generation.

\subsection{Feature Engineering for Recommendations}

As mentioned at the beginning of the section, another group of related works that covers automatic feature engineering. According to Guyon et al., \emph{``feature extraction addresses the
problem of finding the most compact and informative set of
features, to improve the efficiency or data storage and
processing''} \citep{guyon2006feature}. Basic features are a result
of quantitative and qualitative measurements, while new features can
be engineered by combining these or finding new means to generate
additional measurements. In the big data era, the possibilities of
engineering additional features, as well as their potential importance,
have risen dramatically.

Feature engineering (also referred to in the literature as feature
extraction, composition, or discovery) can be performed either
manually or automatically. In the manual method, domain experts
analyze the task for which the features are required, e.g., online
movie recommendation versus customer churn prediction, and conceive
features that may potentially inform the task. The engineering process
involves aggregating and combining features already present in the
data, in order to form new, more informative features. This approach,
however, does not scale well because of the need for a human
expert, the time it takes to compose features, and the sheer
number of possibilities for the new features \citep{domingos2012few}.
Conversely, automatic feature extraction, the process of
algorithmically extracting new features from a dataset, does scale up
well. Many features can be engineered in a short time using a
variety of engineering methods. Coupling automatic feature
engineering with automatic feature selection
\citep{kohavi1997wrappers} (the process of separating between useful and not useful features)
can lead to faster and more accurate recommendation models.

%The importance of automating feature engineering in recommender
%systems approaches is in that they can be crucial to their
%evolution, as mentioned
%in the introduction in a quote by
% \citep{domingos2012few}
%referring to machine learning tasks: \textit{``...there is
%ultimately no replacement for the smarts you put into feature
%engineering.''}. Since recommender systems increasingly rely on
%machine learning methods, this inherently applies to them too.

A basic approach for engineering new features from the existing ones
is to combine them using arithmetic functions. In one study that
evaluated this approach, arithmetic functions, such as min, max,
average, and others, were used \citep{markovitch2002feature}. The
study also presented a specific language for defining features,
where the features were described by a set of inputs, their types,
construction blocks, and the produced output. A framework for
generating a feature space using the feature language as input was evaluated.
The evaluation showed that the framework outperformed legacy feature generation algorithms in terms of accuracy. The main difference between the framework presented at
\citep{markovitch2002feature} and its predecessors was that the
framework was generic and applicable to multiple tasks and machine learning approaches.

Additional automatic feature engineering methods that are
domain-specific were surveyed in
\citep{nixon2008feature,due1996feature} for image recognition and
in \citep{scott1999feature} for text classification purposes. An example of
a feature engineering method for image recognition is quantifying
the amount of skin color pixels in an image in order to
classify whether it contains a human face or not
\citep{garcia1999face}, whereas for text classification a
bag-of-words (frequency of occurrence of each word in a document)
can be generated for every document and used to describe it.

A different suite of methods for eliciting new features, which is
also applicable to recommender systems, is latent features computation.
Methods such as SVD \citep{klema1980singular} and PCA
\citep{wold1987principal} can be used to compute new features and
support the generation of recommendations by decomposing the
available data into components and matching composing factors,
i.e., the latent features. When the data is decomposed and
there exists a set of latent features that can recompose it with a
certain error rate, missing features and ratings can be
estimated \citep{amatriain2011data}. Although it has been shown
that this approach successfully improves the accuracy of the
recommendations \citep{bennett2007netflix}, it is limited in the
interpretability of the latent features found \citep{koren2009matrix}.

The current work defines an automatic and recommendation task agnostic feature engineering process, which is based on graph-based representation of a recommender system data. The details of this process are provided in the following section. 

\section{Graph Based Data Modeling for Recommendation Systems}
\label{sec:graph_based_feature_extraction}
\label{sec:overview_of_suggested_approach}

In this section, an approach for enhancing recommendations based
on representing the data as a graph is presented. This
representation allows a set of graph algorithms to be
applied and a set of graph-related metrics, which
offer a new perspective on the data and allow the extraction of
new features, to be deduced. Following a brief overview of the approach, the structure of recommender system datasets is formalized (Section \ref{sec:recommender_systems_datasets}). Then a detailed
description of porting data from a classical tabular
representation to a graph-based representation is given (Section
\ref{sec:graph_schemes}). An elaboration of methods for
generating multiple graph representations follows (Section \ref{sec:multiple_graph_scheme}) and finally the process of exhaustively distilling graph features from these representations is outlined (Section \ref{sec:graph_feature_extraction}).\footnote{An open source package implementing the approach is released at http://amitti.github.io/GraphRecSys/.}

%\subsection{An Overview of the Suggested Approach}
\begin{figure}[ht!]
    \begin{center}
    \includegraphics[width=0.8\textwidth]{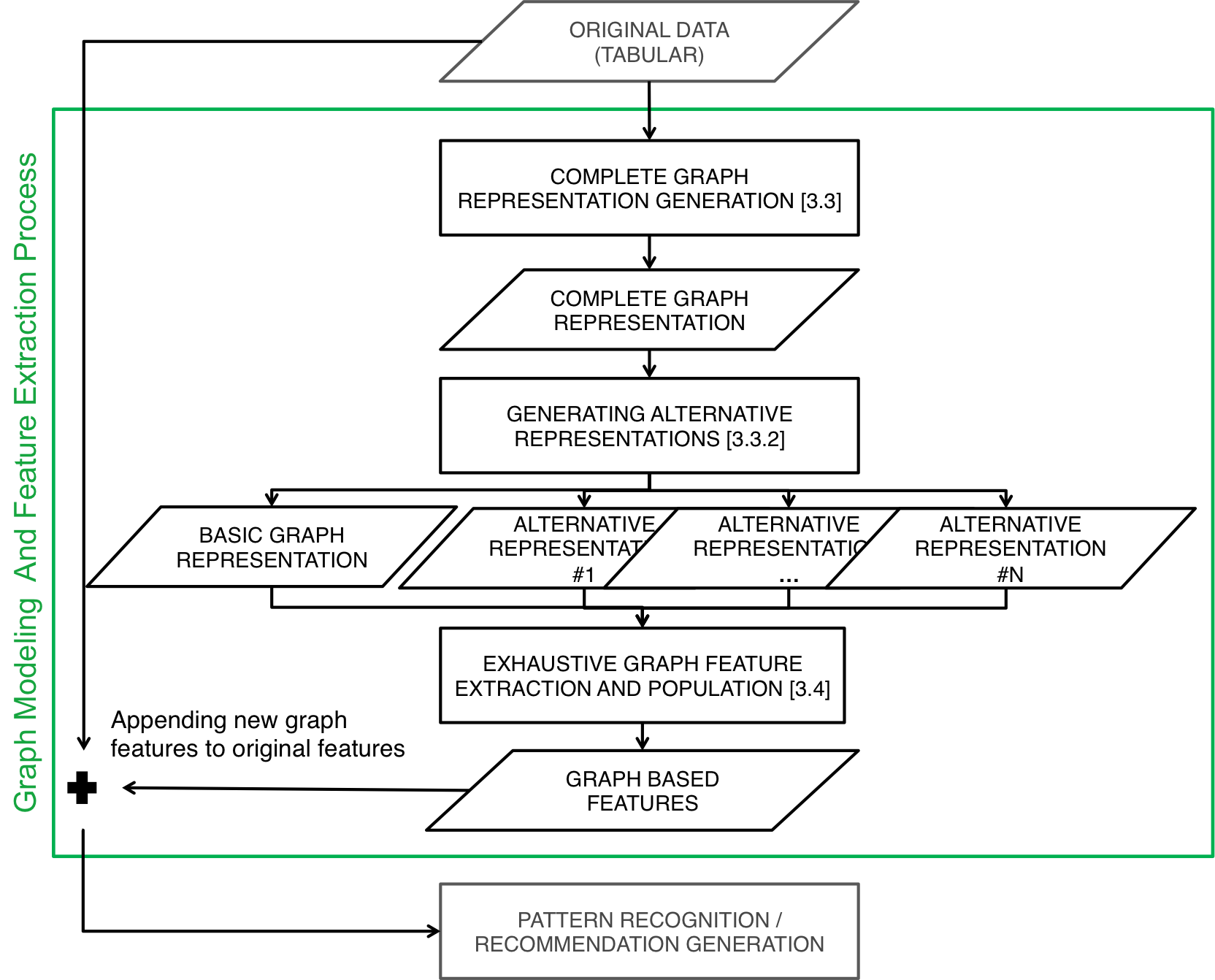}
   \end{center}
    \caption{Graph modeling and feature extraction flow chart}
    \label{fig:process_block_scheme}
\end{figure}

The input to the process (illustrated in Figure \ref{fig:process_block_scheme})
is a tabular recommender system dataset and the output is
a set of graph-based features capturing the relationships
between the dataset entities from the graph perspective.
The first step deals with the generation of a complete graph
representation of the data: the tabular data is converted into a
representation where the dataset entities are nodes, connected
based on their co-occurrence in the data.
%For example, if a row exists the values of which are
%``Pulp Fiction'', ``Quentin Tarantino'', ``Thriller'', matching
%the columns ``movie name'', ``director'', and ``genre'', then
%three nodes, one for each of these values, are
%created\footnote{Some columns may need prior encoding or
%processing, about which more details are provided in the following
%sections.}, and they are interconnected with edges reflecting
%their co-occurrence in the row.
Next, a set of partial representations is derived from the complete graph:
first the basic
representation containing only user and item nodes, and then
additional alternative representations, each with a unique
combination of relationships filtered from the complete graph.
The partial representations are passed to the next step,
where the extraction of the graph features
%, e. Degree Centrality and Average Neighbor Degree, as described in detail in Section \ref{sec:the_extracted_features},
is performed. Finally, the newly
generated graph-based features are used to supplement the original
features available in the dataset and this extended data is fed
into the recommender system for the generation of predictions or
recommendations.\footnote{Note that although the process of selecting features
that are more predictive for the task at hand (i.e., feature selection)
is outside the scope of the propose approach, it is addressed indirectly
by the features extracted from the partial graph representations.} In the
following sub-sections the above steps of the feature extraction
process are elaborated.

\subsection{The Structure of a Recommender System Dataset}
\label{sec:recommender_systems_datasets}

In \citep{burke2007hybrid,ricci2011introduction}, classical
recommendation approaches are categorized into several key groups:
collaborative filtering, content based filtering, demographic,
knowledge-based, community-based, and hybrid approaches. We first
consider the representation of the input data used by these approaches,
which can be converted into a tabular form as follows:

\begin{itemize}
    \item In \textit{collaborative filtering}, the data is represented as a matrix of user feedback on items (matrix dimensions are users$\times$items), where both the users and the items are denoted by their unique identifiers and the content of the matrix reflects the feedback of the users for the items, e.g., numeric ratings or binary consumption logs.
    \item In \textit{content based filtering}, the items are modeled using a set of features, e.g., terms or domain features. Here, the matrix dimensions include the identifiers of the users, as well as the identifiers of the content features, and the values represent the preferences of
        the users for the features. The model also contains a second matrix with item identifiers and the same content features. The values in this matrix represent the weights of the features in each item.
        %(The features are a composing part of the item, while in the ``case-based'' variation detailed below the features can be those that just describe the item and are not necessarily part of its content, e.g., price level). Matching between users and items is then accomplished by evaluating user vectors with item ones.
    \item In \textit{demographic recommenders}, the demographic features of the users are exploited in order to assign them
        to a group with a known set of preferences. Hence, in essence, it is analogous to the representation
        of content-based recommender systems, where a user's demographic features are used instead of individual
        user's characteristics and preferences.
    \item Two variants of \textit{knowledge-based recommenders} -- case-based and constraint-based --
        break the items into weighted features, e.g., the price of a product and the importance of the price for the user.
        %Recommendations are generated by computing a similarity score between users and items based on the users’ ranking of features and the items' characteristic features, e.g., the extent to which they match the user's preferences.
        This model can be represented by two matrices, one contains the items' weighted features and the second contains the users' ranking of the features’ importance. In the items matrix, each column represents a feature, each row represents an item, and the values are the strength, or how representative the feature is of the item. Similarly, in the users matrix, each column represents a feature, each row represents a user, and the values represent the importance of the feature for the given user.
    \item \textit{Community-based recommenders} combine information regarding users' social/trust relations
    with their ratings. Therefore, ratings of a trusted or socially close user are weighted heavier than those of
    a less trusted one. The items’ rating information can be represented in a matrix identically
    to the one described in the collaborative filtering approach. The trust or social relations weights
    between users can be represented by a second matrix, where the rows and columns are represent users and the values quantify the degree of the relationship between them. The values of the matrix diagonal
    are 1, since users fully trust themselves, while the rest of the matrix can be either symmetric or directional.
    %(in the perspective of the diagonal)
    %if the system assumes mutual trust among users, or diagonally asymmetric if user A can trust user B,
    %while the opposite is not true. It is then the recommendation method's specific approach to use the two matrices and create a social/trust based %weighted recommendation.
    %A recommendation may be the result of a function that takes into account users' recommendations weighted by their trustworthiness for the target user.
    \item Finally, \textit{hybrid approaches} combine some of the above stand-alone recommendation models and, therefore,
        can be represented using the matrix representation.
        %For example, a hybridization of a content-based and collaborative filtering recommender (such as the Fab system \citep{balabanovic1997fab}, for instance), computes user-to-user similarity through the correlation of the respective rows of the content-based matrix, but relies on the tabular representation of the recommender's data.
\end{itemize}

The datasets used by the above approaches, which we denote by $D$, contain two key
types of entities. The first refers to the entity for which the recommendations
are generated, i.e., the user; it is referred to as the \textit{source} entity
and denoted by $D_{S}$. The second refers to the entity that is
being recommended, e.g., item, content, product, service, or even
another user. This entity is referred to as the \textit{target}
entity and denoted by $D_{T}$. This notation follows the primary
goal of a recommender system: to recommend a target item to the
source of the recommendation request. \footnote{This definition will
also be useful when moving to a graph representation, where metrics
are defined relative to \textit{source and target} vertices.
An alternative definition of ``target users'' would have led to
confusion and would have broken traditional definitions of metrics,
e.g., shortest path measured from source to target and not vice versa.}
Additional data available in the datasets typically represent the
features of the source and/or the target entity, or the relationships
between the two. The feature
set is denoted by $D_{F}$.

%\begin{sloppypar}
For example, in a movie recommenderation dataset,
$D_{S}$ refers to the system users and $D_{T}$
to the recommendable movies. Any available features
describing either the users or the movies are denoted by $D_{F}$.
User features can be the user's age, gender, and location, while
movie features can be genre, director, language, and length.
A practical assumption is made that in a tabular recommender dataset,
all the features associated with an entity are stored in the same
table as the entity itself. That is, the gender of a user is
stored in the user table rather than in the movie table.
A formal representation of the entities and their features in the above example is $D =
\{D_{S},D_{T},D_{F}\}$, where $D_{S}=user_{id}$, $D_{T}=movie_{id}$, and the features $D_{F}$ are split into $D_{F}=\{D_{FS},D_{FT}\}$ as follows: $D_{FS} =
\{f_{s1}=age,f_{s2}=gender,f_{s3}=location\}$ and $D_{FT}=\{f_{t1}=genre,f_{t2}=director,f_{t3}=language,f_{t3}=length\}$.
%\end{sloppypar}

It should be noted that the source and target entities can have common
features \citep{berkovsky2006decentralized,berkovsky2008mediation}. For example, in
the case of a restaurant recommendation task, the source entity
(user) and target entity (business) can both have the ``location''
feature. The role of the source/target entities and features can
also change according to the recommendation task at hand. In the restaurant recommendation example, when the task
is to recommend restaurants to users, the users are the source entity, the restaurants are the target entity, and
location is a feature of both. However, if the task was to recommend a location, e.g., tourist destinations, for a
user to visit based on the restaurants in that location, then the source entity would still be the users, the target
entity would be the locations, and the restaurants would be the features of the locations.

An important aspect that needs to be considered is the
relationship observed between the entities, e.g., the
fact that a user watched, rated, tagged, or favored a movie.
Relationships can be established not only between a source and a
target entity, but also between two source/target entities.
Examples of relationships between two user entities are the
directional followee-follower relationship or the non-directional
friendship. Relationships between two movies can be established
because they are directed by the same director, are in the same
language, and so forth.
Relationships between entities are defined using the tuple
$(source~\in D_{S}, \{features\}\in D_{F}, target~\in D_{T})$. For
example, the availability of user ratings for a movie is defined
by $rel_{rating}=(user, value, movie)$ and friendship between two
users is defined by $rel_{friend}=(user, \{\varnothing\},
user).$\footnote{Additional friendship features, such as duration
or strength, can also be included.} The set of all possible
relationships in a dataset is denoted by $D_{R}=\{rel_{i}\}$, such
as in the movies example
$D_{R}=\{rel_1=rating,rel_2=friendship\}$.

Given the above formalization of entities, features, and
relationships, a recommendation task implies the prediction of a
relationship between entities. For example, the task of a
movie recommender can be considered as the prediction of the
$rel_{rating}$ relationship. This relationship can be numeric
(star rating) or binary (interested or not interested),
%\footnote{In the case of binary relationships and selecting top-N recommendations, the N selected items are selected arbitrarily from all valid items, e.g., items whose relationship to the user has been predicted as ``interested'', since no ranking could be applied.},
but the recommendations delivered to the users are guided by the
predicted values of $rel_{rating}$. If, on the contrary, the system
is a social recommender that recommends online friends, then the
relationship in question is $rel_{friend}$ and its task is to
recommend a set of candidate friends.

In addition to the original data that is available to the recommender,
more features can be generated and distilled, thus, enriching the dataset.
For example, two popular features frequently computed in rating-based
recommendation datasets are the average rating of a user and the average
rating for an item. These features are associated with the users and items,
stored in the relevant tables, and they are used to refine, e.g.,
normalize, the predicted ratings and improve the quality of the
recommendations \citep{schafer1999recommender}. The question
addressed in this work is whether the availability of additional,
supposedly more complex, features that encompass more information
and stem from graph representation of the data can contribute to
the accuracy of the predictions and the quality of the recommendations.
In the following sub-sections, the details of extracting and populating
features are provided.

\subsection{Transforming a Tabular Representation into a Graph-based Representation}
\label{sec:graph_schemes}

\subsubsection{Basic graph representation for recommender systems data}
\label{sec:basic_graph_scheme}

When moving from the tabular to the graph-based representation of a recommender system dataset, there are multiple graph design considerations.
Three key design questions are:

\begin{enumerate}
\item Should the graph encompass all the available data? What
parts of the dataset are important and need to be represented by
the graph?
\item Which entities from the selected data should be
represented by graph vertices and which entities by graph edges?
\item How should the edges be defined? Should they be directed or undirected?
Should they be labeled? What should the labels be?
\end{enumerate}

Regarding the first question, it is probable that the decision
regarding the data to be represented in the graph is
data-dependent. For some domains, datasets, and recommendation
tasks, certain parts of the data may be more informative than
others. Since the space of possible graph-based data representations
is too large for determining a-priori the most suitable scheme,
a possible alternative is to start with a graph model based on the entire data,
and then, to systematically extract all sub-graph representations and their
features. This leads to automatic coverage of the entire search
space, inherently uncovering the representations that produce the most
effective features. Then, the most informative feature set can be selected.

\begin{figure}[ht!]
    \begin{center}
    \includegraphics[width=0.8\textwidth]{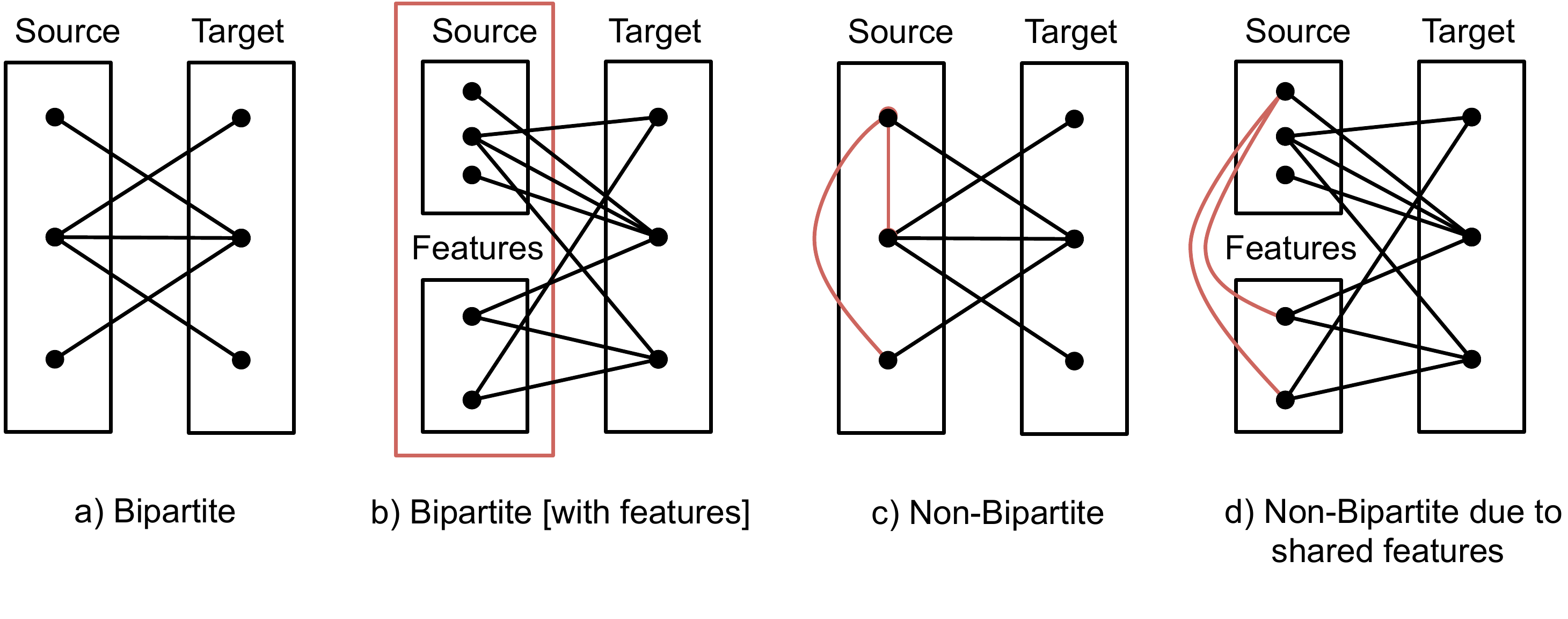}
   \end{center}
    \caption[Recommender systems dataset graph schemes examples]{Examples of two types of graph schemes
    for representing a recommender system dataset: bipartite (a,b) and non-bipartite (c,d).
    In (b) the red block confines the multi-part bipartite graph component. In
    (c) and (d) the red edges break the bipartite structure.}
    \label{fig:graph_schemes}
\end{figure}

To answer the second and third questions, an intuitive modeling approach is used.
Namely, the graph model considers all the source, target, and feature
entities as vertices, while their links and relationships between features
(including user feedback on items) are the edges.
If the information about the relationship is binary, e.g., the
item is viewed or not, the edges are not labeled. Otherwise, the
edges’ labels communicate the information about the relationship,
e.g., rating or type of association. In most cases, the edges are
not directed, as information about a feature connected to an entity
or about an entity connected to a feature is equivalent. Although
this work does not consider directed edges, the proposed approach can
be extended to support this (outlined in Section
\ref{sec:sec_conclusions}).

Based on the above abstraction of recommender systems datasets,
the following basic graph representation emerges. User
and item entities are represented by the graph vertices, and edges
connect a user and an item vertex when an association between the
two is available. This association can be explicit (ratings or likes)
or implicit (content or user view). This graph is called a bipartite
graph \citep{west2001introduction}, because it can be split into two
partitions consisting of the source and target entity vertices, i.e.,
the users and items, respectively (Figure \ref{fig:graph_schemes}-A).

The basic representation can be extended by adding additional
features as new graph vertices and linking them to the existing
vertices. For example, if user locations are provided, each
location can be represented by a vertex and the users associated with the
locations are linked to their vertices. A similar situation may
occur in the target partition of the graph, e.g., the target
entity of movies and a variety of their content features: genre,
actors, keywords, and more (Figure \ref{fig:graph_schemes}-B).
Adding the feature vertices still preserves the
bipartite nature of the graph, but the partition with the added
features gets virtually split into two groups of vertices:
the entities themselves and their features.

The situation changes, however, when adding information within the
source or target partitions, e.g., user-to-user social links or
item-to-item links of the domain taxonomy. This information
introduces new links \textit{within} the partitions, which break the
bipartite structure (Figure \ref{fig:graph_schemes}-C).
Additional information that may break the bipartite structure
is the common features shared between the source and target
partitions. For example, in the movie domain, the items may be
linked to their genres, while the users may also express
their preferences towards the genres. Thus, links to the genre
vertices are established from both the user and item partitions
(Figure \ref{fig:graph_schemes}-D) and the graph is no longer bipartite.
Note that each of the four schemes shown in Figure \ref{fig:graph_schemes}
potentially generates different sets of features and the values of the
features also vary.

Following is an outline\footnote{For readability purposes, the pseudo codes in this paper omit several pre-processing steps and technical optimizations. The exact implementation details can be found in the accompanying library.} of a high-level approach for generating the \textit{complete graph}, which includes all the data and relationships of a recommender dataset.
The algorithm scans all the tables in the dataset, and for each column that is not a source entity column, target entity column, or feedback column (e.g., ratings) it generates a graph node for every unique value appearing in the column. Thus, every unique $user_{id}$ and $movie_{id}$ is assigned to a graph vertex, as well as every actor, director, movie genre, keyword, and so forth. Features that are non-categorical, e.g., movie budget, can be discretized using
a simple binning, e.g., under \textdollar10M, \textdollar10M-\textdollar20M, \textdollar20M-\textdollar30M, etc.
When the range of values is unknown, the discretization can split the values based on their observed distribution,
e.g., four equal-sized quarters, each containing 25\% of the data. Upon discretizing the values in the columns and
creating the nodes, all the nodes matching the values that appear in the same row are connected by edges to the
source and target nodes of the same row, if they are available in the table. The result is a graph that contains all
the values of the features as the graph nodes, which are connected to the source and target entities based on their co-occurrence in the data.

\subsubsection{Multiple sub-graph representations}
\label{sec:multiple_graph_scheme} Despite being included in a
dataset, not all the features are necessarily informative and
contribute to the accuracy of the recommendations. Certain
features may be noisy or bear little information, thus,
hindering the recommendation process. For example, if a feature is
sparsely populated, its values are identical across
users, or it is populated only across a certain
subset of users, then this feature is unlikely to help the
recommender and may not be included in the graph representation.
However, it is hard to assess the contribution of the features in advance with a high degree of certainty.
This leads to the idea of automatically deriving multiple
sub-graph representations from the complete graph and extracting
the graph features for each sub-graph first, and selecting the most informative ones in a later stage.
%\footnote{Given the available cloud computing resources, this is a feasible although computationally intensive task. The approach has been evaluated on public recommender system datasets (detailed in Section \ref{ch:evaluation}) using similar resources.}.
Specifically, all the possible sub-graphs are exhaustively generated
and their features are extracted. Each sub-graph represents a
combination of features influenced by the entities and relationships
included in the graph.
%The best sub-graph feature combinations can be  identified by comparing the accuracy of the recommendations generated using each of them.
The process is presented in detail in Algorithm \ref{algo:subset_graphs_generation}.

\begin{algorithm}[ht!]
\small
\SetAlgoLined
\DontPrintSemicolon
\SetKwInOut{Input}{input}\SetKwInOut{Output}{output}
\BlankLine
\Input{$CompleteGraph$ - complete graph representation of the dataset \\
       $PredEdge$ - edge type of the relationship being predicted }
\Output{$ExtractedGraphFeatures$ - set of features extracted from various sub-graph representations}
\BlankLine
$GraphEdgeTypeCombinations$ $\leftarrow$ \texttt{GenerateEdgeCombinations}(\{$EdgeTypes$\}, $PredEdge$)\;
$ExtractedGraphFeatures$ $\leftarrow$ $\emptyset$\;
\BlankLine
\ForEach{ $EdgeCombination$ $\in$ $GraphEdgeTypeCombinations$ }{
    $SubGraph$ $\leftarrow$ \texttt{RemoveEdgesFromGraph}($CompleteGraph$,  $EdgeCombination$)\;
    $SubGraphFeatures$ $\leftarrow$ \texttt{ExtractGraphFeatures}($SubGraph$,  $PredEdge$)\;
    $ExtractedGraphFeatures$ $\leftarrow$  ($ExtractedGraphFeatures$ $\cup$ $SubGraphFeatures$)\;
    }
\BlankLine return $ExtractedGraphFeatures$\; \caption{Generate
sub-graphs and extract features}
\label{algo:subset_graphs_generation}
\end{algorithm}

The input to the algorithm is the complete graph representation
$CompleteGraph$, which was discussed at the end of Section
\ref{sec:basic_graph_scheme}, and the edge $PredEdge$ representing
the relationship $rel_i$ being predicted. The function
\texttt{GenerateEdgeCombinations} invoked in line 1 returns
all the possible combinations of different types of graph edges.
Note that this function receives also the type of the predicted
edges $PredEdge$. This is done in order to preserve the $PredEdge$
edges in all the sub-graphs. Namely, this type of edges will not
be included in the combinations that are removed from the complete
graph and, therefore, will be present in all the sub-graphs.

Upon generating all the possible edge type combinations, the set is
iterated over and the function \texttt{RemoveEdgesFromGraph} is
invoked to create a sub-graph $SubGraph$ by removing the combination
$EdgeCombination$ from $CompleteGraph$ (line 4). Then,
the function \texttt{ExtractGraphFeatures} is invoked to extract
from $SubGraph$ the set of possible graph features referred to as
$SubGraphFeatures$ (line 5, to be elaborated in Section
\ref{sec:graph_feature_extraction}) and append $SubGraphFeatures$
to the set of features $ExtractedGraphFeatures$ (line 6). Finally,
in line 8 the algorithm returns $ExtractedGraphFeatures$ -- the set
of all the possible graph features from all the possible sub-graphs.

\begin{figure}[ht!]
    \begin{center}
    \includegraphics[width=0.7\textwidth]{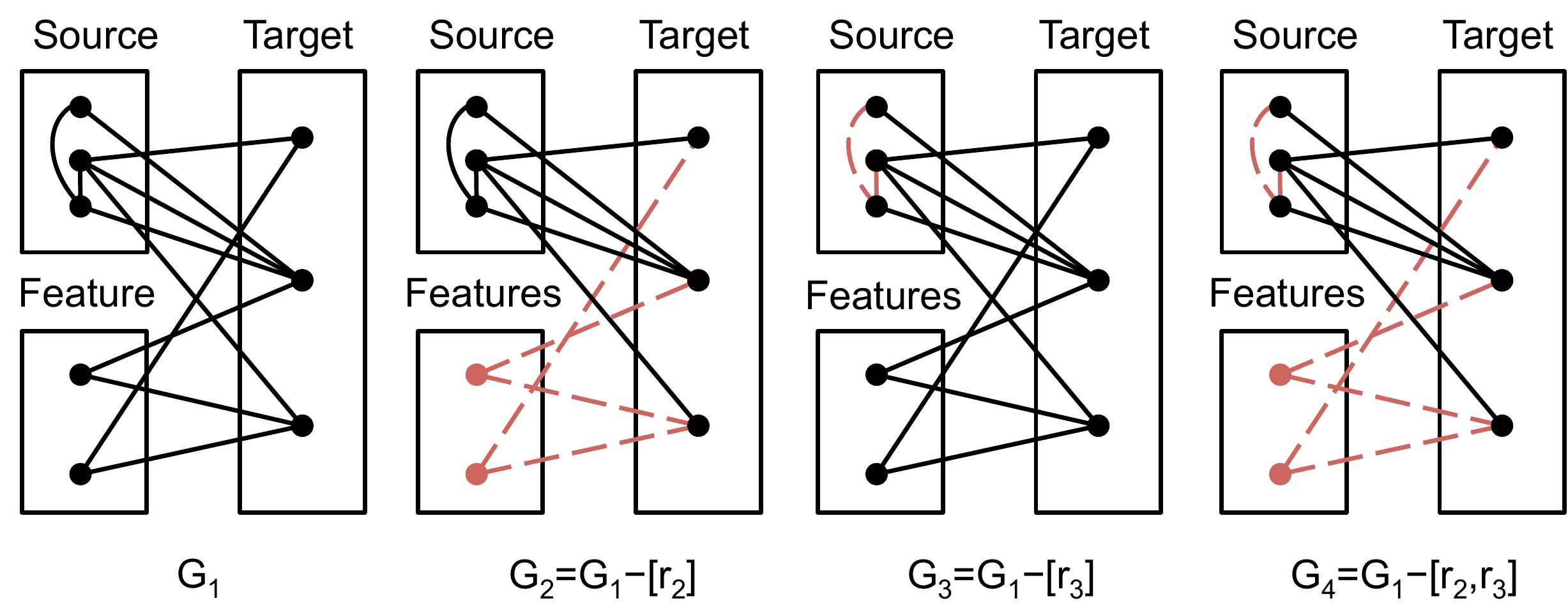}
   \end{center}
    \caption[Subgraphs generation example]{Four sub-graph schemes that are generated from the complete
    schema based on the relationship permutations. Dashed lines represent links removed from the graph.}
    \label{fig:graph_mutation}
\end{figure}

The execution of Algorithm \ref{algo:subset_graphs_generation} is
illustrated by an example in Figure
\ref{fig:graph_mutation}. Consider a graph $G=(V,E)$, where
$V=\{V_{S}\cup V_{T}\cup V_{F}\}$ is the set of vertices of
the source entities $V_{S}=\{V_{S1}, ..., \allowbreak
V_{Sm}\}$, target entities $V_{T}=\{V_{T1}, ...,
\allowbreak V_{Tn}\}$, and domain feature values $V_{F}=\{V_{F1},
..., \allowbreak V_{Sk}\}$. In addition, $E=\{rel_1,
rel_2, rel_3\}$ is the set of graph edges, reflecting three
relationship types: $rel_1$ is the source-target relationship
being predicted; $rel_2$ is the relationship between the target
entities and domain features; and $rel_3$ is the relationship
between the source vertices. In graph terminology, the recommendation
task is to predict the label (or the existence) of an edge
$rel_{1}(i,j)$ between a source vertex $V_{Si}$ and a target
vertex $V_{Tj}$.

%\begin{sloppypar}
For this graph, the set $GraphEdgeCombinations$ created by
\texttt{GenerateEdgeCombinations} includes $Graph\allowbreak
Edge\allowbreak Combinations\allowbreak
=\{\{\varnothing\},\{rel_2\},\{rel_3\},\{rel_2,rel_3\}\}$. These
are the combinations of edges that are removed from the graph
while creating sub-graphs, whereas the predicted relationship
$rel_1$ is preserved in all the sub-graphs. Removing these
combinations of edges, function \texttt{RemoveEdgesFromGraph} generates four variants
of $SubGraph$ shown in Figure \ref{fig:graph_mutation}:
$G_1 \leftarrow CompleteGraph - \{\varnothing\}$, $G_2 \leftarrow
CompleteGraph - \{rel_2\}$, $G_3 \leftarrow CompleteGraph -
\{rel_3\}$, and $G_4 \leftarrow CompleteGraph - \{rel_2, rel_3\}$.
Note that $G_1$ is the complete graph, whereas other sub-graphs have
either $rel_2$ or $rel_3$, or both removed.
For each $SubGraph$, function \texttt{ExtractGraphFeatures} is invoked to
extract the respective feature set $SubGraphFeatures$ and all
the extracted feature sets are appended to
$ExtractedGraphFeatures$.
%\end{sloppypar}

\subsection{Distilling Graph Features}
\label{sec:graph_feature_extraction}

The function \texttt{ExtractGraphFeatures} in line 5 of Algorithm \ref{algo:subset_graphs_generation}
received a sub-graph derived from the complete representation and was invoked to extract a set of graph-based features. Moreover, this function was invoked for all the possible sub-graphs, to
ensure that all the possible graph features are extracted.
The graph-based features are extracted using a number of functions, each calculating a different graph metric. These functions, referred as \emph{generators}, are divided into several, families according to the number of graph vertices they process.

\begin{algorithm}[ht!]
\small
    \SetAlgoLined
    \DontPrintSemicolon
    \SetKwInOut{Input}{input}\SetKwInOut{Output}{output}
    \BlankLine
    \Input{$SubGraph$ - sub-graph derived from the complete graph representation\\
        $PredEdge$ - edge type of the relationship being predicted}
    \Output{$ExtractedSubGraphFeatures$ - set of features extracted from $SubGraph$}
    \BlankLine
    $ExtractedSubGraphFeatures$ $\leftarrow$ $\emptyset$\;
    $SubGraphPredictedEdges$ $\leftarrow$ \texttt{ExtractPredictedEdges}($SubGraph$,$PredEdge$)
    \BlankLine
    \ForEach{($SourceEntity$,$TargetEntity$) of $Edge$ $\in$ $SubGraphPredictedEdges$}{
        \ForEach{\texttt{1-Function} in \texttt{1-VertexGenerators}}{
            $SourceFeatures$ $\leftarrow$ \texttt{1-Function}($SourceEntity$)\;
            $TargetFeatures$ $\leftarrow$ \texttt{1-Function}($TargetEntity$)\;
            $ExtractedSubGraphFeatures$ $\leftarrow$ ($ExtractedSubGraphFeatures$ $\cup$ $SourceFeatures$ $\cup$ $TargetFeatures$)\;
        }
        \BlankLine
        \ForEach{\texttt{2-Function} in \texttt{2-VertexGenerators}}{
            $SourceTargetFeatures$ $\leftarrow$ \texttt{2-Function}($SourceEntity$,$TargetEntity$)\;
            $ExtractedSubGraphFeatures$ $\leftarrow$ ($ExtractedSubGraphFeatures$ $\cup$ $SourceTargetFeatures$)\;
        }
        \BlankLine
        $MultipleEntityCombinations$ $\leftarrow$ \texttt{ExtractEntityCombinations}(\{$VertexTypes$\})\;
        \BlankLine
        \ForEach{$EntityCombination$ $\in$ $MultipleEntityCombinations$}{
            $N$ $\leftarrow$ $|EntityCombination|$\;
            \ForEach{\texttt{N-Function} in \texttt{N-VertexGenerators}}{
                $MultipleEntityFeatures$ $\leftarrow$ \texttt{N-Function}($SourceEntity$, $TargetEntity$, $EntityCombination$)\;
                $ExtractedSubGraphFeatures$ $\leftarrow$ ($ExtractedSubGraphFeatures$ $\cup$ $MultipleEntityFeatures$)\;
            }
        }
        \BlankLine
        return $ExtractedSubGraphFeatures$\;
    }
    \caption{Extract graph features from a sub-graph}\label{algo:feature_extraction}
\end{algorithm}

\begin{figure}[ht!]
    \begin{center}
    \includegraphics[width=0.85\textwidth]{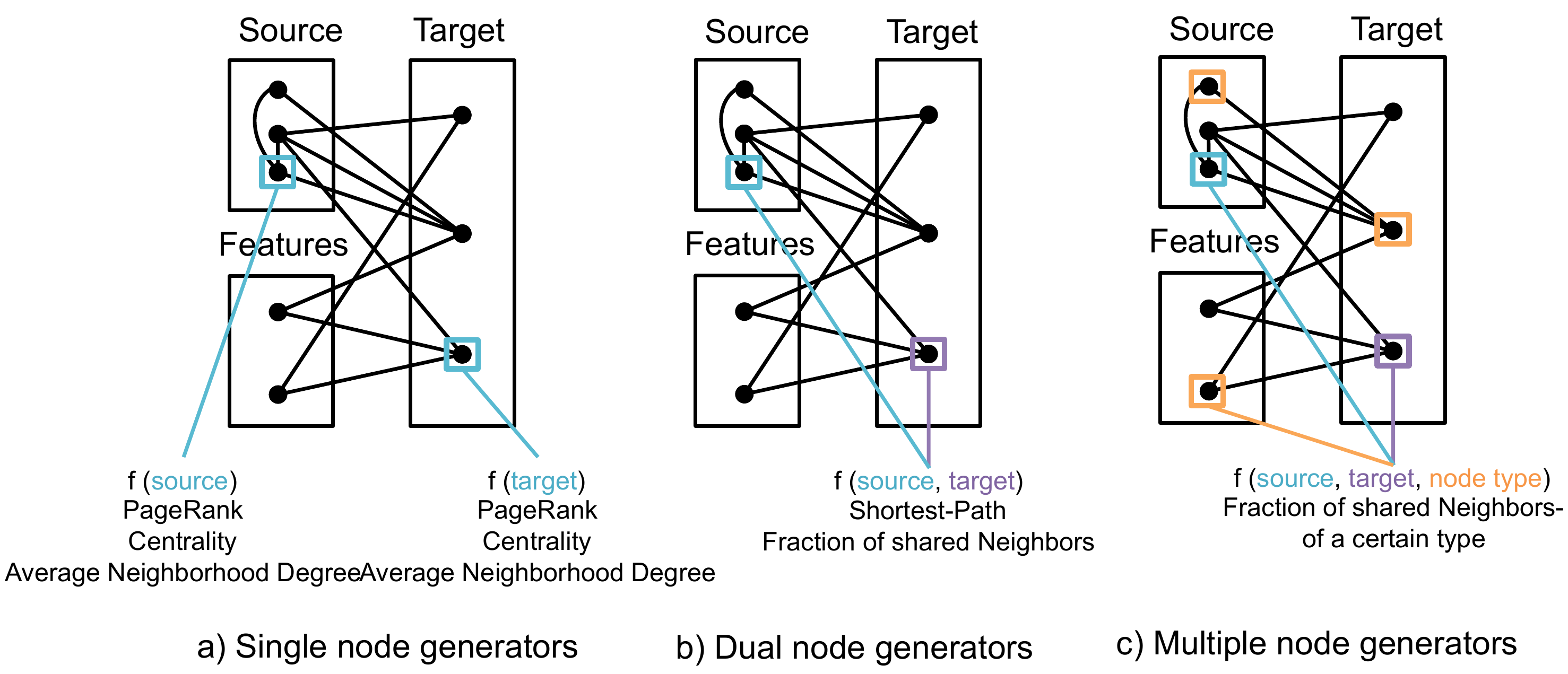}
   \end{center}
    \caption{Key graph-based feature generator families and their instances}
    \label{fig:feature_generator_families}
\end{figure}

The main steps of \texttt{ExtractGraphFeatures} are detailed in Algorithm \ref{algo:feature_extraction}, which uses three types of generators:
\begin{itemize}
\item \texttt{1-VertexGenerators} are
applied to a single graph vertex, either the source or the target
entity, and compute features of this vertex only, e.g., the PageRank
score (Figure \ref{fig:feature_generator_families}-A).
\item \texttt{2-VertexGenerators} are applied
to a pair of vertices, the source and the target entities, and
compute graph-based relationships between the two, e.g., the
shortest path (Figure \ref{fig:feature_generator_families}-B).
\item \texttt{N-VertexGenerators} are applied
to $N \textgreater 2$ vertices, two of which are the source and target entities and the
rest are not. An example function from this family is ``number of vertices of type X,
which are common neighbors of the source and target vertices'' (Figure \ref{fig:feature_generator_families}-C).
\end{itemize}
Section \ref{sec:the_extracted_features} lists the functions from each generator family that were used. Note that these are executed iteratively, in order to generate all the possible graph features. By no means this list of functions is exhaustive; it only exemplifies a number of popular functions that were used, but many more functions can be conceived and added.

%\begin{sloppypar}
At the initial stage of Algorithm \ref{algo:feature_extraction},
edges belonging to the predicted relationship are copied to the
$SubGraphPredictedEdges$ set (line 2). For each $Edge$ in
this set, the generators are invoked as follows. The \texttt{1-VertexGenerators}
functions are invoked in lines 5
and 6, respectively, on the $SourceEntity$ and $TargetEntity$
vertices of $Edge$. Applying these functions to
other vertices is unlikely to produce features that can contribute
to the prediction of the desired relationship, while leading to
significant computational overheads. Hence, \texttt{1-VertexGenerators}
are restricted to these two vertices only. The \texttt{2-VertexGenerators}
are applied in line 10 to the pairs of vertices $SourceEntity$ and $TargetEntity$.
Then, the \texttt{ExtractEntityCombinations} function is invoked in line 13,
in order to create a set of all the possible entity combinations of vertices,
$MultipleEntityCombinations$. These combinations necessarily involve $SourceEntity$
and $TargetEntity$, and in addition any other type of graph vertices. For each
combination $EntityCombination$ of size $N$ (line 15), the relevant
\texttt{N-VertexGenerators} generators are invoked in line 17. Features
extracted by \texttt{1-VertexGenerators}, \texttt{2-VertexGenerators}, and
\texttt{N-VertexGenerators} are all appended to $ExtractedSubGraphFeatures$.
%\end{sloppypar}

Note that the value of $N$ determines the \texttt{N-VertexGenerators} functions
that are invoked and the relationships they uncover. Again, two of the
$N$ vertices are necessarily $SourceEntity$ and $TargetEntity$, whereas
the third vertex can be of any other entity linked to either of them.
For instance, for $N=3$ in the movie recommendation task and entities
of user, item, and location, the relationship can be ``the number of
cinema locations that the user has visited and where the movie is
screened''. The generator considers the user and movie vertices, and then,
scans all the location vertices and identifies those, with edges connected
to both. It should be noted that more complex relationships with a higher
value of $N$ can be considered. Since a broad range of combinations is possible,
the \texttt{N-VertexGenerators} extract a large number of features that
surpasses by far the set of features that can be engineered manually.

\subsubsection{Distilled graph features}
\label{sec:the_extracted_features}
The set of metrics selected for implementation in this work and used for the evaluation of the approach
is now given in detail. The metrics are those that are commonly implemented in widely used graph analysis libraries -- NetworkX \citep{hagberg2008exploring}, igraph \citep{csardi2006igraph}, and Gephi \citep{bastian2009gephi}) -- and used in social network analysis and measurement works \citep{wilson2009user,lewis2008tastes}. It is important to stress that this set of metrics is only a portion of those that could be used and serves only as an example. The space of all graph metrics is large, as can be seen in \citep{costa2007characterization,wasserman1994social,coffman2004graph}, and, thus, could not be exhaustively evaluated within the scope of this work.
%The selected features exemplify the two classes of single and dual vertex feature functions.
%\begin{sloppypar}

The set of \texttt{1-VertexGenerators} functions were implemented and used for evaluation are degree
centrality \citep{borgatti2011analyzing}, average neighbor degree \citep{barrat2004architecture}, PageRank score \citep{page1999pagerank}, clustering coefficient \citep{latapy2008basic}, and node redundancy \citep{latapy2008basic}. These metrics
%were also used in the case studies reported in Section \ref{ch:evaluation} and
are referred to as the \textit{basic} graph features.
%\end{sloppypar}
Following is a brief  of the \texttt{1-VertexGenerators} functions.
\begin{itemize}
\item \emph{Degree Centrality} \citep{borgatti2011analyzing} (or,
simply, node degree) quantifies the importance of a vertex through
the number of other vertices to which it is connected. Hence, in
the bipartite graph, the degree centrality of a user vertex $S_i$
is the activity of $i$, i.e., the number of items with which $S_i$
is associated, and, vice versa, for an item vertex $T_j$ it is the
popularity of $j$, i.e., the number of users who are associated
with $T_j$. In a graph that includes metadata, the number of
metadata vertices associated with either the user or the item
vertex are added to the degree centrality score. The degree of
centrality of a vertex $v$ is denoted by $Deg(v)$.
\item \emph{Average Neighbor Degree} \citep{barrat2004architecture}
measures the average degree of vertices to which a vertex is
connected. In the bipartite graph, this metric conveys for $S_i$ -- the
average popularity of items with which $S_i$ is associated, and
for $T_j$ -- the average activity of users who are associated with
$T_j$. Formally, if $N(v)$ denotes the set of neighbors of a
vertex $v$, then the average neighbor degree is
   \begin{equation} \label{eq:avg_ngh_deg}
    AvgNghDeg(v)=\frac{1}{|N(v)|}\sum_{u\in N(v)}Deg(v)
    \end{equation}
In a graph with metadata, the average neighbor degree of a user/item vertex also incorporates
the popularity of the metadata features with which it is associated.
\item \emph{PageRank} \citep{page1999pagerank} is a widely-used
recursive metric that quantifies the importance of graph vertices.
For a user vertex $S_i$, the PageRank score is computed through
PageRank scores of a set of item vertices \{$T_j$\} with which $S_i$ is
associated and vice versa
%, for a target vertex $T_j$ -- through
%the PageRank scores of the users \{${S_i}$\} who are associated
%with $T_j$.
Thus, the PageRank score of a user vertex $S_i$ can be expressed as
\begin{equation} \label{eq:page_rank}
PageRank(S_i) = \sum_{T_j \in N(S_i)} \frac{PageRank(T_j)}{Deg(T_j)},
\end{equation}
i.e., the PageRank score of $S_i$ depends on the PageRanks of each item vertex $T_j$ connected to $S_i$, divided by the degree of $T_j$. In a graph with metadata, the PageRank scores of
user/item vertices are also affected by the PageRank of the
metadata vertices to which they are connected.
\item \emph{Clustering Coefficient} \citep{latapy2008basic} measures the
density of the immediate subgraph of a vertex as the ratio between
the observed and possible number of cliques of which the vertex
may be a part.
%
%In figure \ref{fig:clust_coef}-left, the sub-graph of $u$ has the potential for three cliques (C1, C2, C3), but actually contains only two (C2 is missing from the actual graph). Thus, the clustering coefficient of $u$ is $ClustCoef(u) = \frac{2}{3}$.
Since cliques of a size greater than two are impossible in the bipartite graph,
$ClustCoef$ measures the density of shared neighbors with respect to the total number of neighbors
of the vertex. The formal definition for the bipartite graphs is
            \begin{equation} \label{eq:clust_coef}
            ClustCoef(v)=\frac{\sum_{u\in N(N(v))}{ \frac{|N(v) \cap N(u)|}{|N(v) \cup N(u)|} }}{\vert  N(N(v)) \vert}
            \end{equation}
%Figure \ref{fig:clust_coef}-right illustrates the bipartite graph. Here, $u$ has only two ``hollow'' neighbors shared with other vertices from its partition, such that $ClustCoef(u) = \frac{2}{3}$.
%
%\begin{figure}[ht!]
%    \begin{center}
%    \includegraphics[width=0.6\textwidth]{gfx/clust_coef.pdf}
%   \end{center}
%    \caption[Clustering coefficient examples]{Examples of clustering coefficient in the non-bipartite (left) and bipartite (right) graph.}
%    \label{fig:clust_coef}
%\end{figure}
\item \emph{Node Redundancy} \citep{latapy2008basic} is applicable
only to bipartite graphs and shows the fraction of pairs of
neighbors of a vertex that is linked to the same other vertices.
This metric quantifies for user vertex $S_a$ - the portion of
pairs of items with which $a$ is associated that are
also both associated with another user $b$. Likewise, for item vertex
$T_x$, it quantifies the portion of pairs of users associated
with $x$ and also both associated with another item $y$. If a
vertex, node redundancy of which is computed, is
removed from the graph, the metric reflects the fraction of its
neighbors that will still be connected to each other through other
vertices. Intuitively, in a bipartite graph $NodeRed(v)$ can be seen as
the portion of connected `squares' of which $v$ is a part, among all the
potential `squares'.
\end{itemize}

%\begin{figure}[ht!]
%    \begin{center}
%    \includegraphics[width=0.6\textwidth]{gfx/node_redundancy.pdf}
%   \end{center}
%    \caption[Node redundancy illustrated]{Node redundancy in the bipartite graph (dashed lines represent missing edges).
%    If the actual graph (first from right) was the union of all the potential ones (R1-R3), then node $u$
%    would have been fully redundant (redundancy score equal to 1). However, since it is missing an edge and
%    does not contain redundancies R1 and R3, $u$'s redundancy score is 0.33.}
%
%    \label{fig:node_red}
%\end{figure}

Next, multiple-vertex generator functions are detailed. Specifically, the following functions from the \texttt{2-VertexGenerators} and \texttt{N-VertexGenerators} families were implemented:
%and shortest path \citep{floyd1962algorithm}
\begin{itemize}
\item \emph{Shortest Path} \citep{floyd1962algorithm}. Unlike the above feature generators that operate on a single vertex, shortest path receives a pair of graph vertices: a source entity and a target entity. It evaluates the distance, i.e., the lowest number of edges, between the two vertices. The distance communicates the proximity of the vertices in the graph, as is a proxy for their similarity or relatedness. A short distance indicates high relatedness, e.g., more items shared between users or more features for items, while a longer distance indicates low relatedness.
\item \emph{Shared Neighbors of Type $X$}. This is one of the \texttt{N-VertexGenerators} functions, which receives three parameters: source entity vertex, target entity vertex, and entity type $X$. It returns the fraction of neighbors shared between the source and target vertices that areof the desired type $X$. The fraction is computed relatively to the union of the source vertex neighbors with the target vertex neighbors. Note that this feature cannot be populated for graphs that do not have a sufficient variety of entities connected to the source and target vertices. For example, the generator is inapplicable for a graph having only the source and target entities.
\item \emph{Complex relationships across entities}. Apart from the above mentioned generators, system designers may define other \texttt{N-VertexGenerators} functions, which could extract valuable features. For example, it may be beneficial for a movie recommender to extract the portion of users, who watched movies from genres $g_1$, $g_2$ directed by person $p$, and released between years $t_1$ and $t_2$. It is clear that it is impossible to exhaustively list all the combinations of such features: this is domain- and application-dependent. Hence, the task of defining these complex generators is left open-ended and invites system designers to use the provided library and develop their own feature generators.
\end{itemize}

To recap, each of the above \texttt{1-VertexGenerators} and
\texttt{2-VertexGenerators} is applied to every source and
target vertex and generates features associated with the vertex
or a pair of vertices. In addition, \texttt{N-VertexGenerators}
is applied to the source and target vertices and all the possible
combinations of other entity types. Recall that this is done for
every sub-graph extracted from the complete graph\footnote{Some
generators should be applied in a different manner to certain
sub-graphs, e.g., $Deg$ and $ClustCoef$ generators in bipartite
and non-bipartite graphs.}
%Hence, there is a need to distinguish between the bipartite and non-bipartite sub-graphs.}
%This is achieved by attempting to ``color'' the graph with only two colors. If the two-coloring is successful, then the graph is bipartite; otherwise, it is non-bipartite \citep{kleinberg2006algorithm}.}
and the complexity of the feature generation task becomes clear.

\subsubsection{Quantifying the number of graph features}
\label{sec:amount_of_features_extracted}

Here, the number of graph features that can be extracted from a recommender system dataset using the
proposed approach is quantified. The quantification illustrates the coverage and computational complexity of the extraction process.
Considering Algorithms \ref{algo:subset_graphs_generation} and
\ref{algo:feature_extraction}, it becomes evident that the number of
extracted features primarily depends on two key elements: the
number of sub-graphs that are extracted from the complete graph
and the number of entities in each sub-graph. Both of these are
derived from the number of entities and relationships in the dataset.

Building on the recommender dataset analysis given in Section
\ref{sec:recommender_systems_datasets}, each dataset $D$ contains
two entities, $D_{S}$ and $D_{T}$, and $|D_{R}|$ relationships.
The features are extracted from the complete graph and also from
the sub-graphs. The latter are generated from any non-ordered
combination of relationships from $D_{R}$, but necessarily contain
the predicted relationship $rel_i$, such that the overall number of
relationships is $M=|D_{R}|-1$. Hence, the number of possible
sub-graph combinations is
\begin{equation} \label{eq:relationship_combinations}
\sum\limits_{x=0}^{M}\binom{M}{x}=\sum\limits_{x=0}^{M} \dfrac{(M)!}{x!(M-x)!},
\end{equation}
the sum of the numbers of combinations of size $x$ that can be
produced, where $x  \in [0,M]$.
%from each subset size from 1 to $N=|D_{R}|-1$ (Eq. 3).

%\noindent\begin{tabularx}{\textwidth}{@{}XXX@{}}
%	
%	\end{equation}
%	&
%	\vspace{0.4mm}
%	\begin{equation} \label{eq:relationship_combinations_simplified} \sum\limits_{x=0}^{N}\binom{N}{x}
%	\end{equation}
%	&
%\end{tabularx}	
	
%Since the complete graph, from which the sub-graphs are produced, can also be used to extract features, that is one additional graph, and leads to Eq. 4 ($\binom{N}{0}=1$).

\begin{figure}[ht!]
	\begin{center}
		\includegraphics[width=0.55\textwidth]{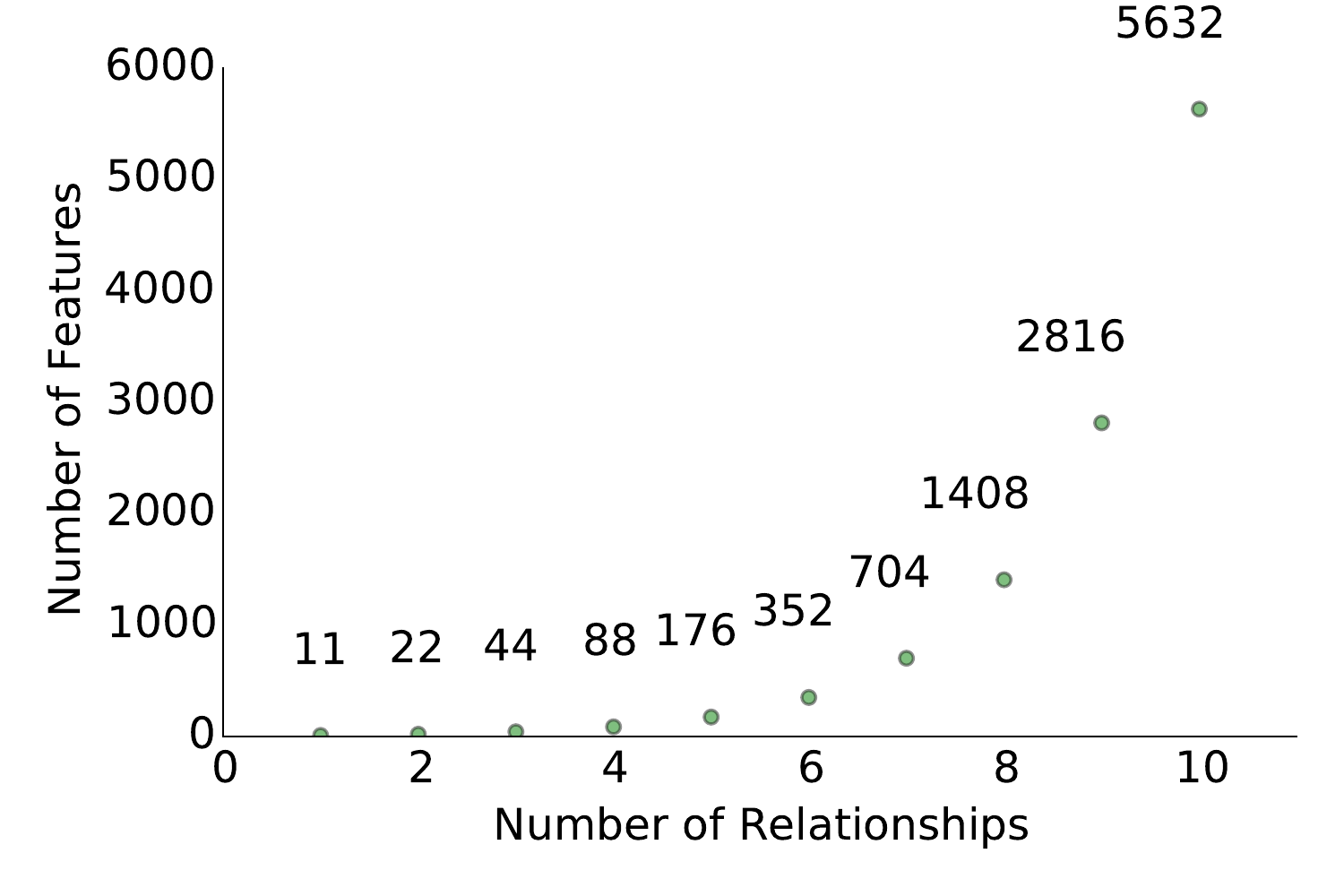}
	\end{center}
	\caption[Number of features as function of number of relationships in dataset]{Number of
		extracted graph features versus the number of relationships in the dataset}
	\label{fig:number_of_features_extracted}
\end{figure}

For each of these sub-graphs, let us assume that $F_1$ features
can be generated by \texttt{1-VertexGenerators} function for $D_{S}$
and $D_{T}$ individually, and $F_2$ features can be generated
by \texttt{2-VertexGenerators} for the pair $(D_{S},D_{T}$).
On top of these, at least $M=|D_{R}|-1$ more complex by
\texttt{N-VertexGenerators} functions
can be applied to every sub-graph, as per the number of
relationships available in the data. This brings the overall
number of generated graph-based features to the order of
\begin{equation}
 \sum\limits_{x=0}^{M} (2F_1 + F_2 + M) \times \dfrac{(M)!}{x!(M-x)!}
\end{equation}

The library that accompanies this work defines $F_1=5$ single node generators
and $F_2=1$ dual node generator. For illustrative purposes only, Figure \ref{fig:number_of_features_extracted} plots the number of extracted features,
which is exponential with the number of relationships $|D_{R}|$. For example,
the number of features extracted from a dataset having $|D_{R}| \leq 4$
relationships is smaller than 100. However, for a dataset with $|D_{R}|=10$
relationships, the number of extracted features exceeds 5,000. Clearly, engineering
all these features manually would require considerable resources, whereas the
proposed approach is fully automated.

%\section{Evaluating the Contribution of Graph Based Features}
\label{sec:case_studies}

\section{Experimental Setting and Datasets}
\label{ch:evaluation}
It is important to highlight that
% contrast to algorithmic recommender systems works, which introduce new recommendation methods,
the product of the presented approach
is graph-based features that help to generate recommendations using
existing recommendation methods. These
features can either be used as stand-alone features, i.e., the
only source of information for the recommendation generation, or be
combined with other features. Hence, the baseline for
comparison in the evaluation part is the performance of
common recommendation methods when applied \textbf{without
the newly generated features}.
%Comparisons of methods that rely on features and those that do not were conducted previously
%\citep{jahrer2010combining,bellogin2013empirical,toscher2009bigchaos}. Their results demonstrated that ensemble methods, e.g., Random Forests and Gradient Boosting, outperform stand-alone approaches such as collaborative filtering. Given those results, similar methods were adopted for the evaluations in this work.

To present solid empirical evidence, the contribution of the graph feature extraction to the accuracy of the recommendations was evaluated using three machine learning methods: Random Forest \citep{breiman2001random}, Gradient Boosting \citep{Friedman00}, and Support Vector Machine (SVM) \citep{gunn1998support}.
%, in order to verify that the contribution is consistent.
Both Random Forest and Gradient Boosting are popular ensemble methods that have been shown to be accurate and won recommendation \citep{koren2009bellkor} and general prediction \citep{yu2010feature} competitions. The methods are also implemented in widely used machine-learning libraries
\citep{pedregosa2011scikit,hall2009weka}, and were shown to perform well in prior recommender systems works \citep{jahrer2010combining,bellogin2013empirical,toscher2009bigchaos}.

In the next section, two case studies showing the contribution of graph-based features are presented. These case studies demonstrate the value of the proposed graph-based approach when applied to a range of recommendation tasks
%(predicting binary and non-binary ratings,  predicting rankings based on implicit feedback, e.g., the number of views/listens. The case studies also include datasets from different
and application domains.
%: movie ratings, music listening habits, business ratings, and interest profiles from social networks.
%The recommendation accuracy metrics used in the case study include: precision at K (P@K) for the ranking and binary rating predictions, RMSE, and MAE for the star rating prediction \citep{ShaniG11}.
Case study \rom{1} evaluates the performance of the
graph-based approach, evaluating its contribution in different
domains and tasks. Case study \rom{2} focuses on the impact of
representing data using different graph schemes on the
recommendations.
%Finally, Case study \rom{3} evaluates the performance of the graph-based approach in cases of sparsity and variability of ratings.
%Dataset \rom{4} could not be used in Case Study \rom{2}, because it lacks the required amount of entities and relationships types, while for Case Study \rom{3}, a single dataset was chosen due to the complexity of the experiment.
Altogether, five datasets were used across the case studies
and the mapping between the datasets and case studies is laid out
in Table \ref{tab:dataset_to_casestudies_mapping}. In the
following sub-sections, a brief characterization of the datasets,
as well as the overview of the recommendation tasks and evaluation
metrics, is provided.
%, followed by the specific performance differences observed in each case.
\begin{table}
\small
	\centering
	%\tiny

	%\resizebox{0.85\columnwidth}{!}{
		\begin{tabular}{lcc}
			\toprule[1pt]
			& \specialcell{Case Study \rom{1} --\\ Overall contribution of \\ graph-based features}& \specialcell{Case Study \rom{2} --\\ Performance of different\\ graph schemes} \\
%& \specialcell{Case Study \rom{3} --\\ Performance in cases of \\ high sparsity/variance}\\
			\midrule[1pt]
			Dataset \rom{1} - Last.fm& $\checkmark$ & $\checkmark$  \\
			\midrule[0.5pt]
			Dataset \rom{2} - Yelp& $\checkmark$ & $\checkmark$ \\ %& $\checkmark$ \\
			\midrule[0.5pt]
			Dataset \rom{3} - Yelp \rom{2}& $\checkmark$ & $\checkmark$  \\
			\midrule[0.5pt]
			Dataset \rom{4} - OSN& $\checkmark$ &   \\
			\midrule[0.5pt]
			Dataset \rom{5} - Movielens & $\checkmark$ & $\checkmark$  \\
			\bottomrule[1pt]
		\end{tabular}
	%}
	\caption{Mapping of datasets to case studies}
	\label{tab:dataset_to_casestudies_mapping}
\end{table}

%\subsection{Datasets}

\subsection{Dataset \rom{1} -- Last.fm}
\label{sec:dataset_lastfm}

The first dataset is of users' relevance feedback provided for
music performers via the Last.fm online service. The
dataset is publicly available\footnote{http://grouplens.org/datasets/hetrec-2011/} and was obtained by
\citep{Cantador:RecSys2011}. The dataset consists of 1,892 users
and 17,632 artists whom the users tagged and/or listened to.
More than 95\% of users in the dataset have 50 artists listed
in their profiles as a result of the method used
to collect the data. There are 11,946 unique tags in the dataset,
which were assigned by users to artists 186,479 times. Each user
assigned on average 98.56 tags, 18.93 of which are distinct.
Each artist was assigned 14.89 tags on average, of which 8.76 are
distinct. The dataset also contains social information regarding
12,717 bidirectional friendship linkss established between Last.fm
users, based on common music interests or real life friendship.

\begin{figure}[ht!]
\centering
	\subfloat[Distribution of the number of friends per user]{
		\includegraphics[width=0.4\textwidth]{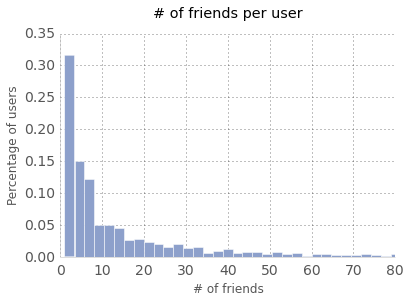}
	\label{fig:characterisation_friends_distributions}
	}
\hspace{1.5em}
	\subfloat[Distribution of the average number of listens per user/artist and overall]{
		\includegraphics[width=0.4\textwidth]{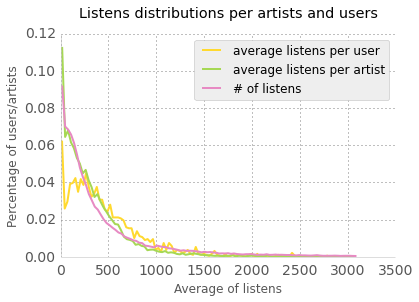}
	\label{fig:listens_distribution}
	}
	\caption{Last.fm data characteristics}
\label{fig:characterisationLast}
\end{figure}

A brief characterisation of the dataset is shown in Figure \ref{fig:characterisationLast}. Figure \ref{fig:characterisation_friends_distributions} illustrates the distribution of the number of friends per user. The average number of user-to-user edges is low, which is illustrated by the vast majority of users having less than 10 friends and about half of users having less than four friends. Intuitively, a friendship edge between two users can be an indicator of similar tastes, and as such, friendship-based features are expected to affect the recommendations. Figure \ref{fig:listens_distribution}, shows the distributions of the number of listens per artist, user, and in total. It can be observed that the overall and per artist distribution are highly similar. The user-based distribution resembles the same behaviour, but drops faster. This aligns with the intuition that the number of users who listen to several hundreds of artists is smaller than the number of artists who are listened by several hundreds of users \citep{haupt2009last}.

There are four relationships in the Last.fm dataset:
\textit{[user, listens, artist]}, \textit{[user, uses, tag]},
\textit{[tag, used, artist]}, and  \textit{[user, friend, user]}.
%\label{sec:lastfm_methodology}
The task defined for this dataset was to predict the artists to whom
a users will listen the most, i.e., the predicted relationship was
\textit{[user, listens, artist]}.
This task requires first predicting
the number of times each user will listen to each artist, then ranking
the artists, and choosing the top K artists.
%, and finally comparing the chosen top K artists to the really listened artists.
Based on the sub-graph generation process detailed in Algorithm
\ref{algo:subset_graphs_generation} and the relationship being predicted,
the data can be represented via eight graph schemes in general. Four graph schemas
that incorporate the source and target entities were evaluated:
%user-to-artist tagging, user-to-user friendships, and, for the purpose of comparison, the schemes in which these were absent (the graphs that were excluded from the analysis are those that did not have edges between users and tags, or tags and artists). The schemes are as follows.
\begin{itemize}
	\item A bipartite graph that includes users and artists only, denoted as the baseline (BL)
    \item A non-bipartite graph that 	includes users, social links, and artists (BL+F)
    \item A non-bipartite graph that includes users, artists, and tags assigned by users to artists (BL+T)
    \item A graph that includes all the entities and relationships: users, tags, artists
	and social links (BL+T+F).
\end{itemize}

\begin{figure}[ht!]
	\centering
	\includegraphics[width=1.0\textwidth,height=0.5\textheight,keepaspectratio]{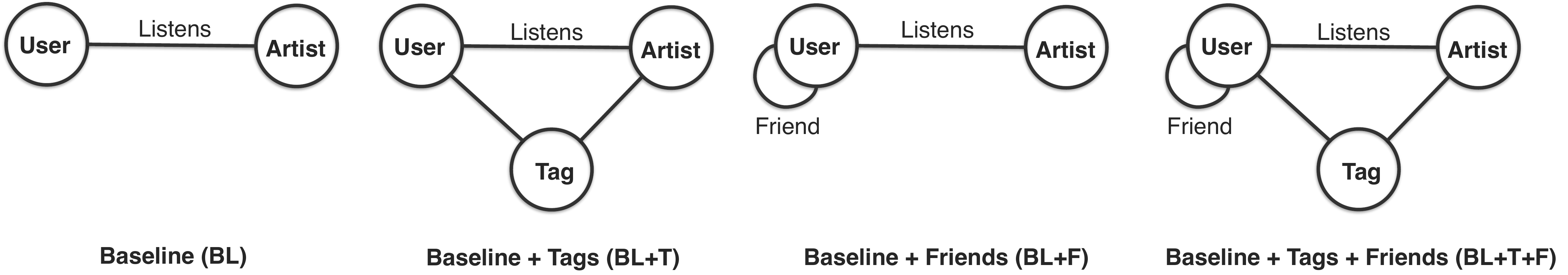}
	\caption{Graph representations for dataset \rom{1} (Last.fm)}
	\label{fig:lastfm_graph_schemes}
\end{figure}

The four graphs are illustrated in Figure \ref{fig:lastfm_graph_schemes}.
For each of the graphs, two sets of features were generated: basic
features, as well as a set of extended features
associated with the auxiliary data being included. The generated
features are used as the input for a Gradient Boosting Decision
Tree regressor \citep{Friedman00}, trained to predict the number
of listens for a given user-artist pair.

A 5-fold cross validation was performed. Users with fewer than
five ratings were pruned, to ensure that every user has at least
one rating in the test set and four in the training set. For each
training fold, a graph was created for each graph model shown
in Figure \ref{fig:lastfm_graph_schemes}.
%In order to evaluate each graph model and its associated features, the performance of the regressor was measured for predicting the number of listens and ranking the artists.
For each user, in the test set, a candidate set of artists was created by
selecting artists out of the set of the artists listened to by the user and
complementing these by randomly selected artists. For example, a
candidate set of 100 artists included 10 artists listened to by
the user and 90 random artists. Three different candidate sizes
were evaluated: 50, 100, and 150.

Then, a regressor was used to predict the
number of listens for each artist in the candidate set, rank the set, and
compute precision at 10 (P@10) as the performance
metric \citep{ShaniG11}.
%That is, the intersection between the top-10 artists in the predicted ranked set and the real 10 artists listened by the user that were included in the candidate set was computed.
If candidate set $CS$ consists of the artists selected from a user's artist set denoted by $UA$ and the randomly selected
artists set $RA$, then P@10 is computed by $P@10 = (UA \cap top\_10\_artists(UA \cup RA))/10$, where top\_K\_artists is the list of top-K artists in $CS$ ranked according to the predicted number of listens.
Finally, an average of the P@10 scores across all the users in the test set is computed.
%This evaluation method is known as \textit{top-n recommendations} \citep{Cremonesi10} and is applied to evaluate recommender systems that use implicit factors, e.g., number of views or behavior logs.
In order to evaluate the significance in the performance of the various graph schemes feature sets, a
two-sided t-test was applied on the results.

\subsection{Dataset \rom{2} -- Yelp (from RecSys-2013)}
\label{sec:dataset_yelp1}

The second dataset is of users’ relevance
feedback given for businesses, such as restaurants, shops, and
services. The dataset was released by Yelp for the RecSys-2013 Challenge
\citep{blomo2013}, and is publicly available.\footnote{https://www.kaggle.com/c/yelp-recsys-2013/data}
For the analysis, users with less than five reviews were filtered out,
which resulted in 9,464 users providing 171,003 reviews and the
corresponding ratings for 11,197 businesses. The average number of
reviews per user is 18.07 and the average number of
reviews per business is 15.27. A key observation
regarding this dataset is the distribution of ratings, which were
almost all positive (more than 60\% of ratings were at least 4
stars on a 5-star scale), and the low variance of ratings across
businesses and users.
This phenomenon is common in star rating datasets, where users tend
to review fewer items that they did not like.
%\citep{ctfm1}.
%Also filtered out from the dataset were user and business metadata features that were highly sparse.

\begin{figure}[ht!]
\centering
	\subfloat[Distribution of the number of reviews per user]{
		\includegraphics[width=0.3\textwidth]{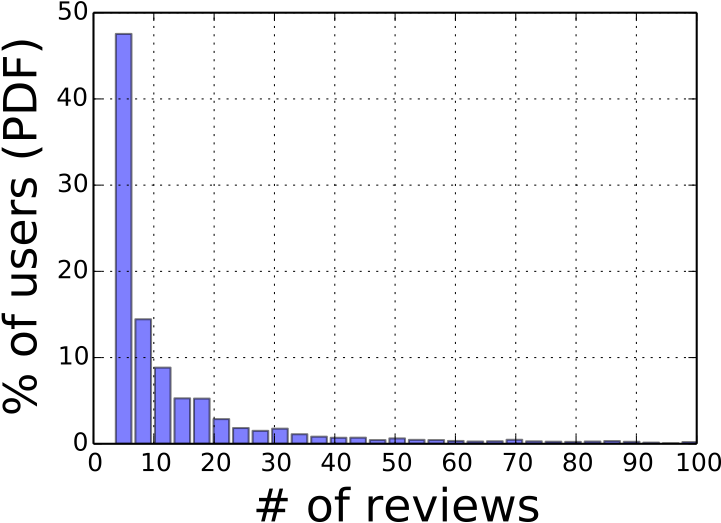}
	\label{fig:userpng}
	}
\hspace{1.5em}
	\subfloat[Distribution of the number of reviews per business]{
		\includegraphics[width=0.3\textwidth]{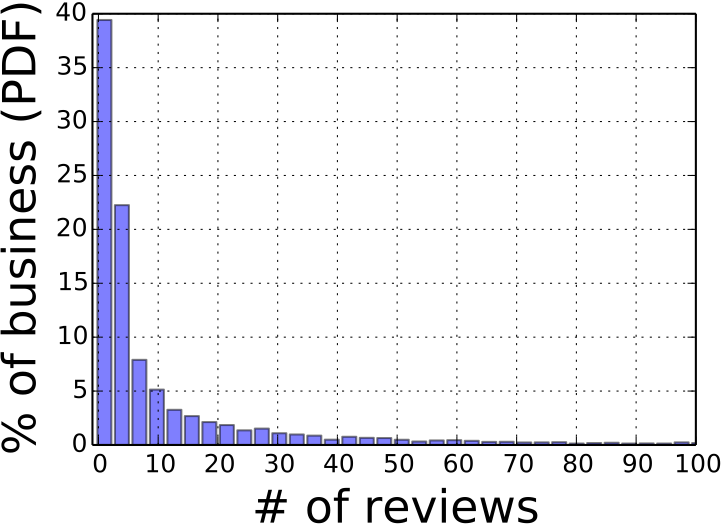}
	\label{fig:buspng}
	}
	\caption{Yelp dataset characteristics}
\label{fig:Yelp1}
\end{figure}

\begin{figure}[ht!]
	\begin{center}
		\includegraphics[width=0.65\textwidth]{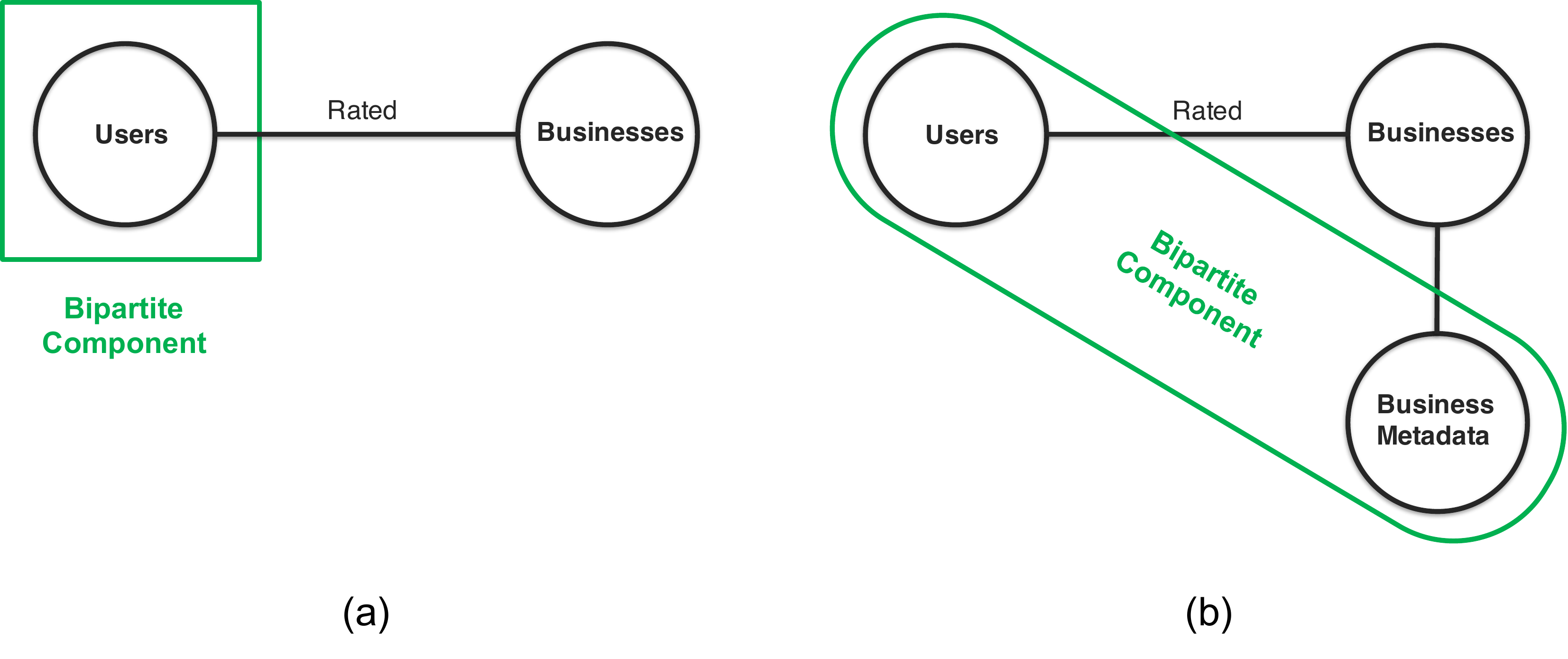}
	\end{center}
	\caption[Yelp - graph schemes]{Graph representations for dataset \rom{2} (Yelp - RecSys-2013 Challenge)}
	\label{fig:yelp_graph_schemes}
\end{figure}

%Table \ref{tab:tbl_data_statistics} and
Figure \ref{fig:Yelp1} summarizes the basic statistics of users and businesses in the Yelp dataset.
Figure \ref{fig:userpng} illustrates the distribution of the number of reviews and ratings per user. A long tail distribution of the number of businesses a user reviewed can be observed, with more than 75\% of the users providing less than 10 reviews. Likewise, we observe in Figure \ref{fig:buspng} the distribution of the number of reviews a business received. Only 24\% of businesses attract more than 10 reviews, while only a few businesses (less than 2\%) have a relatively high number of reviews (more than 100). Despite the high number of categories in the data, the average number of categories with which a business is associated is only 2.68. Every business is also associated with a single location.

%\begin{table}[htbp]
%  \centering
%      \resizebox{0.8\columnwidth}{!}{
%\begin{tabular}{lrrrrrrrr}
%\toprule[1.5pt]
%{} &       mean &        std &  min &  25\% &  50\% &  75\% &  max \\
%\midrule
%Number of reviews per user               &  18.07 &  29.60 &    5 &    6 &    9 &   17 &  588 \\
%Number of reviews per business           &  15.27 &  32.43 &    1 &    3 &    5 &   13 &  528 \\
%Number of categories per business        &   2.68 &   1.14 &    0 &    2 &    2 &    3 &   10 \\
%\bottomrule[1.5pt]
%\end{tabular}
%}
%\caption{Yelp - basic statistics}
%\label{tab:tbl_data_statistics}
%\end{table}

\label{sec:yelp1_methodology} The task defined for this dataset is
the one originally defined for the RecSys-2013 challenge, i.e.,
to predict the ratings a user
will assign to businesses. Two graph models were
implemented and evaluated based on this dataset: a
\emph{bipartite} model with sets of vertices U and B representing users
and businesses and a \emph{tripartite}\footnote{The use of `tripartite'
    is slightly inconsistent with the canonic definition, such that the
    ``bipartite graph with metadata nodes'' notation would be more appropriate.
    For the sake of brevity, the bipartite and tripartite terminology is used.}
model with sets of vertices U, B, and M representing users, businesses,
and metadata items, respectively. The high-level graph representation models
are illustrated in Figure \ref{fig:yelp_graph_schemes}, while the detailed
presentation of the sub-graphs will be given in Section \ref{sec:dataset_yelp2},
in which the follow-up dataset is presented.

The features generated for this dataset were aggregated into three groups:
%Users are denoted by $u$ and businesses by $b$.
\begin{itemize}
%\newpage
	\item \emph{Basic} features that include only the unique identifiers of
	users $\{u_i\} \in U$ and businesses $\{b_j\} \in B$.
    \item \emph{Manual} features that include the
	number of reviews by $u_i$, average rating of $u_i$, number of reviews
	for $b_j$, number of categories $\vert$\{$m$\}$\vert$ with which $b_j$
	is associated, average number of businesses in \{$m$\}, average
	rating of businesses in \{$m$\}, the main category\footnote{Each business in the Yelp dataset is associated with multiple categories, some having an internal hierarchy. The main category is the most frequent root category a business was associated with.}
%For example if a business is associated with the following categories [``Bars'', ``Austrian'',``Bangladeshi'',``Barbeque'',``Casinos''] and the respective root categories are [``Nightlife'', ``Restaurants'', ``Restaurants'', ``Restaurants'', ``Arts \& Entertainment''], then ``Restaurants'' was set as the main category.}
    of $b_j$, average degree of businesses associated with the main category of $b_j$, average degree of businesses in \{$m$\},
%\footnote{Although the last two features, ``business\_main\_category\_degree'' and ``business\_average\_degree\_of\_categories'' are graph features, they were at the time manually engineered outside of the framework's feature extraction process, and hence included in the ``Manual Features'' set and not the ``Graph'' one.},
    and the location of $b_j$.
    \item \emph{Graph} features that include the degree centrality, average
	neighbor degree, PageRank score, clustering coefficient, and node
	redundancy. These features were generated for both user nodes
	$u_{i}$ and business nodes $b_{j}$, whereas an additional shortest
	path feature was computed for the pairs of ($u_{i}$,$b_{j}$).
\end{itemize}

In this case, a Random Forest regression model \citep{breiman2001random}
was applied for the generation of the predictions of users’ ratings for businesses.
At the classification stage, the test data items were run through
all the trees in the trained forest. The value of the predicted rating
%class of a test item was determined by the majority voting of the terminal nodes reached when traversing the trees. In the case of a regression model
%applied for predictions of continuous values rather than discrete class labels, the predicted score
was computed as a linear combination of the scores of the terminal nodes
reached when traversing the trees.
It should be noted that the ensemble of trees in Random Forest and
the selection of the best performing feature in each node
inherently eliminate the need for feature selection. Since every
node uses a single top performing feature for decision making, the
most predictive features are naturally selected in many nodes and
%. Hence, these features have a strong impact on the classifier/regressor, such that
the ensemble of multiple trees virtually replaces the feature selection process.
%A more elaborate presentation of the Random Forest algorithm is provided in \citep{breiman2001random}.

A 5-fold cross validation was performed. For each fold, the predictive
model was trained using both the original features encapsulated in
the dataset and the new graph features. The basic and manual groups of
features were populated directly from the reviews, whereas the graph
features were populated from the bipartite and tripartite graph
representations and augmentrf the former groups of features.
Predictive accuracy of various combinations of features
was measured using the widely-used RMSE metric \citep{ShaniG11}, computed
as $RMSE = {\sqrt { \frac{\sum_n (\hat{y}_{t} - y_{t} )^{2}}{n}}}$,
where $n$ is the number of predictions, $\hat{y}_{t}$ are the predicted
values, and $y_{t}$ are the actual user ratings.  A two-sided t-test
was applied to validate the statistical significance of the results.

\begin{figure}[ht!]
    \centering
%    \subfloat[Distribution of number of reviews per user]{
%        \includegraphics[width=0.35\columnwidth]{gfx/yelp2_user_reviews_count.pdf}
%        \label{fig:yelp2_user_reviews_distribution}
%    }
%    \hspace{0.5em}
%    \subfloat[Distribution of number of reviews per business]{
%        \includegraphics[width=0.35\columnwidth]{gfx/yelp2_business_reviews_count.pdf}
%        \label{fig:yelp2_business_reviews_distribution}
%    }
%    \hspace{0.5em}
    %\subfloat[Distribution of social links in dataset \rom{4}]
    {
        \includegraphics[width=0.75\columnwidth]{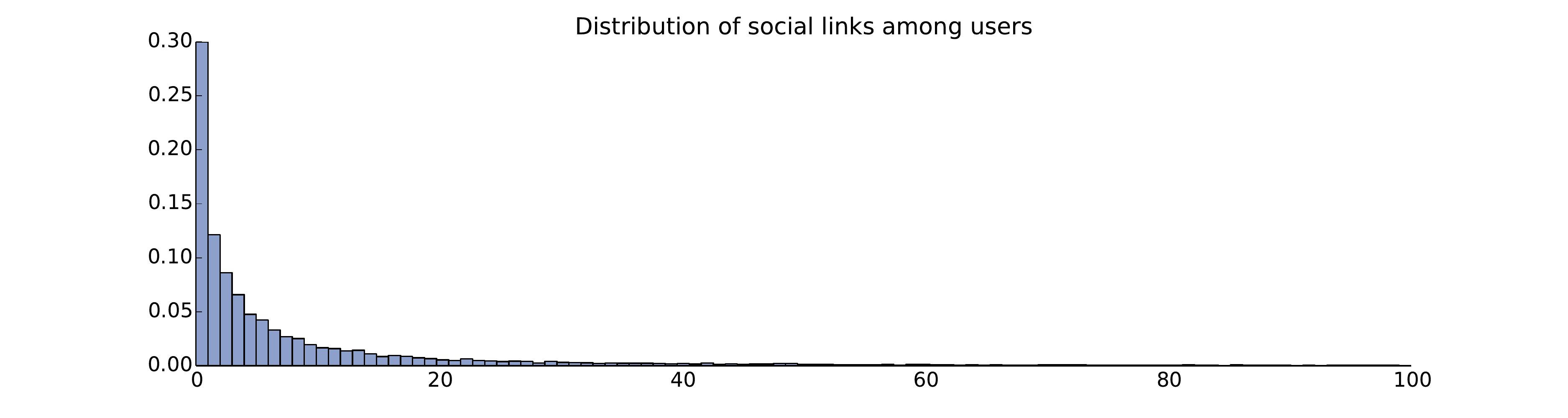}
        \label{fig:yelp2_social_links_distribution}
    }
    \caption{Yelp II Dataset characteristics -- distribution of social links}
    \label{fig:yelp2_characterisation_social}
\end{figure}

\begin{figure}[ht!]
	\begin{center}
		\includegraphics[width=0.95\textwidth]{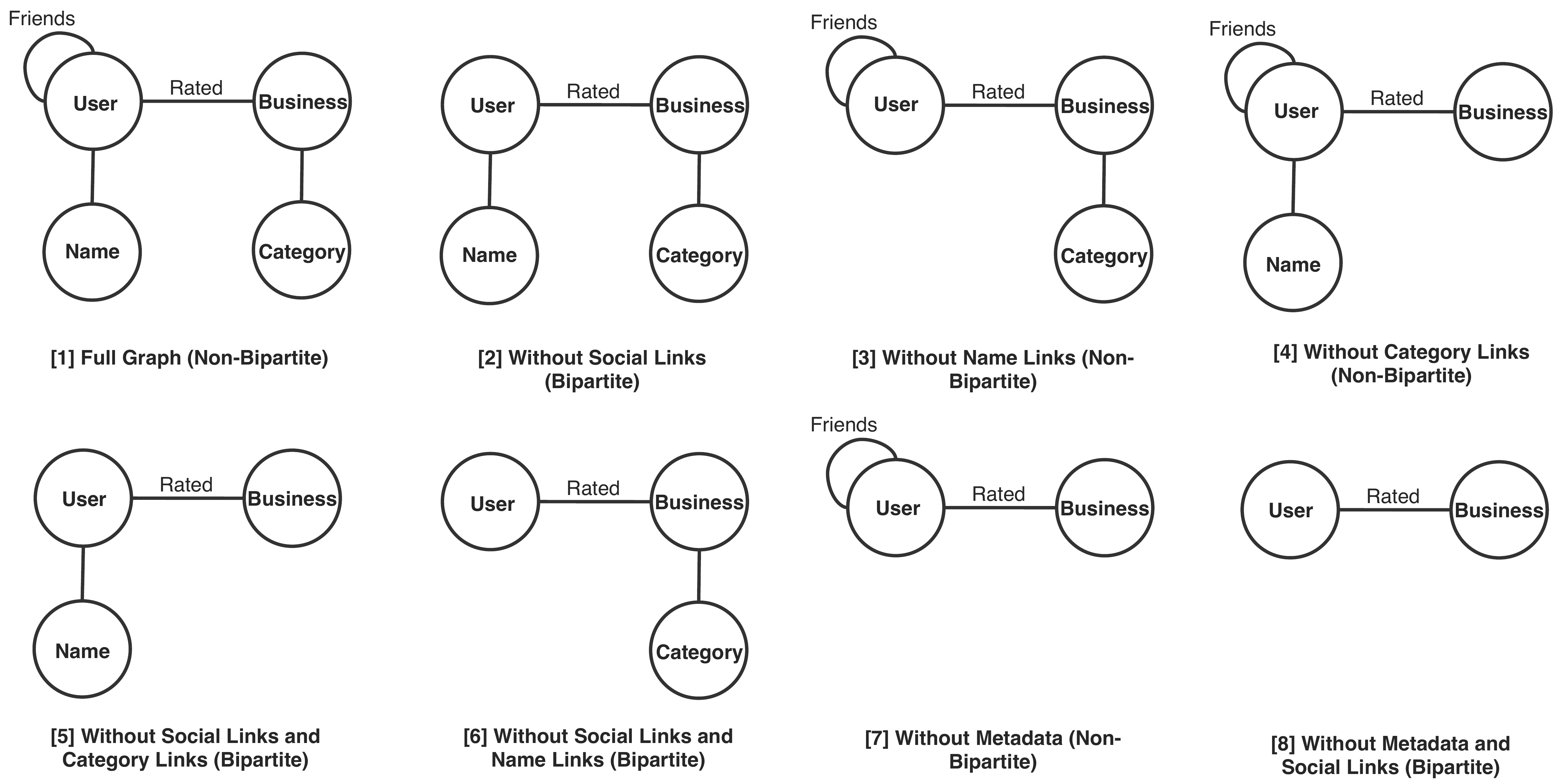}
	\end{center}
	\caption[Yelp \rom{2} - graph schemes]{Graph representations for dataset \rom{3} (Yelp \rom{2})}
	\label{fig:yelp2_graph_schemes}
\end{figure}

\subsection{Dataset \rom{3} -- Yelp \rom{2} (with social links)}
\label{sec:dataset_yelp2}

The third dataset is an extension that was released by Yelp to the previous dataset. The new
version contains more users, businesses and reviews (although their distribution still resembles
the one shown in Figure \ref{fig:Yelp1}), and, more importantly, new information regarding users'
social links.
%The distributions of reviews per users and businesses are provided in Figures \ref{fig:yelp2_user_reviews_distribution} and \ref{fig:yelp2_business_reviews_distribution}. As can be seen, both distributions follow a long tail distribution, where most users and businesses have a low number of reviews.
The distribution of the social links among users is illustrated in
Figure \ref{fig:yelp2_characterisation_social}.
%There were 312 users who had more than 100 social links and were excluded from the figure for readability.
It can be seen that the social links follow a long tail distribution,
where most users have a small number of links: 29\% with no links, 57\% with less than 20 links, and only
a few users with more than 20 links. The social links also break the bipartite structure of the first Yelp dataset,
which influences the generated graph features.

\label{sec:yelp2_methodology}

The task for this dataset is
identical to that of the first Yelp dataset, i.e.,
predicting users’ ratings for businesses. Eight graph models were
generated and evaluated based on this dataset. The models are
illustrated in Figure \ref{fig:yelp2_graph_schemes} and,
depending on the availability of the user-to-user friendship edges,
categorized as bipartite or non-bipartite.
%Users were modeled as nodes in these graphs, in order to evaluate the performance of the graph-based approach on a different type of metadata -- one that does not fall into the social box or ``naturally categorized'' one (such as business categories). The assumption, which is based on previous studies such as \citep{gallagher2008estimating}, is that linking users based on their names could implicitly lead to uncovering ethnic, gender, and even age connections.
The complete graph is shown in the top-left schema. In the following three
schemes one type of edges is missing: either social links, user names, or categories.
In the next three, two types of edges are misisng: social and categories, social and
names, and names and categories. Finally, in the bottom-right graph all three are
missing.

The generated features presented in Section \ref{sec:the_extracted_features}
are referred to in the evaluation of this dataset as the \textit{basic} features. These
features are aggregated into groups, based on the graph scheme from which
they were extracted. For example, all the features extracted from
the graph named ``without category links'' in Figure \ref{fig:yelp2_graph_schemes}
were grouped into a combination having the same name. Another evaluated
combination includes the union of all the features generated from all
the graph schemes, and this is named ``all graph features''. Finally, the union
of ``all graph features'' with the ``basic features'' is referred to as ``all features''.

A 5-fold cross validation was performed. For each fold, the predictive
models were trained using graph features extracted from
each of the above feature sets. The evaluation was
conducted three times, each time training the models using a different
method (Random Forests, Gradient Boosting, and SVM), in order to
evaluate how the choice of method impacts the results. Predictive
accuracy of various feature combinations was
measured using the RMSE metric \citep{ShaniG11}, and a two-sided t-test was applied
to validate statistical significance.

\subsection{Dataset \rom{4} -- OSN}
\label{sec:dataset_osn}

The fourth dataset is an Online Social Network (OSN) profile dataset that was collected from six large networks: Facebook, LinkedIn, Last.fm, Blogger, YouTube, and LiveJournal. The profiles were manually linked and matched to each other by the users themselves, as they mentioned their user names on other OSNs. The lists of user interests were then extracted from the OSNs and categorized into five domains: movies, music, books, TV, and general. The categorization was explicitly made by the users on Blogger, Facebook, and YouTube; all Last.fm interests were categorized as music; no categorization was available on LinkedIn and LiveJournal, so that there interests were treated as general. Users having one interest only and interests mentioned by one user only were filtered, such that the resultant dataset contained 21,880 users with an average of 1.49 OSNs and 19.46 interests per user.

It can be observed in Table \ref{tab:itemsDomain} that the most common
interests in user profiles are from the music and general domains. Table \ref{tab:itemsNetwork}
shows that Facebook is, by far, the OSN with the most listed interests. Table
\ref{tab:availability} shows the number of users who have at least one
interest for a domain and OSN combination, with the right-most column
indicating the total number of users. The OSN with the largest number of
profiles is Facebook, followed by LinkedIn and Last.fm. Music
interests are the most common across the Facebook and YouTube
profiles, while general interests are the most common in Blogger
profiles. Finally, Table \ref{tab:averageItem} shows the average
number of interests a user has in each domain and OSN.

\label{sec:osn_methodology}
The task defined for this dataset was to predict the interests of the users, based on their partial profiles. The data were represented using a single graph model because of the availability of only two entities: users and interests. The model is bipartite graph G = \{U, I, E\}, where users U = \{$u_{i}$\} and interests I = \{$i_{i}$\} are the vertices. User vertices are connected to interest vertices with an edge if the interest is mentioned in one of the available OSN profiles, i.e., E = \{$e_{ij} \mid$ if $u_{i}$ listed $i_{j}$\}). The edges are labeled by the OSN(s), in which the interest was listed.

\begin{table}[ht!]
	\centering
    \small
	%\resizebox{0.8\columnwidth}{!}{
		\begin{tabular}{|l|c|c||l|c|c|}
			\hline
			Domain    & Total   & Unique & Domain  & Total  & Unique \\ \hline
			General interests & 154,245 & 13,053 & Books & 24,404 & 3,789  \\ \hline
			Movies    & 54,382  & 5,190  & TV    & 53,508  & 4,027  \\ \hline
			Musics    & 139,307 & 21,255 & All   & 425,846  & 47,314  \\ \hline
		\end{tabular}
	%}
	\caption[OSN - interests across domains statistics]{Total number of interests and unique interests in each domain}
	\label{tab:itemsDomain}
\end{table}

\begin{table}[ht!]
	\centering
    \small
	%\resizebox{0.8\columnwidth}{!}{
		\begin{tabular}{|l|c|c||l|c|c|}
			\hline
			Network     & Total     & Unique                & Network       & Total     & Unique \\ \hline
			Blogger     & 27,045    & 6,587                 & Livejournal   & 30,924    & 5,198  \\ \hline
			Facebook    & 253,217   & 31,511                & YouTube       & 2,753     & 1,561  \\ \hline
			Last.fm      & 63,952    & 16,483                & All           & 425,846   & 47,314  \\ \hline
			LinkedIn    & 47,955    & 6,325                 & ~             & ~         & ~  \\ \hline
		\end{tabular}
	%}
	\caption[OSN - interests across networks statistics]{Total number of interests and unique interests in each OSN}
	\label{tab:itemsNetwork}
\end{table}

\begin{table}[ht!]
	\centering
    \small
	%\resizebox{0.8\columnwidth}{!}{
		\begin{tabular}{|l|c|c|c|c|c|c|}
			\hline
			~           & General interests         & Movies            & Music                     & Books             & TV    & Total\\\hline
			Blogger     & 2,716    & 1,090              & 1,370                     & 518               &         & 3,136\\ \hline
			Facebook    & 8,391   & 8,922   & 10,453   & 6,565   & 9,619& 11,619\\ \hline
			Last.fm      &                 &                & 7,042             &             &    & 7,042\\ \hline
			LinkedIn    & 7,755    &               &                          &              &    & 7,755\\ \hline
			LiveJournal & 1,494    &              &                        &               &    & 1,494\\ \hline
			YouTube     & 448               & 484               & 650           & 552               &   & 1,548\\ \hline
		\end{tabular}
	%}
	\caption[OSN - interests across users/domains/networks statistics]{Number of users who have at least one interest
		in each domain and OSN}
	\label{tab:availability}
\end{table}

\begin{table}[ht!]
	\centering
    \small
	%\resizebox{0.8\columnwidth}{!}{
		\begin{tabular}{|l|c|c|c|c|c|}
			\hline
			~           & General interests & Movies & Music & Books & TV \\\hline
			Blogger     & 5.913         & 2.976      & 4.676     & 2.577     & \textbf{--}   \\ \hline
			Facebook  & 6.999         & 5.662      & 6.509      & 3.414     & 5.562  \\ \hline
			Last.fm      & \textbf{--}          & \textbf{--}       & 9.082      & \textbf{--}      & \textbf{--}   \\ \hline
			LinkedIn    & 6.183         & \textbf{--}       & \textbf{--}       & \textbf{--}      & \textbf{--}   \\ \hline
			LiveJournal & 20.69         & \textbf{--}      & \textbf{--}       & \textbf{--}      & \textbf{--}   \\ \hline
			YouTube     & 1.288         & 1.278      & 1.395     & 1.177     & \textbf{--}   \\ \hline
		\end{tabular}
	%}
	\caption[OSN - interests across users/domains/networks statistics, continued]{Average number of interests available per user in each domain and OSN}
	\label{tab:averageItem}
\end{table}

From the graph, a set of graph and manually engineered
features was extracted. They can be categorized into two groups:
user features and interest features. Each of these groups can be
split into two sub-groups: basic manual features ($IB$ and $UB$)
and graph-based features ($IG$ and $UG$). The $UB$ features include
the number of OSNs of which the user is a member (${UB}_1$), the number of
user interests in each domain -- books, TV, movies, music, general
(${UB}_2$, ${UB}_3$, ${UB}_4$, ${UB}_5$, ${UB}_6$, respectively), and the total number of
interests (${UB}_7$). The $IB$ features include the number of users who
liked the interest (${IB}_1$), the number of OSNs where the interest
appears (${IB}_2$), Boolean features signifying whether the interest is
listed on each OSN -- Blogger, LinkedIn, Last.fm, LiveJournal,
Facebook, YouTube (${IB}_3$, ${IB}_5$, ${IB}_6$, ${IB}_7$, ${IB}_8$, ${IB}_9$,
respectively), and the domain to which the interest belongs (${IB}_4$).

The graph-based features are identical for users and interests and
contain: Degree centrality (${IG}_1$, ${UG}_2$), Node redundancy (${IG}_2$,
${UG}_1$), Clustering coefficient (${IG}_3$, ${UG}_4$), Average neighborhood degree
(${IG}_4$, ${UG}_3$), PageRank (${IG}_5$, ${UG}_1$), and the Shortest path feature
computing the distance between a user-interest pair.
Additional features defined for this dataset are $IG_{all} = \{IG_{i}\}$, $UG_{all} = \{\cup UG_{i}\}$,
$IB_{all} = \{IB_{i}\}$, and $UB_{all} = \{BG_{i}\}$. Finally, $I_{all} = \{IB_{all} \cup IG_{all}\}$
and $U_{all} = \{UB_{all} \cup UG_{all}\}$.

The experiments using the OSN dataset evaluated the effect of the features on the predictions of the likelihood of a
user to list an interest. A Random Forest classifier was used and trained on the user-interest pairs augmented with
their features. Each pair was classified into the `like' or `dislike' classes. 10-fold
cross validation was applied for evaluation. For each fold, a graph was built, and then, the above features were extracted and fed into the classifier.
Since no real disliked interests were in the data, random interests were selected from the interests not listed by the user.
The number of disliked interests was equal to the number of liked interests for each user. The synthetic disliked interests
were used only to train the classifier and not used in the evaluation.
%A similar approach was used in \citep{mladenic1996personal}, where unvisited URLs were considered as indicators of user lack of interest in a topic.
Precision was the metric chosen to evaluate the quality of predictions for a user: $P=\frac{TP}{TP+FP}$, where
$TP$ is the number of correctly and $FP$ is the number of incorrectly predicted interests \citep{ShaniG11}.

\subsection{Dataset \rom{5} -- Movielens}

\label{sec:dataset_movielens}

\begin{figure}[!htbp]
    \centering
    \subfloat[Distribution of ratings across users]{
        \includegraphics[width=0.4\columnwidth]{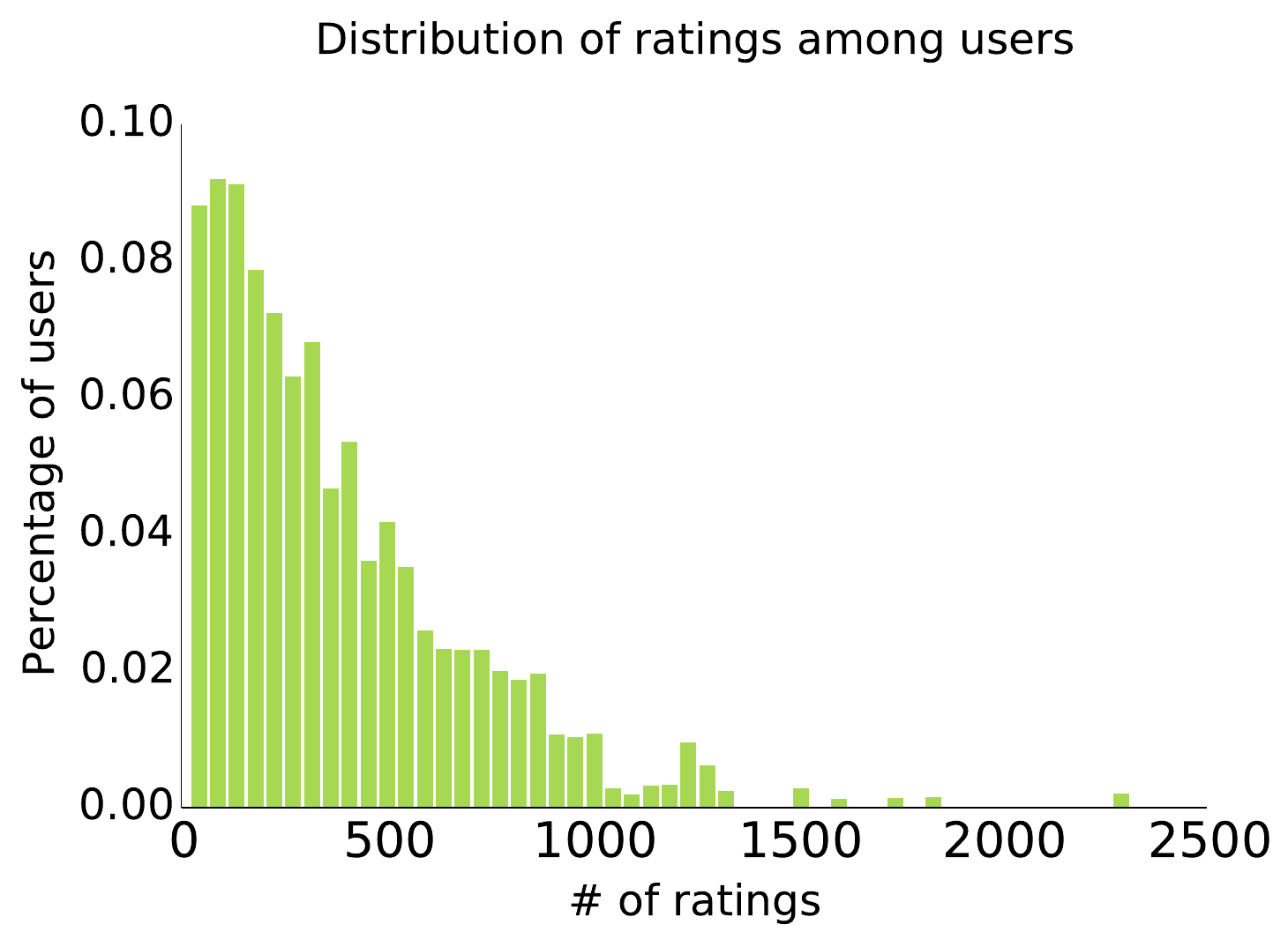}
        \label{fig:movielens_dataset_characterisation_user_ratings_dist}
    }
    \hspace{0.3em}
    \subfloat[Distribution of ratings across movies]{
        \includegraphics[width=0.4\columnwidth]{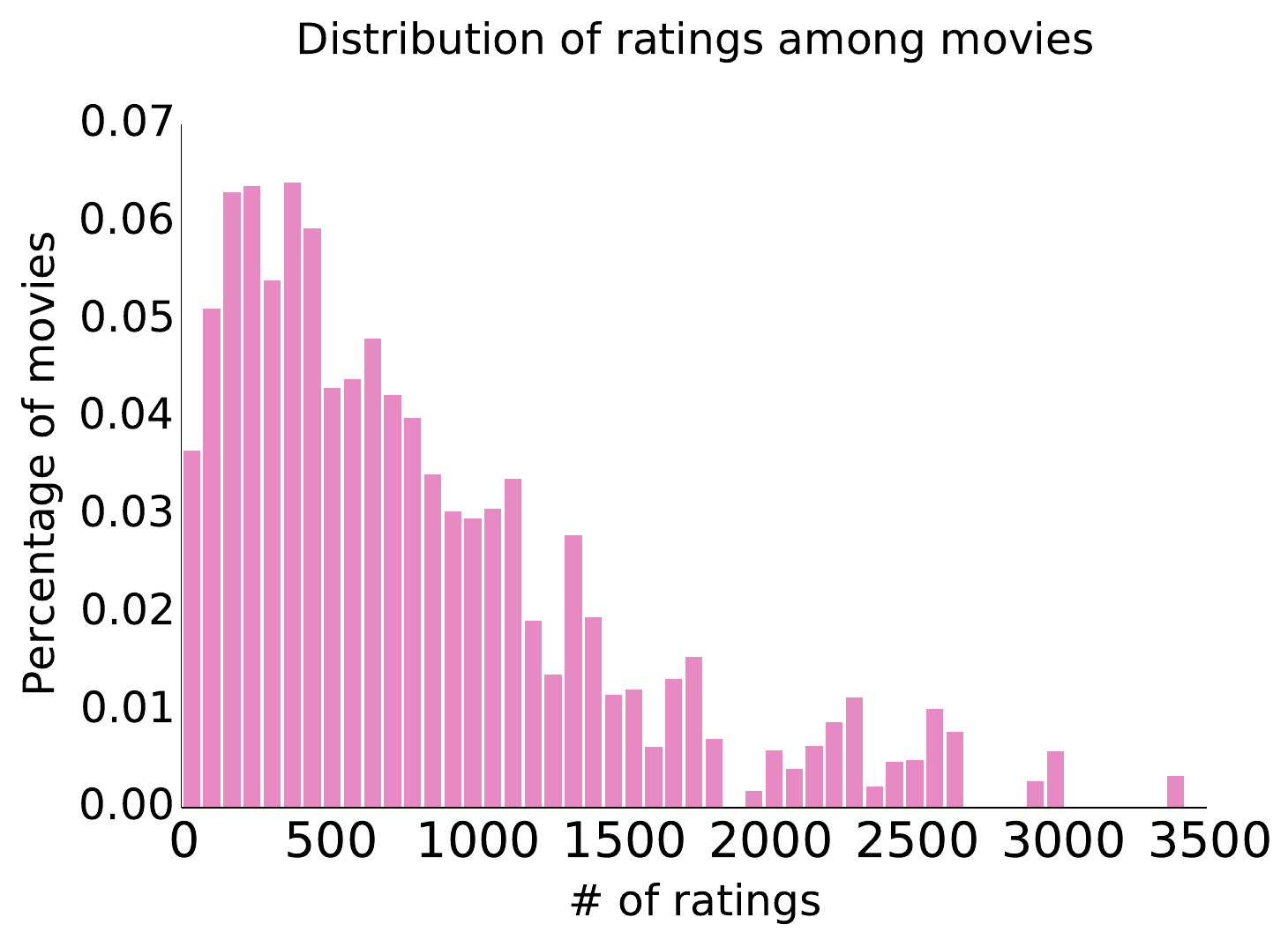}
        \label{fig:movielens_dataset_characterisation_movie_ratings_dist}
    }
    %\hspace{0.3em}
    %\subfloat[Distribution of users across gender]{
    %    \includegraphics[width=0.31\columnwidth]{gfx/movielens_gender_dist.pdf}
    %    \label{fig:movielens_gender_dist}
    %}
    \caption{Movielens dataset characteristics}
    \label{fig:movielens_dataset_characterisation}
\end{figure}

Movielens \citep{lam2012movielens} is a classical recommender system’s dataset studied in numerous prior works. In this work it is used to show that the graph-based approach is as effective on legacy datasets as on more recent datasets including social data. The 1M Ratings Movielens dataset consists of 1,000,209 ratings assigned by 6,040 users for 3,883 movies, on a discrete scale of 1 to 5 stars. Each user in the dataset rated at least 20 movies. The distribution of ratings across users and movies is illustrated in Figures \ref{fig:movielens_dataset_characterisation_user_ratings_dist} and
\ref{fig:movielens_dataset_characterisation_movie_ratings_dist}, respectively.
The dataset contains metadata of both users and movies. The user metadata includes the gender, occupation, zip code area, and age
group, while the movie metadata contains the genre(s) of the movies.

\begin{figure}[!ht]
	\begin{center}
		\includegraphics[width=0.8\textwidth]{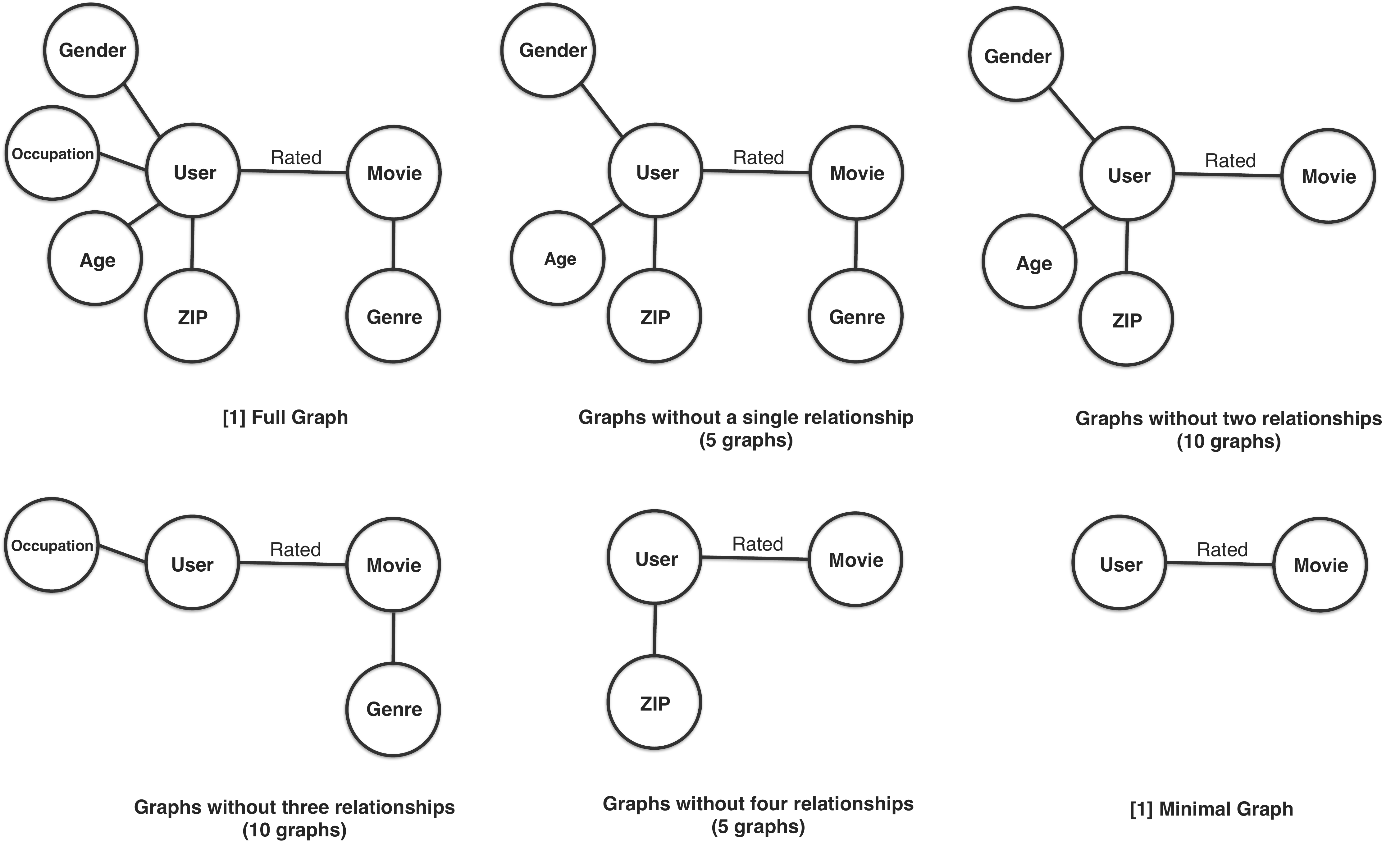}
	\end{center}
	\caption[Movielens - graph schemes]{Graph representations for dataset \rom{5} (Movielens).
		Each graph is an example of the sub-graphs in the group.}
	\label{fig:movielens_graph_schemes}
\end{figure}

\label{sec:movielens_methodology} The task defined for this dataset was to predict what ratings would users assign to movies. Based
on the above description of the dataset, 32 graph schemes were generated and evaluated (see Figure \ref{fig:movielens_graph_schemes}).
The schemes are categorized based on the number of relationships
that were removed from the complete graph that contains all the
entities and relationships. As can be seen, there are four
categories: schemes with a single node type removed, containing 5
sub-graphs, schemes with 2 node types removed containing 10 sub-graphs,
schemes with 3 node types removed containing 10 more sub-graphs, and
finally, schemes with 4 node types removed containing 5 graphs.
The minimal graph scheme is the one from which all the entities and
relationships were removed, except for the source and target
entities and the predicted `rating' relationships.

A 5-fold cross validation was performed. For each fold, the predictive models were trained using graph features extracted from each of the above graph schemes. The evaluations were conducted twice, training the models using the Random Forest and Gradient Boosting approaches, in order to evaluate how the choice of the learning method impacts the results. The predictive accuracy of various combinations of the above feature sets was measured again using the RMSE and MAE predictive accuracy metrics \citep{ShaniG11}, and a two-sided t-test was applied to validate statistical significance.

\subsection{Summary of the datasets, features, and metrics}
\label{sec:datasets_summary}
Table \ref{tab:dataset_summary} summarizes this section and presents the experimental datasets, number of source and target entities, various sub-graph schemes investigated, number of extracted feature sets, groups of features, and evaluation metrics exploited. The five datasets contain large numbers of users and items and cover a broad range of data types, application domains, and recommendation tasks. The datasets also contain both legacy and recently collected datasets, such that the evaluation presented in the following section offers solid empirical validity.

\begin{table}[htbp!]
	\centering
	
	\resizebox{1.05\textwidth}{!}{
		
	\begin{tabular}{lllcclll}
	\toprule[1pt]
	Dataset & \shortstack[l]{Source and \\ Target Entities} & Graph Schemes & \shortstack[l]{Graphs} & \shortstack[l]{Feature Sets} & \shortstack[l]{Extracted\\ Features} & \shortstack[l]{Learning Method} & \shortstack[l]{Evaluation\\ Metric} \\
	\midrule[1pt]
	Last.fm & \shortstack[l]{1,892 (users) \\ 17,632 (artists)} & \shortstack[l]{Bipartite + Non-bipartite \\(w/\ social links, w/\ tags, \\ w/\ social links+tags)} & 4 & 7 & \shortstack[l]{Basic graph features, \\extended graph features} & \shortstack[l]{Gradient Boosting}& P@K \\
	\midrule[0.5pt]
	Yelp & \shortstack[l]{9,464 (users) \\ 11,197 (businesses)} & \shortstack[l]{\\Bipartite + \\ Bipartite with metadata} & 2 & 13 & \shortstack[l]{Basic graph features, \\ manually engineered} & \shortstack[l]{Random Forest} & RMSE\\
	\midrule[0.5pt]
	Yelp \rom{2} & \shortstack[l]{13,366 (users) \\ 14,853 (businesses)} & \shortstack[l]{Bipartite + \\ Non-bipartite (w/\ social links),\\ with and without metadeta} & 8 & 13 & Basic graph features & \shortstack[l]{Random Forest, \\ Gradient Boosting, \\ Support Vector Machine} & RMSE \\
	\midrule[0.5pt]
	OSN & \shortstack[l]{21,880 (users) \\ 47,314 (interests)} & Bipartite & 1 & 8 & \shortstack[l]{Basic graph features, \\ manually engineered} & \shortstack[l]{Random Forest} & Precision\\
	\midrule[0.5pt]
	Movielens& \shortstack[l]{6,040 (users) \\ 3,883 (movies)} & \shortstack[l]{\\Bipartite + \\ Bipartite with metadata} & 32 & 36 & Basic graph features & \shortstack[l]{Random Forest, \\ Gradient Boosting} & \shortstack[l]{RMSE, \\ MAE} \\
	\bottomrule[1pt]
	\end{tabular}

	}

	\caption{Summary of datasets characteristics}
	\label{tab:dataset_summary}
	
\end{table}
\section{Results and Analysis}
\label{results}

\subsection{Case Study \rom{1}: Overall Contribution of the Graph-based Approach}
\label{ch:evaluation_1}

This case study answers the broad question: \textit{How does the use of graph features affect the performance of rating predictions and recommendation generation in different domains and tasks?} Each of the above datasets was represented by graphs and graph-based features were extracted from the graphs using the approach
detailed in Section \ref{sec:graph_based_feature_extraction}. For each dataset, a matching recommendation task was defined as follows: for the Last.fm dataset the task was to predict the artists to which users will listen; for the two Yelp datasets the task was to predict user ratings for business; for the OSN dataset, to predict the interests in user profiles; and, for the Movielens dataset, to predict user ratings for movies.
The tasks were performed and evaluated under three conditions:
\begin{itemize}
	\item Prediction with versus without the newly extracted graph features
    \item Prediction with user-related versus item-related graph features
    \item Prediction using features of a bipartite graph versus extended graph schemes, e.g, containing metadata.
\end{itemize}

All the evaluations were conducted using the N-fold cross validation methodology \citep{kohavi1995study}, with $N=5$ folds in the Last.fm, both Yelps, and Movielens datasets, and $N=10$ in the OSN dataset. For each fold, the complete graph representation was generated based on the entities from both the training and test sets, except for the relationships being predicted in the test set.
%in order to prevent leakage of information from the test set to the graph metrics.
%\begin{sloppypar}
%Training the prediction model on different feature subsets in each evaluation led to differences in the results. In order to evaluate whether the differences are significant,
A two-sided t-test was conducted with the null hypothesis of having identical expected values across the compared prediction sets. The tests assumed that the predicted ratings using feature set A and predicted ratings using feature set B were taken from the same population.
%This assumption holds for all evaluations, since the training target values used in each evaluation, e.g, Movie ratings in the Movielens training sets, were identical for all subsets evaluated.
The threshold used for a statistically significant difference was p=0.05.
%, and figures such as \ref{fig:yelp_ttest_significance} and  \ref{fig:yelp2_ttest_results} portray the significance of the differences between feature sets performance in each evaluation.

\begin{figure}[ht!]
    \begin{center}
    \includegraphics[width=0.4\textwidth,height=0.6\textwidth,keepaspectratio]{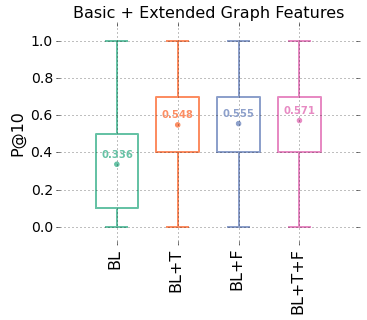}
   \end{center}
	\caption{Precision of feature combinations using the four graphs - Last.fm dataset.}
    \label{tab:tbl_lastfm_feature_combinations_precision}
\end{figure}

\subsubsection{Dataset \rom{1} -- Last.fm Results}
\label{sec:eval1_dataset1_lastfm}

Four graph schemes were generated for the Last.fm dataset, as per the structure in Figure \ref{fig:lastfm_graph_schemes}. For each graph scheme the set of \textit{basic} graph features listed in Section \ref{sec:the_extracted_features} was extracted and populated. The basic features encapsulate only the user-artist listening data and denoted by $\hat{F}$. In addition, when the social and tagging data is available, namely, in the BL+T, BL+F, BL+T+F schemes, the set of \textit{extended} features can be extracted. These features are denoted by $F$, e.g., ${F}_{BL+T}$ denotes the set of extended features extracted from the graph with the tagging data. Note that for the BL schema in Figure \ref{fig:lastfm_graph_schemes}, having neither social nor tagging data, the basic and extended feature sets are identical, i.e., ${F}_{BL} = \hat{F}_{BL}$.

The P@10 results obtained for the extended feature sets extracted from the four schemes are summarized in Figure \ref{tab:tbl_lastfm_feature_combinations_precision}. The boundaries of the boxes represent the 25th and 75th percentile of the obtained P@10, and the average P@10 is marked by the dot inside the boxes. The values of the average P@10 are also given.
The baseline for comparison in this case is the performance of
the graph features extracted from the bipartite scheme ${F}_{BL}$,
which scored P@10=0.336.
%The same set of features extracted from the graph scheme also including social tags nodes and links, $F_{BL+T+F}$, scored P@10=0.444, which is an improvement of 32.14\%.
A notable improvement, between 63\% and 70\%, was observed when the extended
feature sets were extracted. For instance, ${F}_{BL+F}$, scored P@10=0.555, which is
an improvement of more than 65\%. A combination of the extended features using the graph
that includes both social tags and friendships,
%(BL+T+F in Figure \ref{fig:lastfm_graph_schemes}, and
${F}_{BL+T+F}$ is the best performing feature. This scored the highest
P@10=0.571 and improved the baseline P@10 by as much as almost 70\%.

%\begin{table}[ht!]
%	\centering
%	\resizebox{0.6\columnwidth}{!}{
%		\begin{tabular}{lrrrr}
%			\toprule
%			Graph Scheme & $F$ & Improvement & $\hat{F}$ & Improvement\\
%			\midrule
%			BL+T+F & 0.444 & 32.14 \% & 0.571 & 69.94 \%\\
%			BL+F & 0.497 & 47.91 \%& 0.555 & 65.17 \%\\
%			BL+T & 0.498 & 48.21 \%& 0.548 & 63.09 \%\\
%			\rowcolor{LightGray}
%			BL & 0.336 &  & N/A & \\
%			\bottomrule
%		\end{tabular}
%	} \caption[Last.fm.fm - results summary table]{P@10 of graph schemes and
%	feature combination for basic [$F$] and extended [$\hat{F}$] features.
%	The baseline performance of $F_{BL}$ is highlighted in light gray.
%, improvement 	relative to baseline. BL is the BaseLine graph, +T denotes with Tag vertices and links, and +F denotes with Friendship %links.)
%    }
%\label{tab:tbl_lastfm_feature_combinations_precision}%
%\end{table}%

In order to evaluate the significance of the results, a paired t-test was performed with each group of features, using the P@10 values obtained for each of the four graphs. The results show that among the extended feature sets, all the differences were significant, p$<$0.05. Thus, the inclusion of auxiliary tagging  and friendship data improved the accuracy of the prediction, while their combination including both components led to the most accurate predictions. More importantly, the extraction of graph-based features was shown to consistently and significantly boost the performance of the recommender, in comparison to the variant not using the extracted features.
%of which artists a user will listen to. This improvement was by 32.14\%-69.94\% in P@10 relative to a baseline case where only basic graph features of a minimal graph scheme (bipartite, users and artists graph) were used.
%The conclusion can be drawn that the friendship data does not affect the performance differently from the tags data, when only the basic graph features are used. This is, however, different in the case of the extended features, where all the performance results across all graphs are significant when using the extended graph features.

\begin{table}[ht!]
\small
	\centering
	%\resizebox{0.6\columnwidth}{!}{
		\begin{tabular}{lllrr}
			\toprule
			& Features combination & Features & RMSE & Improvement\\
			\midrule
			1& All\_Features &Basic$\cup$Manual$\cup$Graph&1.0766&8.82\%\\
			2& AllExcept\_Tripartite & Basic$\cup$Manual$\cup$Bipartite&1.0775&8.75\%\\
			3& AllExcept\_Basic & Manual$\cup$Graph &1.0822&8.35\%\\
			4& Manual\_and\_Bipartite & Manual$\cup$Bipartite &1.0850&8.11\%\\
			5& AllExcept\_Bipartite & Basic$\cup$Manual$\cup$Tripartite &1.0896&7.72\%\\
			6& Manual\_and\_tripartite & Manual$\cup$Tripartite&1.1073&6.22\%\\
			7& AllExcept\_Manual & Basic$\cup$Graph &1.1095&6.04\%\\
			8& All\_Graph & Bipartite$\cup$Tripartite &1.1148&5.59\%\\
			9& AllExcept\_Graph & Basic$\cup$Manual&1.1175&5.36\%\\
			10& Bipartite &  &1.1188&5.25\%\\
			11& Tripartite &  &1.1326&4.09\%\\
			\rowcolor{LightGray}
			12& Basic &  &1.1809& N/A\\
			13& Manual &  &1.1853&-0.37\%\\
			\bottomrule
		\end{tabular}
	%}
	\caption[Yelp - results summary table]{RMSE of selected feature combinations - Yelp dataset (baseline combination in light gray).}
	\label{tab:tbl_feature_combinations_rmse}%
\end{table}

\subsubsection{Dataset \rom{2} -- Yelp Results}
\label{sec:eval1_dataset2_yelp1}

Improvements due to the use of the graph-based approach were also evident in experiments using the second dataset (Yelp).
As per the description in Section \ref{sec:dataset_yelp1}, basic (user and business identifiers), manual (number of reviews,
average rating, business categoriy and location), and graph-based features were extracted and populated.
The latter were further broken down into the bipartite and tripartite features. In this dataset, the performance of the basic
features related to the user-to-business associations serves as the baseline. Table \ref{tab:tbl_feature_combinations_rmse}
presents the full results for all the feature combinations.

The largest improvement in the RMSE of business ratings prediction was an 8.82\% decrease
obtained for the combination of graph features with basic and manually engineered ones (row 1).
The similarity of the RMSE scores obtained by the various combinations is explained primarily
by the low variance of user ratings in the dataset. Since most ratings are similar, they are
highly predictable using simple methods and there is only a limited space for improvement.
A combination containing only the graph features (row 5) outperformed the baseline performance
by 5.59\%. On the contrary, the use of manual features (row 13) slightly deteriorated the accuracy
of the predictions. This demonstrates the full benefit of the graph-based approach: extracting
the graph features took less time than crafting the manual ones, and the graph features also
outperformed the manual ones.

\begin{figure}[ht!]
    \begin{center}
    \includegraphics[width=0.45\textwidth]{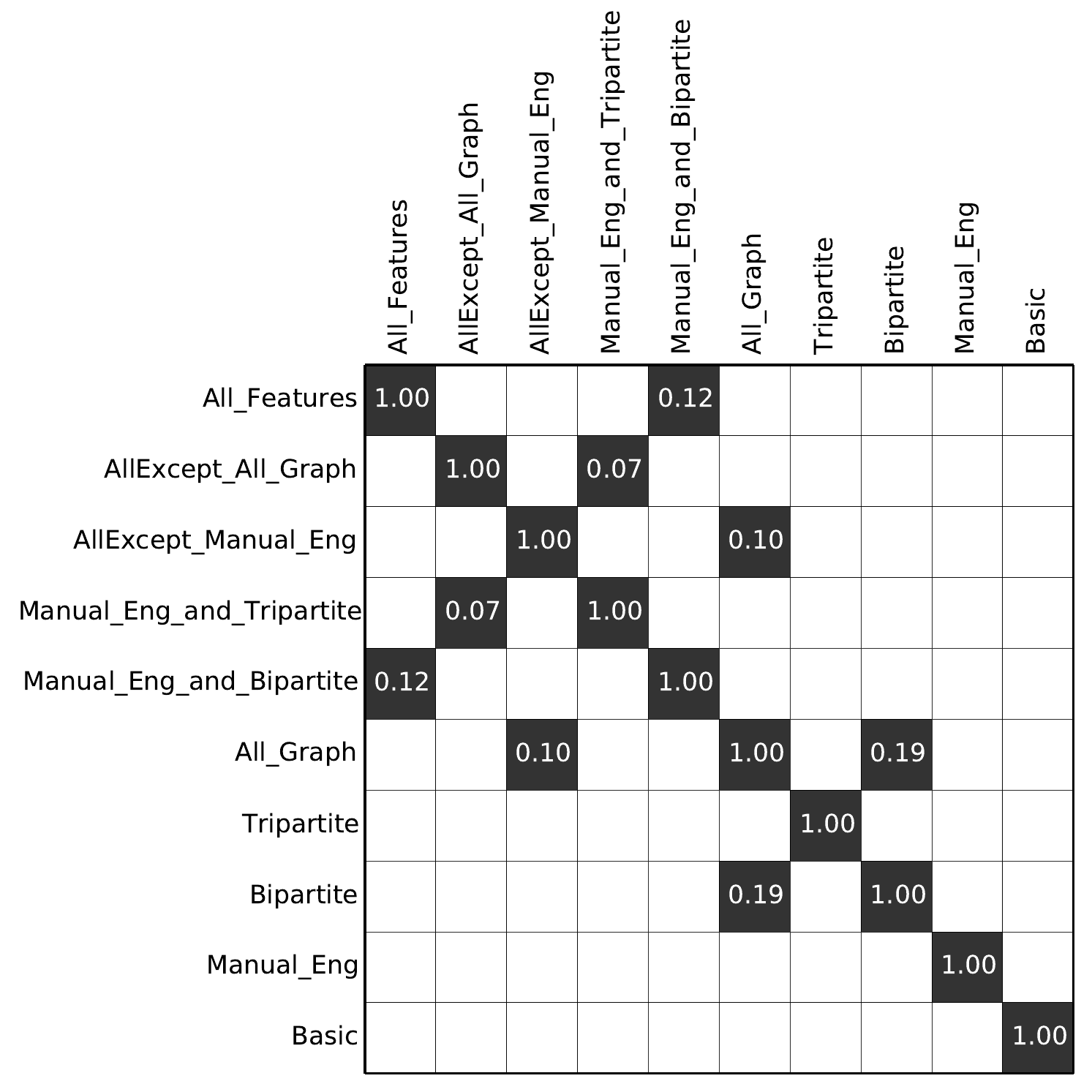}
   \end{center}
	\caption{Significance of the differences between feature combinations in the Yelp dataset.
            White cells - significant, dark cells - not significant, p-value given.}
    \label{fig:yelp_ttest_significance}
\end{figure}

%\begin{figure}[ht!]
%\centering
	%\subfloat[Yelp - results significance test]{
		%\includegraphics[width=0.4\textwidth]{gfx/ttest_significance.pdf}
	%\label{fig:yelp_ttest_significance}
	%}
	%\hspace{0.3em}
	%\subfloat[Yelp \rom{2} - results significance test]{
    %
	%		\includegraphics[width=0.45\textwidth]{gfx/yelp4_ttest_significance.pdf}
    %
	%	\label{fig:yelp2_ttest_results}
	%}
	%\caption{Significance of the differences between feature combinations in the Yelp dataset.
            %\rom{2} (Yelp) and \rom{3} (Yelp \rom{2}).
     %       White cells - significant, dark cells - not significant, p-value given.}
%\end{figure}

An examination of the differences in the accuracy of the results obtained when combining various groups of features
revealed a number of findings.
% that support the hypothesis that graph features overall contribute to the accuracy of the recommendations.
An analysis of the performance of each group of features shows that the bipartite
and tripartite feature sets performed noticeably better than the basic and manual feature
sets (rows 10 and 11 versus rows 12 and 13). A combination of graph features (row 8) still
outperforms slightly, although significantly, the combination of the basic and manually
engineered features (row 9).
To analyze the impact of the feature groups, each group was excluded from the overall set of
features and the change with respect to the All\_Features combination (row 1) was measured.
%These variants are denoted by AllExcept\_{group}
%, where \emph{group} is the combination of features that is excluded from the computation (rows 2, 3, 5, 7, 9 in Table \ref{tab:tbl_feature_combinations_rmse}).
When the graph features were excluded (row 9), the predictions were less accurate than when
the basic (row 3) or manual features (row 7) were excluded. This indicates that the graph
features provide the most valuable information, which is not covered by the basic and manual
features.

The paired t-test performed using the RMSE values revealed that the majority of differences
were significant, p$<$0.001. The insignificant differences are highlighted in Figure \ref{fig:yelp_ttest_significance}.
Three conclusions can be drawn from the insignificant pairs: (1) the bipartite features are
comparable to all the graph features, which indicates the low contribution of the tripartite
features; (2) all graph features are comparable to all features except for manual (i.e., graph
and basic features), which indicates the low contribution of the basic features; and, (3) the
combination of manual and bipartite features is comparable to the combination of all the features,
which indicates that the former two are the most informative features in this case.

\subsubsection{Dataset \rom{3} -- Yelp \rom{2} (with social links) Results}
\label{sec:eval1_dataset4_yelp2}
The results of the evaluation using the extended Yelp \rom{2} dataset that includes social links
between users are in line with the results of the original Yelp dataset. Table \ref{tab:yelp2_feature_combinations_performance}
lists the results of this evaluation for a selected set of feature combinations:
basic user and business features, feature of the complete graph, features of all the
sub-graphs, and the union of all the available features.

\begin{table}[ht!]
\small
	\centering
	%\resizebox{0.6\columnwidth}{!}{
		\begin{tabular}{llrr}
			\toprule
			& Features Subset & RMSE & Improvement\\
			\midrule
			1 & All Features &1.1416&1.73\%\\
			2 & All Graph Features &1.1417&1.73\%\\
			%3 & Without Name Links &1.1450&1.45\%\\
			3 & Complete Graph &1.1450&1.45\%\\
			%5 & Without Social Links &1.1463&1.33\%\\
			4 & Business Features &1.1465&1.32\%\\
			%7 & Without Social and Name Links &1.1465&1.32\%\\
			%8 & Without Category Links &1.1508&0.95\%\\
			%9 & Without Metadata &1.1508&0.94\%\\
			%10 & Without Social and Category Links &1.1519&0.85\%\\
			%11 & Without Metadata and Social Links &1.1523&0.82\%\\
			5 & User Features &1.1580&0.33\%\\
			\rowcolor{LightGray}
			6 & Basic Features &  1.1619 & N/A\\
			\bottomrule
		\end{tabular}
	%}
	\caption[Yelp \rom{2} - results summary table]{RMSE of selected feature combinations - Yelp \rom{2} dataset (baseline combination in light gray).}
	\label{tab:yelp2_feature_combinations_performance}%
\end{table}

The results show that the
combinations including graph features generally outperform the basic feature sets.
The best performing combination of graph-based features only, using all the features
from all the sub-graph schemes (row 2, RMSE=1.1417), achieves a 1.73\% improvement
over the baselines. When adding the basic features to all the graph-based features,
a slightly lower RMSE=1.1416 (row 1) is obtained. Another noticeable difference is
between the business-related features, which achieve RMSE=1.1465 and the user-related
features, which achieve RMSE=1.158 (rows 4 and 5, respectively). This intuitively
indicates that the predicted ratings assigned to the businesses being predicted are more
informative than the ratings of the target user.
%The influence of the ``Name'' vertices and social links is not evident, while the ``Category'' vertices do have an effect, as discussed below.
Again, the achieved improvements are generally modest, primarily due to the low
variance of ratings in the Yelp \rom{2} dataset.

The performance differences between the evaluated combinations are mostly significant,
p$<$0.01, except for two pairs of feature sets. The difference between business-related features
and complete graph features is borderline, with p=0.07. Also the difference between `All Features'
and `All Graph Features' is expectedly insignificant. This shows that the most important
contribution to the predictive accuracy comes from the graph features, while the
addition of the basic features improves the prediction only a little.

%The performance differences between the combinations are
%statistically significant, as can be seen in Figure
%\ref{fig:yelp2_ttest_results}, which shows the t-test results of
%any two feature combinations. Two major ``blocks'' can be noticed
%in Figure \ref{fig:yelp2_ttest_results}, which split the features
%combinations space into two groups (excluding the All Features and
%Basic Features combinations): a group of combinations of graphs
%without the ``Category'' vertices and combinations that include
%features from graphs that included the ``Category'' vertices. This
%shows the importance of the business category metadata information
%in the dataset as compared to the other graph elements: name
%vertices and social links, which are less contributory, probably
%because of their sparsity, which was discussed in Section
%\ref{sec:dataset_yelp2}.

\subsubsection{Dataset \rom{4} -- OSN Results}
\label{sec:eval1_dataset3_osn}
In the fourth dataset of online social networks, user mentions of interests in their profiles were predicted. Table \ref{tab:osn_precision_per_feature} shows the precision scores achieved by
individual features listed in Section \ref{sec:dataset_osn}, as well as by a number of their
combinations. These features include the individual interest- and user-focused graph features IG and UG; basic interest and graph features IB and UB; their unions IG\_All, UG\_All, IB\_All, and UB\_All; as well as I\_All = IG\_All $\cup$ IB\_All and U\_All = UG\_All $\cup$ UB\_All.
The baseline here is the prediction using the available user-interest features only.

\begin{table}[ht!]
\small
	\centering
	%\resizebox{1.0\columnwidth}{!}{
		\begin{tabular}{l|c||l|c||l|c}
			\hline
			Feature/Group & Precision & Feature/Group & Precision & Feature/Group & Precision\\
			\hline
			\cellcolor[gray]{0.8}All     & 0.6455  &  IB4     & 0.5287 & UB5       & 0.4529\\
			\cellcolor[gray]{0.9}IG\_All & 0.5821  &  IB5     & 0.5230 & UG1       & 0.4514\\
			IG1     & 0.5745  &  IB6     & 0.5221 & UB6       & 0.4507 \\
			IG2     & 0.5734  &  IB7     & 0.5215 & \cellcolor[gray]{0.9}UG\_All   & 0.4465\\
			IG3     & 0.5691  &  IB8     & 0.5206 & UG3       & 0.4440  \\
			IB1     & 0.5687  &  IB9     & 0.5205 & UG2       & 0.4430  \\
			\cellcolor[gray]{0.9}I\_All  & 0.5642  &  \cellcolor[gray]{0.8}Baseline      & 0.5128 &  UB7       & 0.4405\\
			\cellcolor[gray]{0.9}IB\_All & 0.5599  &  SP        & 0.5107 &  UG4     & 0.4401 \\
			IG4     & 0.5589  &  UB1       & 0.4771 &  UG5       & 0.4400 \\
			IG5     & 0.5560  &  UB2       & 0.4736 &  \cellcolor[gray]{0.9}UB\_All   & 0.4391 \\
			IB2     & 0.5482  &  UB3       & 0.4646 &  \cellcolor[gray]{0.9}U\_All    & 0.4376 \\
			IB3     & 0.5307  &  UB4       & 0.4632 &            &      \\
			\hline
		\end{tabular}
	%}
	\caption{Average precision for individual features and feature combinations in the OSN dataset \emph{IG} - Interests Graph features, \emph{UG} - Users Graph features, \emph{IB} - Interests Basic (manual) features, \emph{UB} - Users Basic (manual) features.}
	\label{tab:osn_precision_per_feature}%
\end{table}

\begin{figure}[ht!]
	%\begin{minipage}[h]{0.475\textwidth}
		\centering
		%\resizebox*{1.0\textwidth}{!}{
    \label{fig:cdf_precision}
    \includegraphics[width=0.45\textwidth]{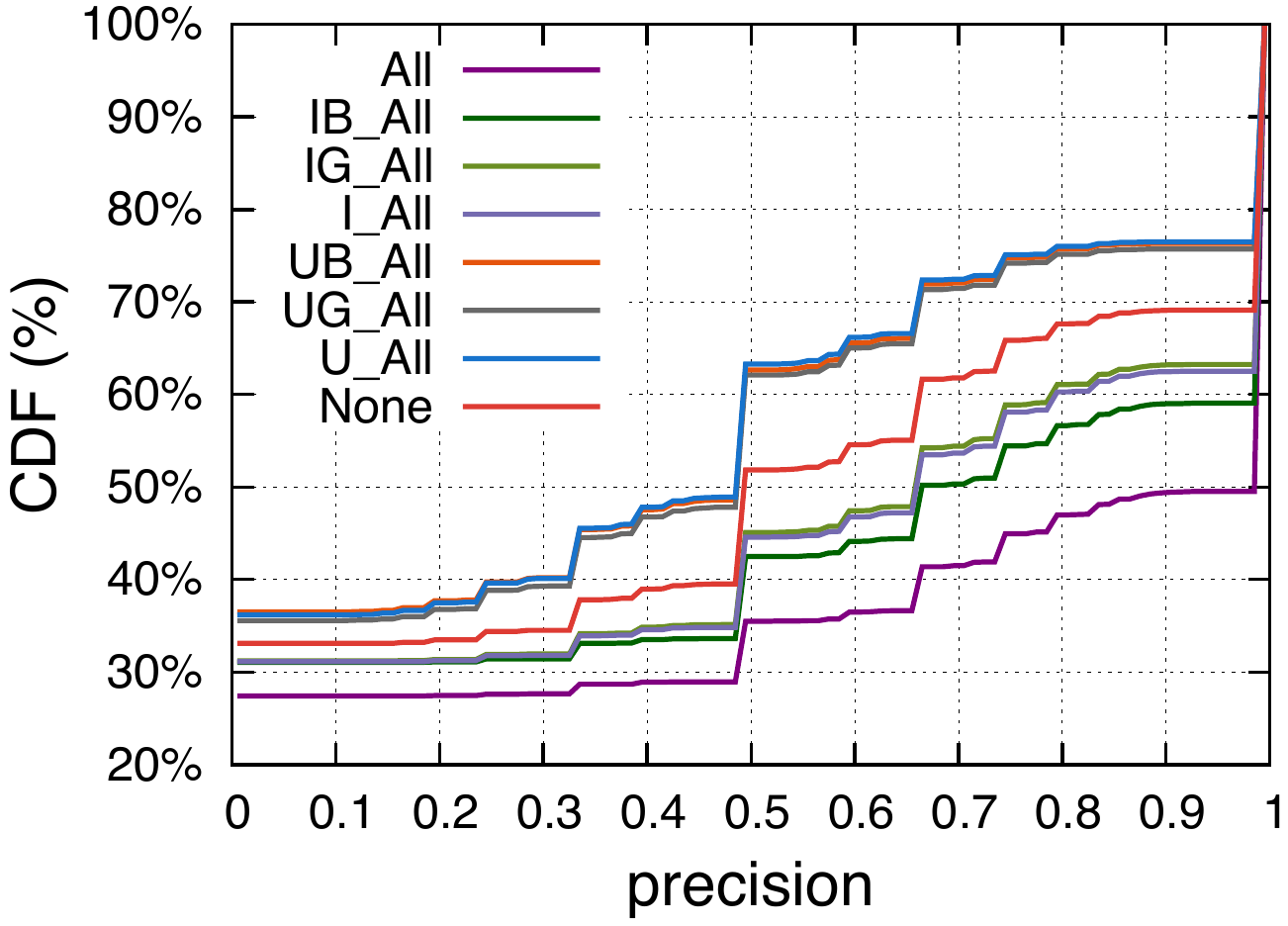}
    %}
	%\end{minipage}
	%\begin{minipage}[h]{0.475\textwidth}
	%	\centering
	%	\resizebox*{1.0\textwidth}{!}{\label{fig:shortest_path}\includegraphics
	%		{gfx/cdf_shortest_path_distance.pdf}}
	%\end{minipage}
	\caption[OSN - results graphed]{Precision CDF of the various feature combinations - OSN dataset.}
	\label{fig:osn_dataset_statistics}
\end{figure}

%, i.e., whether a user will be interested in
%a subject or not. Using interest-related graph-based features
%(basic graph features) resulted in a precision that is 9.8\% higher than
%that achieved using only user/interest IDs (precision of 0.56
%versus 0.51, ``IG\_All'' versus ``Baseline'' in Table
%\ref{tab:osn_precision_per_feature}). Combining the graph features
%with manually engineered ones produced the highest precision,
%0.65, ``All'' in Table \ref{tab:osn_precision_per_feature} (the
%full list of features composing ``All'' is provided in Section
%\ref{sec:osn_methodology}).

Overall, item-related features were again seen to improve the precision of the
predictions more than user-related features. In fact, the accuracy of all the
item features is above the baseline precision of 0.56, while the accuracy of all
the user features is below the baseline. This can also be seen through the comparison
of the union of all item features I\_All with the union of user features U\_All,
which shows the superiority of the former. Zooming on the item features, it can
be observed that the graph-based item features IG outperform most of the basic item
features IB, except for IB1 (the number of users who mentioned the interest),
which turns out to be a reliable predictor. As a result, the union of item-related
graph features IG\_All obtains a higher precision than its basic feature
counterpart IB\_All, 0.58 vs 0.56. Combining the graph features with the
basic ones produced the highest precision, 0.65. Hence, graph-based features
resulted in an improvement of the the interest predictions.

%Drilling down to specific features, several observations can be made. Figure \ref{fig:osn_dataset_statistics}b demonstrates the change in prediction precision relative to the distance (shortest path/minimum number of edges) between the source vertices (users) and target vertices (predicted interests). For interests at a distance of 5, 90\% of predictions are made with zero precision, while for interests at a distance of 3 more than 90\% of the predictions have a precision higher than 0.5, and ~75\% a precision higher than 0.7.

The cumulative distribution functions of the results obtained by selected feature combinations
is illustrated in Figure \ref{fig:osn_dataset_statistics}. The combined feature combination
`All', performs best by having the highest precision over large portions of the data.
The item-related graph-based feature combination is third best, and it is very close to the
combined graph-based feature set that comes second. The performance of all user-related feature
sets is visibly lower, which shows another argument in favor of the extraction of the graph-based features.

\subsubsection{Dataset \rom{5} -- Movielens Results}
\label{sec:eval1_dataset5_movielens}

\begin{table}[ht!]
\small
	\centering
	%\resizebox{0.75\columnwidth}{!}{
		\begin{tabular}{llrrrr}
			\toprule
			& Features Subset & RMSE & Improvement & MAE & Improvement\\
			\midrule
			1 & All Features     & 1.0272 & 4.20\% & 0.8303 & 6.06\% \\
			2 & All Graph Features   & 1.0362 & 3.36\% & 0.8349 & 5.53\% \\
			%3 & graph w/[Age, Gender, Genre, Zip] subset     & 1.0369 & 3.29\% & 0.8353 & 5.48\% \\
			%4 & graph w/[Age, Gender, Occupation, Zip] subset    & 1.0373 & 3.25\% & 0.8357 & 5.44\%\\
			%5 & graph w/[Gender, Genre, Occupation, Zip] subset  & 1.0384 & 3.16\% & 0.8365 & 5.35\%\\
			3 & Movie Features & 1.0400 & 3.01\% & 0.8380 & 5.18\% \\
			\rowcolor{LightGray}
			4 & Basic Features  & 1.0722 & N/A & 0.8838 & N/A \\
			5 & User Features & 1.0895 & -1.61\% & 0.8967 & -1.46\% \\
			\bottomrule
		\end{tabular}
	%}
	\caption[Movielens - results summary table]{Performance of selected features combinations - Movielens dataset (baseline combination in light gray, rows are sorted by RMSE).}
	\label{tab:movielens_feature_combinations_performance}%
\end{table}

Finally, the experimentation with the Movielens dataset re-affirms the contribution of graph-based
feature extraction to the recommendation generation. The task in this dataset was to predict
movie ratings, whereas the predictions were evaluated using the MAE and RMSE predictive accuracy metrics. Table \ref{tab:movielens_feature_combinations_performance} summarizes the perofrmance of a selected
group of features. The basic user-item pairs are compared here with the user and item features
used individually, all the extracted graph-based features, and the union of all of them,
denoted by `All Features' (row 1).

The already discussed superiority of item features over user features (row 3 versus
row 5) can be clearly seen again. In this case, the former improve the accuracy of
the predictions by 3-5\%, while the latter only deteriorate it. The extraction of the
graph-based features (row 2) also leads to an improvement of 3.36\% and 5.53\% relative
to the baseline, using the RMSE and MAE metrics, respectively. When combined with other
features, the graph features achieve the best result, which is RMSE=1.0272, or a 4.20\%
improvement over the baseline. Those performance differences were statistically evaluated
and found significant.
%As in the previous evaluations, the feature set containing the aggregation of all graph features from all sub graphs outperforms any individual feature subset of a single scheme (row 2 versus rows 3-34, and 36 in Table \ref{tab:movielens_feature_combinations_performance}). This
%supports the importance of generating multiple graph schemes from
%a dataset, since each scheme provides its unique perspective.
%Additionally, two key insights regarding the performance of
%specific entities in the dataset can be derived. First, the
%movie-related graph features, which is the subset aggregating all
%single node movie graph features extracted from all graph subsets,
%performed better than the equivalent subset of user nodes features
%(1.04 versus 1.0895 RMSE, 4.54\% difference, rows 6 and 36,
%Table \ref{tab:movielens_feature_combinations_performance}).
%The second entity-related insight is that feature sets that were
%extracted from graph schemes containing the ``Genre'' vertices
%outperformed schemes that did not have this entity (rows 3-5 \&
%7-20 versus rows 21-34). This leads to the conclusion that the
%Genre entity encodes important information in the case of this
%dataset and task. Within the two groups of subgraphs (those with and those
%without the ``Genre'' entities), the difference in the results is
%mostly statistically insignificant.

\subsubsection{Performance across Learning Methods, Datasets, and Metrics}
\label{sec:performance_across_regressors_and_classifiers}

This case investigated the the impact of the graph-based
features affect on the accuracy of the recommendations.
Although all the evaluations reported so far show that using the
graph-based features improves the accuracy of the recommendations,
the results cannot be fully corroborated yet, as the conducted
experiments use different learning methods, datasets, and evaluation
metrics (see Table \ref{tab:dataset_summary}). To confidently
address the resarch question, the design of the evaluation has
overlaps in these factors, so that the contribution of the graph
features can be singled out.

The analysis below aims to establish whether the observed improvements should
be attributed to the information contributed by the graph features or to the
differences in the experimental settings, i.e., learning method, dataset, and
metric.The results of the experiments used in the analysis are summarized in
Table \ref{tab:summary_of_all_experiments_and_results}.
In all the cases, the performance of the baseline approaches not using the
graph features, which were highlighted in light gray in all the tables, is compared
to the performance of all the graph-based based features, i.e, row 8 in
Table \ref{tab:tbl_feature_combinations_rmse}, row 2 in Table \ref{tab:yelp2_feature_combinations_performance},
and row 2 in Table \ref{tab:movielens_feature_combinations_performance}.

		\begin{table}[ht!]
        \small
			\centering
			%\resizebox{0.7\textwidth}{!}{
				\begin{tabular}{llllrrr}
					\hline
					\toprule[2pt]
					\multicolumn{4}{c}{} &                 \multicolumn{3}{c}{Results} \\
					\cmidrule(l){5-7}
					& Dataset & Metric    & Method & Baseline & Graph Features & Improvement \\
					\cmidrule(l){2-2}
					\cmidrule(l){3-3}
					\cmidrule(l){4-4}
					1& Yelp I & RMSE & Random Forest & 1.1809 & 1.1148 & 5.59\% \\
					2&Yelp II & RMSE & Random Forest & 1.1619 & 1.1417 & 1.74\% \\
					3&Yelp II & RMSE & Gradient Boosting & 1.2480 & 1.1715 & 6.13\% \\
					4&Yelp II & RMSE & SVM & 1.1818 & 1.1783 & 0.30\% \\
					5&Movielens & RMSE & Random Forest & 1.1667 & 1.0268 & 11.90\% \\
					6&Movielens & RMSE & Gradient Boosting & 1.0722 & 1.0362 & 3.36\% \\
					7&Movielens & MAE & Random Forest & 0.9144 & 0.8157 & 10.79\% \\
					8&Movielens & MAE & Gradient Boosting & 0.8838 & 0.8349 & 5.53\% \\
					\bottomrule[2pt]
				\end{tabular}
                %}
				\caption[Case study \rom{1} results summary]{Summary of experiments and results for Case Study \rom{1}.}
				\label{tab:summary_of_all_experiments_and_results}
			\end{table}

%\footnote{The 	results of datasets \rom{1}, Last.fm, are omitted from this table, since there was no comparison between all 	generated graph features and basic ones. The comparison was between basic and enhanced graph schemes}.
	
Included are the results of
experiments using the Yelp, Yelp \rom{2}, and Movielens datasets, which were
discussed in sections \ref{sec:eval1_dataset2_yelp1}, \ref{sec:eval1_dataset4_yelp2},
and \ref{sec:eval1_dataset5_movielens}, respectively. That said, results
in rows 3, 4, 5, and 7 of Table \ref{tab:summary_of_all_experiments_and_results}
are presented here for the first time. This is due to the fact that previously
reported Yelp experiments (both datasets) used Random Forest as their learning
method, while the Movielens experiments used Graduate Boosting. Here, new Yelp \rom{2}
results with Gradient Boosting and SVM, and new Movielens results with Random
Forest are also presented. It should also be highlighted that experiments using the
Last.fm and OSN datasets are excluded from the analysis, since, unlike the other
three experiments, they use classification accuracy metrics. As such, they differ
in two factors, dataset and evaluation metric, and are not comparable with the
other experiments.

			In order to demonstrate that the improvement is not due to the
			selected dataset, the metric and learning method were fixed, while the
			approaches using different datasets were compared.
%, as illustrated in Figure \ref{fig:experimental_method_fixed_dataset}.
            Two evaluations sets are applicable to this scenario: (1) Random
            Forest predictions evaluated with the RMSE metric, using the Yelp, Yelp \rom{2},
            and Movielens datasets (rows 1, 2, and 5), and (2) Gradient Boosting
            predictions also evaluated with RMSE, but using the Yelp \rom{2} and
			Movielens datasets (rows 3 and 6). The results of these experiments
			show an improvement of 1.74\% to 11.90\%, which allows to eliminate
            the selected dataset as a possible reason for improvement.

					To demonstrate that the improvement is also not due to the selected
					machine learning method, the dataset and metric were fixed, while the
                    approaches using different learning methods were compared.
            %as illustrated in Figure \ref{fig:experimental_method_fixed_method}.					
					Three evaluation sets are applicable to this scenario: (1) RMSE of
                    business predictions using the Yelp \rom{2} dataset, where the learning methods
                    are Random Forest, Gradient Boosting, and SVM (rows 2, 3, and 4), (2)
                    RMSE of movie rating predictions using the Movielens dataset, where the methods
                    are Random Forest and Gradient Boosting (rows 5 and 6), and (3) MAE of
                    movie rating predictions using the the Movielens dataset, where the methods are
                    Random Forest and Gradient Boosting (rows 7 and 8). The results of these
                    experiments show an improvement across all experiments, ranging from 0.30\% to
                    6.13\% for the Yelp \rom{2} dataset, and from from 3.36\% to 11.90\% for the
                    Movielens dataset. The low variance of ratings in the Yelp datasets, which
                    was discussed earlier, is the main reason for the low improvement observed.
                    This is particularly noticeable with the SVM method, which struggles to
                    linearly separate businesses with moderate ratings. Thus, the learning
                    method cannot be the reason for the accuracy improvement.

				Finally, to demonstrate that the improvement is not due to the selected
                evaluation metric, the dataset and method were fixed, while the performance
				of approaches using different metrics was compated.
				Two evaluation sets are applicable to this scenario: (1) Random Forest movie
                rating predictions using the Movielens dataset, evaluated using RMSE and MAE
                (rows 5 and 7), and (2) Gradient Boosting movie rating predictions also using
                the Movielens dataset, and also evaluated using the RMSE and MAE metrics (rows
                6 and 8). The results of these experiments show a clear improvement across,
                ranging from 3.36\% to 11.90\%, allowing to eliminate the selected evaluation
                metric as a possible reason for improvement.

                Summing up this causal analysis, all three hypotheses that the improved performance is
                driven by the differences in the experimental settings (dataset, learning method,
                and evaluation metric) were rejected. Thus, it can be concluded that the reason for
                the observed improvement lies in the inclusion of graph-based features, contributing
                new information to the recommendation process.

\subsection{Case Study \rom{2}: Different Graph Schemes and their Impact on Recommendations}
\label{ch:evaluation2}

As mentioned in the Section \ref{sec:graph_based_feature_extraction},
various sub-graphs and graph schemes can be generated for each dataset.
The feature extraction process will, thus, yield a number of graph schemes,
corresponding feature sets, and even the values of the same graph features.
This leads to to the second research question: \textit{How are the recommendations affected by the sub-graph and its representation used to generate the graph features?}
In order to answer this question, another set of experiments was conducted.

In these experiments, the accuracy of recommendations when using various graph schemes
was evaluated using four datasets: Last.fm, both Yelp datasets, and Movielens. The OSN
dataset was not used here because it was represented as a simple bipartite graph,
lacking the desired number of entities and relationships. The recommendation tasks were
identical to the previous experiments, i.e., to predict listened artists in the Last.fm
dataset, user ratings for businesses in the two Yelp datasets and user ratings for movies
in the Movielens dataset. An N-fold cross validation methodology similar to the one
reported in Section \ref{ch:evaluation_1} was followed. Also, the same two-sided t-test
statistical significance testing was carried out.

\begin{figure}[ht!]
    \begin{center}
    \includegraphics[width=0.55\textwidth,height=0.5\textheight,keepaspectratio]{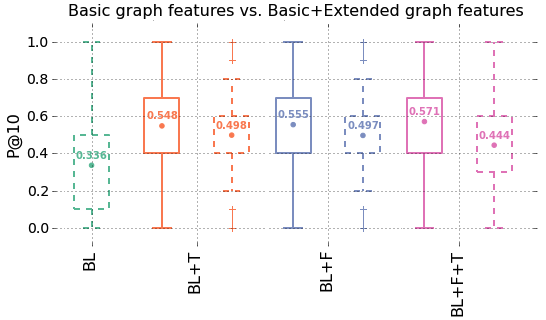}
   \end{center}
	\caption{Last.fm results: Precision of the extended (solid) versus the basic (dashed) feature sets.}
    \label{fig:effect_on_graph_metrics}
\end{figure}

\subsubsection{Dataset \rom{1} -- Last.fm Results}
\label{sec:eval2_dataset1_lastfm}

%\begin{figure}[ht]
	%\subfloat[Average precision for the four graph schemes,
	%when using all generated graph features -- basic and extended]{
%	    \centering
	%    \includegraphics[width=0.37\textwidth,height=0.6\textwidth,keepaspectratio]{gfx/lastfm_p@10.png}
    %
	%    \label{fig:effect_of_social_link_features}
	%}
	%\hspace{1.5em}
%	\subfloat[Precision of the extended
%	(solid boxes) versus the basic (dotted boxes) feature sets for the four graph representations]{
%	    \centering
%	    \includegraphics[width=0.54\textwidth,height=0.5\textheight,keepaspectratio]{gfx/boxplots_bl_bl+delta_addon50_cropped.png}
%
%	    \label{fig:effect_on_graph_metrics}
%	}
%	\caption{Last.fm results}
%\end{figure}

The evaluations using the Last.fm dataset
focused on the influence of the social elements, i.e., friendship
links and tags, on the obtained recommendation accuracy. In this dataset,
the results of recommendations based on the bipartite user-artist graph
representation (BL in Figure \ref{fig:lastfm_graph_schemes}) were compared
with those of three non-bipartite schemes, BL+T, BL+F, and BL+T+F,
including, respectively, the tags assigned by the users to the artists,
social friendship links between the users, and tags and friendship links
alike.
As mentioned in Section \ref{sec:eval1_dataset1_lastfm}, two sets of graph
features were extracted for each schema: a set of basic features $\hat{F}$ and a
set of extended features $F$. Although the basic feature set $\hat{F}$ is
shared across all the schemes, their values may change due to the presence
of additional graph nodes. The extended feature set $F$ is composed
of the basic features along with new features that were extracted from
the social links and tags available in each schema. A detailed
discussion of the extended feature set can be found in \citep{tiroshi2014graph}.
%The entities on which the extended features are based differ from one scheme to another\footnote{The process of extracting the extended features
%is automatic and detailed in Algorithm \ref{algo:feature_extraction} lines 13-20.
%The categorization of the features into ``basic'' and ``extended'' was done manually for the purpose of the evaluation.}.
%Specifically, for the BL+T graph the extended features included in addition to the basic graph features: ``ratio\_of\_shared\_neighbors'' (ratio
%between the intersection of neighbors a user and artist have, and the unification set of those neighbors) and
%``ratio\_of\_shared\_neighbors\_of\_type'' (same as previous
%feature only with a filter on the type of neighbors) with \emph{Type=Tags}. The BL+F graph extended features were identical to those of BL+T except with %\emph{Type=Users} in the ``ratio\_of\_shared\_neighbors\_of\_type'' feature. And last, the BL+T+F extended features set had all three extended feature types: %``ratio\_of\_shared\_neighbors'', ``ratio\_of\_shared\_neighbors\_of\_type'' with \emph{Type=Tags}, and ``ratio\_of\_shared\_neighbors\_of\_type'' with %\emph{Type=Users}. The general trend that was noticed in this experiment was that adding more types of entities to the graph improved the results.

Figure \ref{fig:effect_on_graph_metrics} shows the obtained P@10 scores
averaged over all the users in the test set, when using both the basic and extended feature sets.
%\footnote{For the BL schema, only the basic features are used, since the graph has no auxiliary data for the extended features.}
For each representation, the solid boxes on the left denote the results
obtained with the extended features $F$, whereas the dotted boxes on the right present the results obtained with the basic features $\hat{F}$.
%The boundaries of all the boxes represent the 25th and 75th percentile of the obtained P@10, and the average P@10  is marked by the dots inside the boxes. The values of the average P@10 are given.
First, it can clearly be observed that the inclusion of the social
auxiliary data of either the assigned tags or friendships links
substantially improves P@10. When both the tags and friendship
links are included in the BL+T+F model, the highest average P@10
is observed.
%An improvement of 69.94\% with respect to the BL model that includes no auxiliary data is statistically significant (P@10=0.571 versus P@10=0.336). A modest improvement of the BL+T+F model with respect to the models that included either the tag (4\%, P@10=0.571 versus P@10=0.548) or the friendship data (2.8\%, P@10=0.571 versus P@10=0.555) should be noted. This shows that including both types of social auxiliary data further improves the accuracy of the recommendations.
Both in the basic and the extended feature sets, the BL+T and BL+F models
obtain comparable P@10 scores, showing the effect of the inclusion of
auxiliary data in the graph schemes. However, as noted in Section \ref{sec:dataset_lastfm},
the tag data includes more than 186K tag assignments, whereas the
friendship data consists of only 12K user-to-user links. Since the
obtained precision scores are comparable, a single friendship link
is more influential than a single artist tag and yields a greater
improvement in the recommendation accuracy. Looking at the significance
tests conducted within the basic and extended feature sets, significant
differences, p$<$0.05, were observed between all the pairs of extended
features and all the pairs of basic features except for the $\hat{F}_{BL+T}$
and $\hat{F}_{BL+F}$ pair.

\begin{table}[ht!]
\small
    \centering
    %\resizebox{0.6\columnwidth}{!}{
        \begin{tabular}{lllrr}
            \toprule
            & Features combination & Features & RMSE & Improvement\\
            \midrule
            8& All\_Graph & Bipartite$\cup$Tripartite & 1.1148 & 5.59\%\\
            10& Bipartite &  & 1.1188 & 5.25\%\\
            11& Tripartite &  & 1.1326 & 4.09\%\\
            \rowcolor{LightGray}
            12& Basic &  &1.1809& N/A\\
            \bottomrule
        \end{tabular}
    %}
    \caption[Yelp - results cutout for graph schemes performance]{Yelp results: RMSE of the bipartite versus the tripartite feature sets.
    Full results are given in Table \ref{tab:tbl_feature_combinations_rmse}.}
    \label{tab:tbl_yelp_feature_combinations_rmse_cutout}%
\end{table}

When comparing the performance of the extended graph features to the performance of the corresponding basic features (solid boxes versus dashed boxes in
Figure \ref{fig:effect_on_graph_metrics}, it can be seen that the extended sets consistently outperformed the basic sets across all the four graph schemes, and the difference within the
pairs was statistically significant, p$<$0.05.
In the BL+T scheme, the extended graph features from improved on the basic features extracted from it by 10\%, P@10=0.548 versus P@10=0.498, while in the BL+F scheme the improvement was by 11.6\%, P@10=0.555 versus P@10=0.497. The largest improvement was noted in the BL+T+F scheme, where the extended graph features outperformed the basic features by as much as 28.6\%, P@10=0.571 versus P@10=0.444. Surprisingly, when the basic feature set $\hat{F}_{BL+T+F}$ set was found achieve a lower P@10 than $\hat{F}_{BL+T}$ and $\hat{F}_{BL+F}$. A possible explanation for this can be that including both types of social data but not extracting and populating the extended features leads to redundancy in the graph and degrades the performance of the recommender.

\subsubsection{Dataset \rom{2} -- Yelp Results}
\label{sec:eval2_dataset2_yelp1}

For the Yelp dataset and the task of business rating prediction,
two graph schemes were compared: a pure
bipartite graph that contained only the users and businesses, and
a tripartite graph that, on top of user and
business nodes, also contained metadata nodes describing the
businesses. The two graph schemes are illustrated in Figure
\ref{fig:yelp_graph_schemes}. The
reason these were the only graph schemes created is that sparse
features having a small number of unique features, were
filtered from the dataset. These features would have resulted
in most of the nodes of a group, e.g., users, being connected to a
single node, which would render it meaningless. For example,
adding three ``gender'' nodes, male, female, and unspecified,
would have resulted in all users being connected to either one of
the three, essentially creating three large clusters in the graph.

\begin{table}[ht!]
\small
    \centering
    %\resizebox{0.5\columnwidth}{!}{
        \begin{tabular}{llrr}
            \toprule
            & Features Subset & RMSE & Improvement\\
            \midrule
            2 & All Graph Features &1.1417&1.73\%\\
            \midrule
            7 & Without Name Links &1.1450&1.45\%\\
            3 & Complete Graph &1.1450&1.45\%\\
            8 & Without Social Links &1.1463&1.33\%\\
            9 & Without Social and Name Links &1.1465&1.32\%\\
            \midrule
            10 & Without Category Links &1.1508&0.95\%\\
            11 & Without Metadata &1.1508&0.94\%\\
            12 & Without Social and Category Links &1.1519&0.85\%\\
            13 & Without Metadata and Social Links &1.1523&0.82\%\\
            \midrule
            \rowcolor{LightGray}
            6 & Basic Features &  1.1619 & N/A\\
            \bottomrule
        \end{tabular}
    %}
    \caption[Yelp \rom{2} - results cutout for graph schemes performance]{Yelp \rom{2} results: RMSE of various sub-graph feature sets.}
    \label{tab:yelp2_feature_combinations_performance_cutout}%
\end{table}

The complete set of graph features was generated for both the bipartite and
tripartite representations. The results in Table \ref{tab:tbl_yelp_feature_combinations_rmse_cutout}
show the RMSE scores obtained for these feature sets. Note that
these results are essentially extracted from the results presented
in Table \ref{tab:tbl_feature_combinations_rmse} and their original
row numbers are preserved. The experiments showed that the bipartite
schema, not including the metadata nodes, performed slightly but
significantly better than the tripartite schema with metadata,
RMSE=1.1188 versus RMSE=1.1326. The relative improvement with
respect to the baseline recommendations was 1.16\% higher. This
difference in the performance of the schemas led to their
unified feature set, which is the All\_Graph, to outperform
the two feature sets individually. However, the superiority of
All\_Graph was statistically significant only when compared to
the tripartite schema, as can be seen in Figure
\ref{fig:yelp_ttest_significance}.

\subsubsection{Dataset \rom{3} -- Yelp \rom{2} (with social links) Results}
\label{sec:eval2_dataset4_yelp2}

\begin{figure}[ht!]
    \begin{center}
    \includegraphics[width=0.55\textwidth]{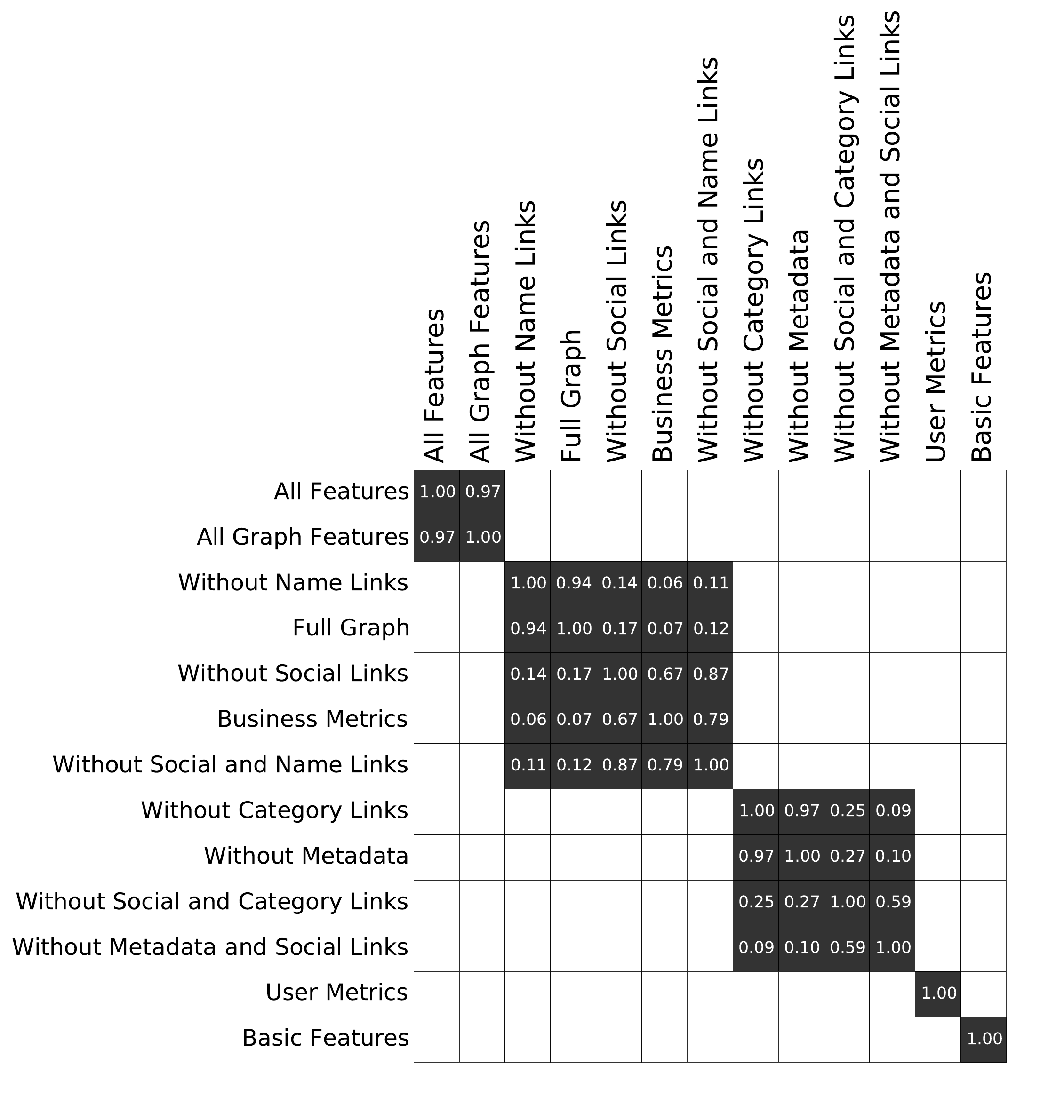}
   \end{center}
	\caption{Significance of the differences between feature combinations in the Yelp \rom{2} dataset.
            White cells - significant, dark cells - not significant, p-value given.}
    \label{fig:yelp2_ttest_significance}
\end{figure}

The richer information provided by the Yelp \rom{2} datasets allowed for the
creation of a larger set of sub-graphs. These are illustrated in Figure \ref{fig:yelp2_graph_schemes},
where various combinations of entities are removed from the complete graph.
Thus, in addition to the complete graph, seven sub-graph representations can
be created and the performance of the feature sets extracted from these
can be compared. The results of this experiment are presented in Table \ref{tab:yelp2_feature_combinations_performance_cutout}.
The complete graph and the seven sub-graphs are compared to the basic feature
set and the union of all the graph features, which were, respectively, the
baseline and best performing combination in Table \ref{tab:yelp2_feature_combinations_performance}.
The numbering of rows already presented in Table \ref{tab:yelp2_feature_combinations_performance}
is preserved (rows 2, 3, and 6), while the rows of all the sub-graphs from
Figure \ref{fig:yelp2_graph_schemes} are numbered 7 to 13. The significance
of the differences between the sub-graphs is shown in Figure \ref{fig:yelp2_ttest_significance}.

As can be clearly seen, the results of the various sub-graphs fell into two groups,
based on the significance tests. The groups were: sub-graphs containing the `category'
relationship (``Without Name Links'', ``Complete Graph'', ``Without Social Links'',
and ``Without Social and Name Links'') and sub-graphs not containing the `category'
relationship (``Without Category Links'', ``Without Metadata'', ``Without Social and
Category Links'',  and ``Without Metadata and Social Links''). The former group of
sub-graphs (rows 7, 3, 8, and 9 in Table \ref{tab:yelp2_feature_combinations_performance_cutout})
performed significantly better than the latter (rows 10, 11, 12, and 13), which
highlights the importance of business categories in predicting the business ratings.
This is also in line with the dominance of business features over the user features
that was already observed in Table \ref{tab:yelp2_feature_combinations_performance}.
The union of all the graph-based features extracted from all the sub-graphs (``All
Graph Features'' in row 2) expectedly outperformed all other sub-graphs and feature
sets. This highlights the strength of the proposed approach in producing all the possible
features from all the possible sub-graph representations of the data rather than
identifying the optimal sub-graph and dealing with feature selection.

\subsubsection{Dataset \rom{5} -- Movielens Results}
\label{sec:eval2_dataset5_movielens}

\begin{table}[ht!]
\small
	\centering
	%\resizebox{0.75\columnwidth}{!}{
		\begin{tabular}{llrrrr}
			\toprule
			& Features Set & RMSE & Improvement & MAE & Improvement\\
			\midrule
			%1 & All Features     & 1.0272 & 4.20\% & 0.8303 & 6.06\% \\
			2 & All Graph Features   & 1.0362 & 3.36\% & 0.8349 & 5.53\% \\
			6 & graph w/[Age, Gender, Genre, Zip]      & 1.0369 & 3.29\% & 0.8353 & 5.48\% \\
			7 & graph w/[Age, Gender, Occupation, Zip]     & 1.0373 & 3.25\% & 0.8357 & 5.44\%\\
			8 & graph w/[Gender, Genre, Occupation, Zip]   & 1.0384 & 3.16\% & 0.8365 & 5.35\%\\
			%3 & Movie Features & 1.0400 & 3.01\% & 0.8380 & 5.18\% \\
			9 & graph w/[Genre, Occupation]    & 1.0410 & 2.91\% & 0.8391 & 5.06\% \\
			10 & graph w/[Age, Genre, Zip]      & 1.0411 & 2.90\% & 0.8386 & 5.12\% \\
			11 & graph w/[Genre, Occupation, Zip]   & 1.0411 & 2.90\% & 0.8385 & 5.12\% \\
			12 & graph w/[Age, Genre, Occupation]      & 1.0412 & 2.90\% & 0.8390 & 5.07\% \\
			13 & graph w/[Age, Gender, Genre]      & 1.0412 & 2.89\% & 0.8392 & 5.05\% \\
			14 & graph w/[Age, Gender, Genre, Occupation] & 1.0413 & 2.89\% & 0.8390 & 5.07\% \\
			15 & graph w/[Age, Genre, Occupation, Zip]     & 1.0413 & 2.89\% & 0.8388 & 5.10\%\\
			16 & graph w/[Genre]   & 1.0413 & 2.89\% & 0.8393 & 5.04\% \\
			17 & graph w/[Gender, Genre, Occupation]   & 1.0413 & 2.88\% & 0.8392 & 5.04\% \\
			18 & graph w/[Age, Genre]      & 1.0414 & 2.88\% & 0.8395 & 5.02\% \\
			19 & graph w/[Age, Gender, Genre, Occupation, Zip]  & 1.0414 & 2.88\% & 0.8390 & 5.08\% \\
			20 & graph w/[Genre, Zip]      & 1.0415 & 2.87\% & 0.8388 & 5.09\% \\
			21 & graph w/[Gender, Genre]   & 1.0416 & 2.86\% & 0.8396 & 5.00\% \\
			22 & graph w/[Gender, Genre, Zip]      & 1.0416 & 2.85\% & 0.8391 & 5.06\% \\
			23 & graph w/[Age]     & 1.0425 & 2.77\% & 0.8413 & 4.81\% \\
			24 & graph w/[Zip]     & 1.0426 & 2.77\% & 0.8407 & 4.88\% \\
			25 & graph w/[Age, Zip]    & 1.0426 & 2.76\% & 0.8409 & 4.86\% \\
			26 & graph w/[Age, Occupation]     & 1.0426 & 2.76\% & 0.8412 & 4.82\% \\
			27 & graph w/[Age, Gender]     & 1.0427 & 2.76\% & 0.8413 & 4.81\% \\
			28 & graph w/[Age, Gender, Zip]    & 1.0427 & 2.75\% & 0.8408 & 4.87\% \\
			29 & graph w/[Age, Occupation, Zip]    & 1.0427 & 2.75\% & 0.8408 & 4.86\% \\
			30 & graph w/[Occupation, Zip]     & 1.0427 & 2.75\% & 0.8410 & 4.84\% \\
			31 & graph w/[Occupation]      & 1.0428 & 2.75\% & 0.8414 & 4.80\% \\
			32 & graph w/[Gender, Occupation, Zip]     & 1.0428 & 2.74\% & 0.8409 & 4.85\% \\
			33 & graph w/[Gender, Zip]     & 1.0431 & 2.72\% & 0.8411 & 4.83\% \\
			34 & graph w/[Gender]      & 1.0431 & 2.71\% & 0.8418 & 4.75\% \\
			35 & graph w/[Gender, Occupation]      & 1.0432 & 2.70\% & 0.8417 & 4.77\% \\
			36 & graph w/[Age, Gender, Occupation]     & 1.0433 & 2.70\% & 0.8417 & 4.77\% \\
			\rowcolor{LightGray}
			4 & Basic Features  & 1.0722 & N/A & 0.8838 & N/A \\
			%5 & User Features & 1.0895 & -1.61\% & 0.8967 & -1.46\% \\
			\bottomrule
		\end{tabular}
	%}
	\caption[Movielens - results summary table]{Performance of selected features combinations - Movielens dataset (baseline combination in light gray, rows are sorted by RMSE).}
	\label{tab:movielens2_feature_combinations_performance}%
\end{table}

\begin{figure}[ht!]
    \begin{center}
    \includegraphics[width=0.85\textwidth]{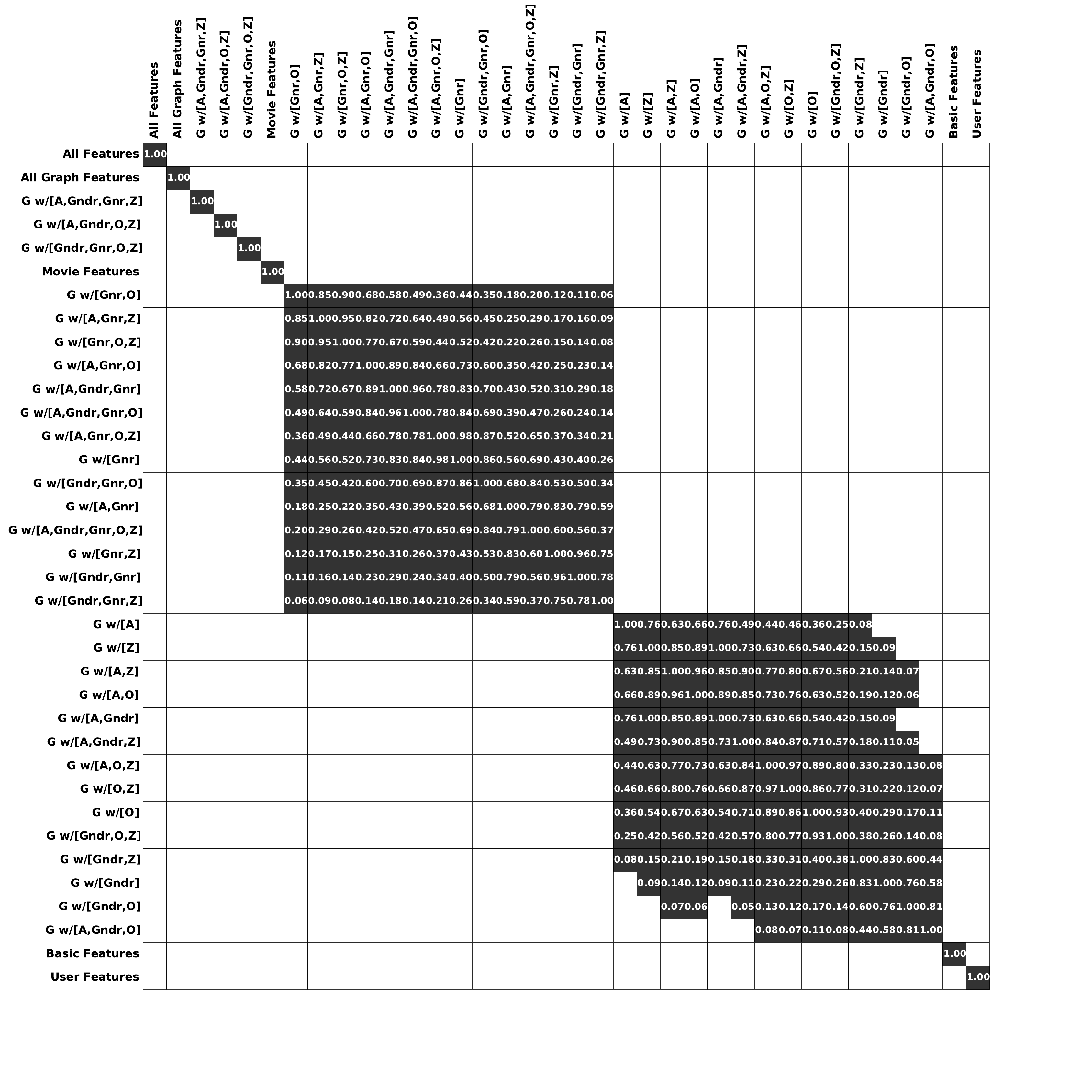}
   \end{center}
	\caption{Significance of the differences between feature combinations in the Movielens dataset.
            White cells - significant, dark cells - not significant, p-value given. ``G w/[]'' denotes
            sub-graphs that contain the listed entity types, where: A=Age, Gndr=Gender, Gnr=Genre, O=Occupation, and Z=Zip.}
    \label{fig:ml_ttest_significance}
\end{figure}

The Movielens dataset offered an even richer information about users and items and allowed for the
extraction of 32 sub-graph schemes. Only a small sample of these is illustrated in Figure \ref{fig:movielens_graph_schemes}.
The MAE and RMSE scores obtained for the 32 sub-graphs are listed in Table \ref{tab:movielens2_feature_combinations_performance}
and the significance test results are given in Figure \ref{fig:ml_ttest_significance}.
The sub-graphs are compared to the basic feature set and the union of all the graph features, which were
presented in Table \ref{tab:movielens_feature_combinations_performance} (rows numbered 2 and 4). The rows
corresponding to the various sub-graph representations are numbered 6 to 36.
For the sake of clarity, the sub-graphs are denoted by the entity types \textit{included} rather than excluded.
For example, ``graph w/[Age, Genre, Zip]'' denotes the sub-graph with the `Age', `Genre', and
`Zip' entities, which is identical to the complete graph with the `Occupation' and `Gender' entities excluded. In
Figure \ref{fig:ml_ttest_significance}, the names of the included entities are further abbreviated, as detailed
in the caption.

The significance test shows that the `genre' relationship in Movielens sub-graphs plays a similar role to
the ``category'' relationship in Yelp. Sub-graphs containing this relationship (rows 6 to 22) outperformed
those, where it was excluded (rows 23 to 36), and the differences between the groups are significant.
A common link between the `category' relationship in Yelp \rom{2} and the `genre' relationship in MovieLens
is that they both divide the item space -- be it businesses or movies -- into connected groups, which
affects values of the item features.
%Yelp's \rom{1} results, however, do not support this, because the ``Tripartite'' graph scheme features did
%not perform as well as the bipartite ones. The ``Tripartite'' scheme included relationships such as ``Location'' and
%``Categories'' that divided the items space too; perhaps, the distribution across values was the problem.}.
In agreement with previous results, the feature set that unifies all the graph features from all the
sub-graph schemes (``All Graph Features'', row 2) achieves the highest accuracy and outperforms any other feature
set. Again, this is attributed to the broad coverage of the proposed feature extraction mechanism, which
produces and aggregates promising feature combinations.

\subsubsection{Summary}
The purpose of this analysis was to analyze the differences driven by the
sub-graphs that are used for the feature extraction. To recap the results
obtained using the four datasets, the following was established.

\begin{itemize}
\item Features extracted from different graph schemes performed
differently, not following a certain pattern tied to the entities
or relationships included in the sub-graph. This means that it was
not possible to conclude which relationships lead to better results
if included in the graph. We posit that this is dataset-specific
and may be affected by additional factors, such as density of
a specific feature, distribution of its values, domain-specific
considerations, and so forth. This finding comes through in the
`category' and `genre' relationships in the Yelp \rom{2} and Movielens
datasets, but not in the Yelp \rom{1} dataset. Notably, the social
links had a major contribution in the Last.fm dataset, but not in the
Yelp \rom{2} dataset, possibly due to the sparsity of the latter.
\item Features extracted from the complete graph representations,
i.e., those containing all the relationships and entities in the dataset,
were not necessarily the best performing feature sets. A negative example
can be seen in the basic features of the BL+F+T schema in Figure \ref{fig:effect_on_graph_metrics}
that are dominated by the basic feature of BL+F and BL+T alike. Having said
that, the feature set that aggregated (that is, unified) all the graph features
from all the sub-graph schemes performed the best in the other three
scenarios in which it was evaluated: Yelp, Yelp \rom{2}, and Movielens.
We consider this to be a strong argument in favor of using the proposed
approach, as its exhaustive nature allows to cover a range of features and
necessarily uncover the most informative ones, as well as their best
combination.
\end{itemize}

The differences across the obtained results do not allow to generalize and determine a priori the best performing sub-graph
and feature set. Due to this, the suggested approach of generating sub-graphs, populating features from each of them, and
then aggregating the features in the feature sets is more likely to uncover the best performing feature combination.
%This approach is also supported by the statement in \citep{domingos2012few} \textit{``One way this is often done today
%is by automatically generating large numbers of candidate features and selecting the best by (say) their information gain with
% respect to the class''}, where the selection process in this study was embedded in the regressors used, e.g., Random Forest.
Note that this trades off with computational overheads and potential scalability issues in large-scale datasets
(discussed in detail in Section \ref{sec:discussion_computational_complexity}). We believe that future research may unveil
rating patterns or characteristics of datasets, which may predict the contribution of certain sub-graph, data entities, or
even types of features.

%\input{sec_case_study3}

%\section{Discussion}
\section{Discussion and Conclusions}
\label{ch:summary}

\subsection{Discussion}
\label{sec:sec_discussion}
The effectiveness of the graph-based approach for improving recommendations was demonstrated in the previous sections. It has been shown that precision and accuracy gains can be achieved by representing tabular data by graphs and extracting new features from them. This contrasts and complements prior approaches that improved recommendations by enhancing the recommendation techniques themselves. Also established are the benefits of the graph-based approach across recommendation domains, tasks, and metrics. These findings show that the graph representation exploits indirect latent links in the data, which lead to an improved recommendation accuracy. Finally, the approach is generic and it can be applied to many recommender system datasets.

The suggested process is automatic and can be run end-to-end, from data representation to feature extraction, without human intervention, unlike manual feature extraction methods, which are often time consuming and requires domain expertise. Using the proposed graph-based approach, rich features, based on intricate relationships between various data entities and sub-graph scheme variations, can be systematically extracted from a dataset. This allows for a better coverage of the features space with a considerable lower effort, as discussed in detail in Section \ref{sec:graph_based_feature_extraction}. In the following sub-sections, the key limitations and challenges encountered in the experiments and case studies are discussed.

\subsubsection{Overfitting}
\label{sec:discussion_overfitting}
Regarding concerns referring to possible
overfitting due to the newly generated features, as long as the
volume of available data greatly exceeds the number of extracted
features, there is little risk that the features will be the
cause of overfitting. The high diversity of unique data
characteristics can hardly be captured in full by a smaller subset of
features. Recommender system datasets tend to be in the medium to
large scale (tens of thousands to millions of data points), while the
number of features generated by the proposed approach is still in the
scale of tens to hundreds.

Additionally, machine learning methods such as Random Forests have
internal mechanisms for feature selection and can filter out
features that overfit. They do so by training on a sample of the
dataset and evaluating the performance of the features on the rest
of the data. A feature that performs well on the sample but
underperforms on the test data is ranked low. In the evaluations,
cross validation was used with at least N=5 folds, showing that
the models and features on which they are built are in fact
generalizable. Moreover, it was shown that in cases of sparse
data, which require a higher degree of generalization, the
graph features still outperformed other features.

\subsubsection{Scalability}
\label{sec:discussion_computational_complexity} A possible
disadvantage of the proposed approach is that some graph-based
computations, e.g., PageRank, are iterative and may take a long
time to converge. In the age of Big Data, recommender system
datasets are getting large and this limitation may become a hurdle.
The representation of the datasets results in large graphs and the
computation issue becomes a bigger problem. A general approach for
handling this issue in a deployed system would be to extract the
graph-based features offline, say, on a nightly basis, and use the
pre-computed values for real-time predictions. This may resolve the
problem under the reasonable assumption that the values do not change
substantially too frequently. Another means to overcome the computational
latency is through using a distributed graph feature computation library.
Such a library, e.g., Okapi\footnote{http://grafos.ml/okapi.html}, can
use distributed tools in order extract the graph features.

Another factor that adds to the computational complexity of the
approach is the exhaustive search for new features. It should be
noted that the complexity of the process of generating every possible
sub-graph and populating the matching feature combinations is
exponential. The number of relationships in current recommendation
datasets (as surveyed in Section
\ref{sec:recommender_systems_datasets}) is still manageable, and
can be accommodated by the proposed approach. However in the future,
with additional data sources being integrated for recommendation purposes,
this might become unsustainable and will require a long-term solution. Two
possible approaches for handling this issue are parallelization, e.g., each
sub-graph being processed by a different machine, and heuristics for
pruning less relevant sub-graph representations.

\subsubsection{Initial Transition to the Graph Model}
\label{sec:discussion_transition_to_graph_model}
Another possible disadvantage of graph-based features is the possible need for human intervention when generating the initial complete graph. Non-categorical feature values, e.g., income or price, may generate a large number of vertices, which would lead to a low connectivity of the graph, since not many users or items would share the exact value of the feature. This would lead to a very sparse graph and will need to be addressed by a manual intervention by a domain expert, who can determine how the non-categorical values can be grouped and categorized, e.g., by creating appropriate income or price buckets. A naive solution for this might be to attempt to auto-categorize such features based on the observed distribution of their values, e.g., first quarter, second quarter, and so on. This may, however, mask the differences between fine-grained groups and cause information loss.

Also to be acknowledged in this context is the historic human contribution that was required in order to conceive the graph methods exploited in this work for the generation of the various basic graph features: shortest path, degree, PageRank, etc. Indeed, these methods took a considerable amount of time and effort to evolve; however, they are reusable for generations and the overheads related to their development have been shared across many subsequent applications, while manually engineered features would usually not be highly reusable. Overall, when weighting the ease, quantity, and the possible contribution of the graph-based features to the accuracy of the generated recommendations against the above mentioned disadvantages, it can be concluded that it is worth to generate and populate such features, when designing a recommendation engine.

\subsection{Conclusions and Future Work}
\label{sec:sec_conclusions}

In this work, a new approach for improving recommendations was
presented and evaluated. Unlike many previous works, which focused on
addressing the recommendation problem by making improvements to
the recommendation algorithms, the presented approach does so by
suggesting a different way of looking at the dataset used for recommendation. It
proposed representing the datasets using graphs and then to extract
and populate new features from those graphs, all in a systematic fashion, and feed
the new features into existing recommendation algorithms. New features and
relationships that were not visible in the original tabular form can
be thus uncovered. In this manner, applying this approach may compliment
classical recommendation approaches and further enhance them.

The methodology, implementation, and analysis of the approach were
described in detail and the approach was evaluated from two main
perspectives: overall contribution to recommendations and impact
of various graph representations. The evaluation encompassed a
number of datasets, recommendation tasks, and evaluation metrics.
Furthermore, the datasets belonged to four application domains (movies,
music, businesses, and personal interests) that in part included metadata
and in part included social links. The recommendation tasks varied from
binary link predictions to star rating predictions. A number of
state-of-the-art classifiers and regressors were used for the generation
of the predictions. All in all, the presented evaluations examined the impact of the graph representations and showed
that the approach had a profound effect on the accuracy of the
recommendations.

The graph-based representation and features were shown to lead to
the generation of more accurate recommendations.
%The contributions to certain scenarios of high variability and high sparsity were also demonstrated. Finally,
The variations in performance across various graph schemes and the justification
for systematically extracting them, due to that, was established. The approach
presented was implemented in a library and is being provided as open source
software for the community to use and build on-top. Given such a
library, the cost of generating additional features that can
improve recommendations becomes substantially lower, in terms of
computation time and effort. It can be adopted as a natural
first resort, when given a dataset and recommendation task, or
as a complementary aid to enhance the standard manual feature
engineering.

The conducted evaluations suggested and demonstrated the potential of the proposed
approach in improving the recommendations by exploiting the benefits of links between
entities and characteristics of entities extracted from the graph representations.
Therefore, this work lays the foundations for further exploring how graph-based
features can enhance recommender systems and automatic feature engineering in the
more general context. Several variables were investigated in this work but many
more require additional attention. The following paragraphs identify several directions
of exploration, which were identified as possible research directions
in future works.
%Laying the foundations for engineering new features from a dataset by representing it as a graph opens a wide space of possibilities to be mapped. Three aspects that were identified during this study as possible next interesting routes to explore in future works follow.
\begin{itemize}
\item{Temporal Aspects}. Given a dataset that includs dated actions that are not sparse, the time aspect can be used to build a different type of graphs. Each graph will represent a snapshot in time and will either contain or exclude a link between vertices based on whether it was available in the dataset at that time. A combination of two temporally adjacent graphs will reflect the evolution of the data over that period of time. The main question in this setting is how such temporal graphs will affect the values of features extracted from them and how a recommender systems that use these features will perform in their respective recommendation tasks.
\item{Weighted and Labeled Graphs}. Several features in a dataset can be used to populate the edge labels when constructing the graph based representation. The labels, once set, can be taken into consideration in some graph features being extracted. One example would be to calculate a weighted PageRank score that will have jumps from a vertex to its neighbors based on a skewed probability correlated with the weight on the edge linking to the neighbor. This could lead to further improvement in the recommendations; however, this requires fine-tuning of initial weights on edges that do not naturally have them, e.g., social relationship edges in the Last.fm dataset.
\item{Directed Graphs}. Similarly, in cases where the direction of the edges can be important, the process can be extended to include this aspect by generating additional graph representations, with various combinations of the edge directions. For example, in one variant, edges will be directed from the source vertex to the target vertex, in another, in the opposite direction, and in a third one there will be no direction. This will guarantee coverage in terms of expressing the direction of the edges, and the performance of the features in the various scenarios can be evaluated.
\end{itemize}

The effects of these modifications on the scalability of the approach can be handled using the previously suggested methods,
either by scaling the computations (e.g., computing the features of each subgraph in a separate process), or using distributed
graph computation libraries, or identifying heuristics for pruning the feature and subgraph space.



%\begin{thebibliography}
\small
\bibliographystyle{apalike}
\bibliography{bibtex}

%\small
%\begin{thebibliography}{10}
%\bibliographystyle{elsarticle-harv}
%\providecommand{\url}[1]{{#1}}
%\providecommand{\urlprefix}{URL }
%\expandafter\ifx\csname urlstyle\endcsname\relax
%  \providecommand{\doi}[1]{DOI~\discretionary{}{}{}#1}\else
%  \providecommand{\doi}{DOI~\discretionary{}{}{}\begingroup
%  \urlstyle{rm}\Url}\fi

%\end{thebibliography}

\end{document}